\documentclass[usenatbib,psfig]{mn2e}

\usepackage{epsfig}
\usepackage{graphics}
\usepackage{url}
\usepackage[percent]{overpic}
\usepackage{xcolor}

\newcommand{\kms}{km~s$^{-1}$}
\newcommand{\Ha}{H$\alpha$}
\newcommand{\vrec}{V$_{rec}$}

\begin{document}


\title[Ca-rich and SN2002cx-like transients]{Environment-derived constraints on the 
progenitors of low-luminosity type I supernovae\thanks{Based on observations made with the Isaac Newton Telescope operated
on the island of La Palma by the Isaac Newton Group in the Spanish Observatorio
del Roque de los Muchachos of the Instituto de Astrofisica de
Canarias, observations made with the Liverpool Telescope operated on the
island of La Palma by Liverpool John Moores University in the Spanish
Observatorio del Roque de los Muchachos of the Instituto de Astrofisica de
Canarias with financial support from the UK Science and Technology Facilities
Council, and observations made with the 2.2m MPG/ESO telescope
at La Silla, proposal ID: 084.D-0195.}
}
\author[J. D. Lyman et al.]
{\parbox{\textwidth}{J. D. Lyman$^1$\thanks{E-mail:
jdl@astro.livjm.ac.uk }, 
P. A. James$^1$,
H. B. Perets$^2$,
J. P. Anderson$^3$,
A. Gal-Yam$^4$,
P. Mazzali$^1$
and
S. M. Percival$^1$}\vspace{0.4cm}\\
$^1$Astrophysics Research Institute, Liverpool John Moores University, Liverpool, L3 5RF, UK\\
$^2$Technion - Israel Institute of Technology, Haifa 32000, Israel\\
$^3$Departamento de Astronom\'ia, Universidad de Chile, Casilla 36-D,
Santiago, Chile\\
$^4$Benoziyo Center for Astrophysics, Weizmann Institute of Science, 76100 Rehovot, Israel\\
}

\date{Accepted . Received ; in original form }

\pagerange{\pageref{firstpage}--\pageref{lastpage}} \pubyear{2011}

\maketitle

\label{firstpage}

\begin{abstract}
We present a study of the properties of the host galaxies of unusual
transient objects of two types, both being sub-luminous compared with
the major classes of supernovae.  Those of one type exhibit unusually
strong calcium features, and have been termed `Ca-rich'. Those
of the second type, with SN2002cx as the prototype and SN2008ha as the 
most extreme example to date, have some
properties in common with the first, but show typically lower ejecta
velocities, and different early spectra.  We confirm important differences in the environments of
the two types, with the Ca-rich transients preferentially occurring in
galaxies dominated by old stellar populations.  Quantitatively, the
association of the the Ca-rich transients with regions of ongoing star
formation is well matched to that of type Ia supernovae.  The
SN2002cx-like transients are very different, with none of the present
sample occurring in an early-type host, and a statistical association
with star-formation regions similar to that of type II-P supernovae, and therefore 
a delay time of 30-50 Myrs. 
\end{abstract}


\section{Introduction}

Recent years have seen radical developments in the understanding of supernovae (SNe), driven by larger samples, higher quality and better-sampled spectroscopy and lightcurves, and better control of selection systematics through dedicated SN searches.  One result of this has been that the traditional empirically-motivated classification system of SNe has faced a series of challenges, with the finding of many transients that do not fit into any of the existing classifications.  Indeed, in the case of the transients that are discussed in the present paper, it is not yet clear whether they lie within the broad class of core-collapse supernovae (CCSNe), or should be considered as a subset of the type Ia SNe class (SNeIa), with long-lived progenitors and a final explosion mechanism involving a white dwarf (WD) primary. 

The aims of this paper are to constrain the progenitor systems of two of these putative new classes of supernova; one termed `Ca-rich' on the basis of the relative strength of calcium lines in spectra observed during the nebular phase \citep[also called `SN2005E-like' after the prototypical event;][]{pere10}, and another possibly related class that includes SNe 2002cx \citep{li03} and 2008ha, termed `SN2002cx-like'. In their overall spectral properties, the Ca-rich transients quite closely resemble CCSNe of type Ib (i.e. lacking hydrogen, but showing strong helium features) which led to the claim by \citet{kawa10} that one of the members of the class, SN2005cz, could indeed be a core-collapse object with a 10~M$_{\odot}$ zero-age progenitor.  This would be a surprising discovery, given that the host galaxy of SN2005cz, NGC~4589, is an elliptical galaxy with a `classical E2 morphology' \citep{sand94}, and a corresponding expectation of a predominantly old stellar population.  Simultaneously, the even more extreme environment of SN2005E, the prototypical member of the Ca-rich class that occurred far from the disc plane of an early type S0/a galaxy, NGC~1032, led \citet{pere10} to conclude that these explosions are likely to arise from the accretion of helium on to an old, low-mass progenitor, probably a WD. Modelling was used to show that such a progenitor can reproduce the observed properties, with ejecta that has high velocities but low masses, and a composition that is dominated by the products of helium burning. \citet{pere11} extended this analysis to SN2005cz in NGC~4589, again preferring a low-mass, long-lived progenitor, in contradiction to \citet{kawa10}. 

The spectroscopic and environmental properties of the general class of these Ca-rich transients have been investigated by \citet{pere10} and \citet{kasl12}. The former identified eight SNe in this group  (SN2000ds, SN2001co, SN2003H, SN2003dg, SN2003dr, SN2005cz, SN2005E and SN2007ke) and the latter identified three additional objects in this class from the Palomar Transient Factory survey \citep[henceforth PTF]{law09, rau09}. \citet{kasl12} combined these three new objects (PTF09dav, PTF10iuv (SN2010et) and PTF11bij) with two of the better observed earlier events (SN2005E and SN2007ke) which share common properties of low peak luminosities, fast photometric evolution, high ejecta velocities, strong Ca emission lines and locations in the extreme outskirts of their host galaxies.  They follow \citet{pere10, pere11} in preferring long-lived, low-mass progenitors, pointing out that the core-collapse objects with the lowest generally-accepted progenitor masses, those of type II-P, are almost never found at the extreme outlying locations that characterise these five Ca-rich events.  

\citet{vale13} have reported on another possible member of the Ca-rich class, SN2012hn, that was discovered by the Catalina Real-Time Transient Survey. This was initially classified as a peculiar type Ic supernova \citep{beni12}, but \citet{vale13} conclude from analysis of later spectroscopic and light-curve data that SN2012hn much more closely resembles members of the Ca-rich class, with a low peak luminosity and rapid evolution. This is supported by its location in the outskirts of an early type (E/S0) galaxy (discussed further in this paper).  However, it should be noted that \citet{vale13} find some detailed spectral differences between SN2012hn and other members of the Ca-rich class.

A very recent study by \citet{yuan13} has investigated the progenitors of the Ca-rich class by comparing their host galaxy locations to results from cosmological simulations. By comparison to the simulated metallicity distribution in hosts, they find the progenitors are likely to be of low metallicity and, tied with their remote locations compared to the bulk of the host stellar mass, consequently of old age ($\sim$10~Gyr). They conclude that a massive star origin for such events is disfavoured.

Some similarities exist between the Ca-rich events and the unusual transient SN2008ha \citep{vale09, fole09, fole10a}, in particular the extremely low luminousity and the inferred low ejecta-mass, and some similaritites in the late spectra. SN2008ha, however, does not show evidence for helium (it is classified as a SN type Ia event) and has extremely low photospheric velocity ($\sim$2000~km~s$^{-1}$ cf. 6000-11000~km~s$^{-1}$ for the Ca-rich transients). \citet{fole13} have recently linked SN2008ha and similar objects, including the prototypical example SN2002cx \citep{li03}, to a proposed new class of stellar explosion, that they term `Iax'.  These differ from normal type Ia SNe in having lower maximum-light ejecta velocities (2000-8000~km~s$^{-1}$) and lower peak luminosities for a given light-curve shape.   SN2008ha then appears as probably the most extreme object in this class identified to date, with the lowest peak luminosity, and ejecta velocities at the bottom end of the range for this class.  \citet{fole13} infer high rates, with $\sim$30 for every 100 SNeIa in the local Universe. Given the still debated/unknown origin of these events we will generally use the term `transients' rather than `supernovae' throughout this paper. 

Various models were suggested for the origin of these transients including complete thermonuclear deflagration of a WD \citep{li03,bra04}, failed detonation of a C/O WD \citep{jor12} or possibly a peculiar type of CCSN event \citep{vale09}.  \citet{fole13} suggested the progenitors to be C/O WDs that accrete material from a He-star, and therefore consider some possible connections between SN2002cx-like and Ca-rich transients, where both type of events arise from a He-shell detonation scenario.  However, one of the major differences between the two types is their environment, as first noted by \citet{pere10}. The Ca-rich events occur in all galaxy types (with a large fraction in early-type galaxies), and/or far from the centres of host galaxies \citep{kasl12}, whereas SN2002cx-like transients preferentially occur in late-type, star-forming galaxies, indicating a possibility for having younger progenitor systems.  \cite{fole13} suggest that the difference might originate from a different origin of the accreted He in the two cases, i.e. SN2002cx-like events arise from accretion from a He-rich non-degenerate donor star, whereas the Ca-rich events originate from accretion from a degenerate He-WD. 

\citet{vale09} discuss the class of SN2002cx-like events in general, and SN2008ha specifically, and conclude that these may be low-luminosity CCSNe, with progenitors that are either high-mass (25-30~M$_{\odot}$) Wolf-Rayet stars, or stars from the low-mass limit of CCSNe (7-9~M$_{\odot}$). However, \citet{eldr13} have recently discussed SN2008ha in the context of a study of the rates of CCSNe, and on the balance of evidence decide in favour of a thermonuclear interpretation. They thus exclude it from their study, although they warn that the evidence is far from conclusive, and that further study of SN2008ha and other SN2002cx-like transients is clearly required.

It is clear from the above discussion that the association with different types of stellar environment is of key importance in distinguishing between these different types of luminous transients, and in constraining the possible progenitor objects.  However, much of the environmental information, e.g. the association of the Ca-rich transients with old populations and SN2002cx-like transients with young, lacks quantification and in many cases is little more than anecdotal.  Host galaxy classifications give some useful information, but they are notoriously subjective and, even if free from actual errors, they do not give precise information on the stellar population at the location of the transient event.  For example, even a late-type spiral may have a bulge, or extreme outer disc, that is entirely composed of old stars. In this paper, we will make use of both host galaxy types and quantified measures of star-formation activity, local to the sites of events within their host galaxies, applied specifically to the known samples of Ca-rich and SN2002cx-like transients, to determine whether they appear to rise from the same progenitor populations, and to compare these populations with the same measures for other types of SN (including `normal' SNeIa, and core collapse types SNe Ib, Ic and II-P). 

\section{Methods}
\label{sect:methods}
The methods we employ are explained in detail in \citet{jame06}; these have been previously applied to large samples of supernovae in two subsequent papers \citep{ande08, ande12}.  The last of these three papers provides the main comparison sample for the current work.

Following previous work, each transient is assigned a normalised cumulative rank (NCR), based on pixel statistics of a continuum-subtracted \Ha{} image of the host (taken either prior to, or long after the transient), as a measure of the degree of association of the transient with recent star formation within its host.

The continuum subtracted \Ha{} images are trimmed to contain the host and transient location and then binned $3\times3$ such that the pixel location of the transient given by the WCS forms the centre of a $3\times3$ `super-pixel'. A pixel in our binned images represents $\sim0.9$ arcsecond across the various instruments used, or $\sim260$ pc at the mean galaxy distance. Star residuals and artefacts arising from saturation in the subtracted images are masked using a local median. Pixel values in this binned image are sorted, cumulatively summed and then normalised by the total sum of pixel values. In this way each pixel now has an associated NCR value between 0 and 1 (any negative values are set to 0). Any pixel with NCR~$= 0$ is considered a background pixel, i.e. there is no \Ha{} flux at that position. Positively valued pixels are then ranked within the NCR method such that low values have an association with weak emission, and high values are coincident with the brightest \Ha{} emitting regions of the host. Specifically, the NCR value is the fraction of host galaxy flux that is below the level of flux at the location of the transient, i.e. NCR~$= 1$ means the transient location is at the site of the most intense star-formation activity within its host galaxy.

Using these methods, \citet{ande12} find a clear separation of the CCSN subtypes, with types  II-P, Ib and Ic forming a clear sequence of increasing strength of association with current sites of star formation, and high mean NCR values.  This is most simply interpreted in terms of a sequence of increasing mean progenitor mass, and hence decreasing progenitor lifetime.  

\citet{crow13} has looked at the progenitor constraints that can be drawn from association of SNe with ongoing star formation, using a smaller sample than \citet{ande12} with higher spatial resolution, and employing rather different statistical methods based on distance to the nearest region of H$\alpha$ emission.  \citet{crow13} finds very similar results to \citet{ande08} and \citet{ande12} in terms of the difference of strength association between SNeII and SNeIbc, which he interprets in terms of a large fraction of SNeII outliving their natal star-formation regions.  \citet{crow13} argues that the complications involving lack of resolution of individual SF regions should obscure any differences between the correlation strengths for shorter-lived, higher-mass progenitors than those of the SNeII, but this argument seems hard to reconcile with the clear statistical differences found for the populations of SNe Ib and Ic investigated by \citet{ande12}.

\Ha{} was chosen as a star formation tracer since there already exists large samples of NCR values for the more common SN types which can be compared to. The typical duration of \Ha{} emission from HII regions is comparable to that of the ages of the middle-to-lower mass end of CCSNe. \citet{kun13} show the evolution of the \Ha{} equivalent width for a single burst in Starburst99, which weakens strongly after 5 Myr, falling to very low values after $\sim$15 Myr (roughly the lifetime of a 14 M$_\odot$ star). This, however, is a lower limit since a typical star formation region will not form stars in a delta-function manner. \Ha{} imaging thus allows us, through the NCR method, to distinguish between transients whose progenitor ages fall entirely within, or overlap with, this limit. Since each transient's NCR value is normalised to its own host, we are not sensitive to absolute calibtration issues of \Ha{} as a star-formation rate tracer \citep{lee09,bott12}.

The NCR method is particularly reliant on the \Ha{} filter used for observations. Its transmission profile must allow for detection of \Ha{} over a reasonable velocity range so as to detect all host galaxy emission, whilst being narrow enough to allow for accurate subtraction of the underlying continuum light. Clearly, if a filter fails to transmit \Ha{} emission from some regions of the host, this will affect the NCR value of the transient.
As such, transients that are potentially well separated from their hosts in recession velocity (\vrec) provide a problem of filter choice, especially when \vrec{} cannot be determined for the transient itself. In the present study, for all cases except PTF09dav, the filter with a central wavelength best matching the host-\Ha{} wavelength was chosen; for PTF09dav, the redshift of the transient was used to find the best matched filter as its host is anonymous. Given the widths of the filters (typically $\sim$2000--3000~\kms), this meant \Ha{} over a broad range of host velocities would be detected, giving confidence that we are not missing some regions of \Ha{} emission in the host or, importantly, at the location of the transient.

\begin{table*}
 \centering
 \begin{minipage}{140mm}
  \caption{\Ha{} narrow-band filter properties.}
  \begin{tabular}{llcc}
  \hline
Filter name  & Telescope & Wavelength limits & \vrec{} limits \\
             &           & \AA               &  km/s                \\      
\hline            
`Halpha'     & INT      & 6522--6614  & $-$1865--2357     \\
`Ha 6657'\footnote[1]{No scanned transmission profile is available for this filter so the limits are based on manufactured specification.}    & INT      & 6618--6697  & 2400--6100       \\
`H-alpha-100' & LT:RATCam & 6517--6617 & $-$2093--2478      \\
`Ha\_6566'   & LT:IO    & 6522--6610  & $-$1865--2164       \\
`Ha\_6634'   & LT:IO    & 6608--6662  & 2080--4520    \\
`Ha\_6705'  & LT:IO    & 6680--6733  & 5349--7764     \\ 
`Ha\_6755'  & LT:IO    & 6729--6783  & 7595--10047    \\ 
`Ha\_6822'   & LT:IO    & 6798--6849  & 10747--13097  \\
`665/12'     & MPI-2.2  & 6598--6713  & 1616--6857    \\
\hline
\end{tabular}
\label{tab:filter_props}
\end{minipage} 
\end{table*}

\section{Transient samples and observations}

The samples of transients analysed here are inevitably somewhat eclectic and subject to selection biases, and thus cannot be considered in any sense to represent a statistically complete sample of objects of either type.  This is unavoidable for classes of transient objects that are both relatively rare (although the global rates are highly uncertain) and substantially fainter than the main SN types.  Thus, in order to compile the samples of Ca-rich and SN2002cx-like transients presented here, we have used a variety of sources.  Most of the Ca-rich transients are listed in \citet{pere10} and \citet{kasl12}, alongside SN2012hn \citep{vale13}. For a complete recent compilation of the SN2002cx-like transients, see \citet{fole13}.

We stress here that although we are investigating two classes of transients, their unknown nature, and the lack of detailed observations for some, means that there is potential contamination in each sample by transients of different origin or potential for diversity within the each sample. We will discuss progenitor constraints for each sample as a whole since we are already limited by small numbers, however it may be true that some specific events differ from these conclusions due to their erroneous classification. 

New imaging observations presented here were made using the Isaac Newton Telescope (INT) and Liverpool Telescope (LT) at La Palma and the MPI2.2 at ESO. For each transient, exposures were taken in the R band, to characterise the continuum light, and a narrowband \Ha{} filter. Details of the \Ha{} filters used are given in Table \ref{tab:filter_props}, where wavelength and corresponding \vrec{} limits are defined as the 50 per cent transmission limits of the filter. Exposure times were 300 seconds for R band and 900 seconds for \Ha{}, this corresponds to a limiting \Ha{} flux of $\sim3.8\times10^{-16}$~erg~s$^{-1}$~cm$^{-2}$ (see \citealt{ande12} for a discussion of star formation limits using this method). Images taken with the LT were reduced using the automated pipeline; standard bias and overscan subtraction and flat fielding was performed for other data. Typical seeing was 1--2 arcseconds (see Tables \ref{tab:carich_obs} and \ref{tab:2008ha_obs}). Subtraction of the R band images from the \Ha{} images was performed using a version of the ISIS code \citep{alar00}.

Data for the Ca-rich and SN2002cx-like transients in the present study are given in Tables \ref{tab:carich_props} and \ref{tab:2008ha_props}, respectively. These list the International Astronomical Union (IAU) supernova name for all transients except PTF09dav and  PTF11bij, which are not on the IAU list; the host galaxy name, classification and recession velocity from the NASA Extragalactic Database (NED)\footnote{\url{http://ned.ipac.caltech.edu/}}, the absolute discovery magnitude (taken from the Asiago Supernova Catalog\footnote{\url{http://heasarc.gsfc.nasa.gov/W3Browse/all/asiagosn.html}}, using distance modulus values for the host taken from NED); and the classification of the supernovae from the IAU database.

Details of the observations and NCR values are given in Table \ref{tab:carich_obs} for the Ca-rich transients and Table \ref{tab:2008ha_obs} for SN2002cx-like transients. Velocity limits from Table \ref{tab:filter_props} are shown for the \Ha{} filter used --- the bulk of the detected light in our continuum-subtracted images will come from emission within these velocity limits (although the filters also have non-negligible transmission for a few hundred \kms{} outside these limits). Whether we detect any recent star formation (i.e. \Ha{} emission) in our observations is also noted.

Images of the of the twelve Ca-rich hosts are shown in Figure \ref{fig:carich_imgs}, showing the R band and continuum-subtracted \Ha{} exposures with the location of the transient marked. Of these, six (NGC~2768, NGC~1032, NGC~4589, NGC~1129, IC~3956 and NGC~2272) are early-type galaxies, and hence should have no recent star formation. Indeed, we find no star formation as traced by \Ha{} at the location of the transients in these early hosts or anywhere else in the hosts. The only apparent emission in the subtracted images arises from the very centre of these galaxies; due to the difficulties in obtaining a clean subtraction on such extremely bright regions, this is most likely to be artefacts arising from the image subtraction process and saturation effects rather than real \Ha{} flux, although we cannot rule out either conclusively. It is not clear which galaxy hosted the very isolated transient SN2010et, as we discuss below. The remaining five hosts all display varying levels of star formation.

Figure \ref{fig:2008ha_imgs} shows the corresponding images for the SN2002cx-like sample. All these hosts are late type, and all display strong ongoing star formation with prominent HII regions. 

Further discussion of the hosts of the two samples is given in Sect. \ref{sect:hostdiscuss}.

As a check on the presence and nature of emission lines at the locations of these events, we also obtained long-slit optical spectroscopy (with a slit width of 1.5$^{\prime\prime}$) of two of the host galaxies in our samples (Fig. \ref{fig:2000ds2003H_spec}).  The observations were taken on the INT in January 2013 using the IDS spectrograph with the R632V grating.  The slit was positioned to include both the galaxy nucleus, as a positional reference, and the location of the transient. The spectral range covered included the location of any potential \Ha{} emission. NGC~2768 (SN2000ds) was observed at an airmass of 1.3 in seeing of 1.8$^{\prime\prime}$, the corresponding values for NGC~2207 (SN2003H) were 1.57 and 0.8$^{\prime\prime}$.

\begin{table*}
 \centering
 \begin{minipage}{140mm}
  \caption{Properties of the Ca-rich transients and their host galaxies.}
  \begin{tabular}{lllccl}
  \hline
SN name  & Host galaxy & Host type        & \vrec{} & Discovery Abs. mag & IAU classn. \\
         &             &                  & (\kms{})  & (unfiltered mag) & \\
\hline            
2000ds   & NGC~2768	 & E6             & 1373  & $-13.59$ & Ib/c \\
2001co   & NGC~5559      & SBb            & 5166  & $-15.69$ & Ib/c \\
2003H    & NGC~2207      & SABbc          & 2741  & $-14.16$ & Ib/c \\
2003dg   & UGC~6934      & Scd (edge-on)  & 5501  & $-15.31$ & Ib/c \\
2003dr   & NGC~5714      & Scd (edge-on)  & 2237  & $-15.06$ & Ib/c \\
2005E    & NGC~1032      & S0/a (edge-on) & 2694  & $-15.86$ & Ib/c \\
2005cz   & NGC~4589      & E2             & 1980  & $-16.36$ & Ib   \\
2007ke   & NGC~1129      & E              & 5194  & $-15.71$ & Ib   \\
PTF09dav & Anon          & Sb\footnote{Classified by PAJ based on our imaging}             & 11123 & $-14.7$ & --   \\
2010et & Uncertain       & --             &       & $-13.8$  & --   \\
PTF11bij & IC~3956       & E       & 10406 & $-15.9$\footnote{$M_R$ at discovery taken from \citet{kasl12}} & --  \\
2012hn   & NGC~2272	 & SAB0           & 2130  & $-16.0$\footnote{$M_R$ at peak taken from \citet{vale13}} & I-p \\
\hline
\end{tabular}
\label{tab:carich_props}
\end{minipage} 
\end{table*}

\begin{table*}
 \centering
 \begin{minipage}{210mm}
  \caption{Observations of the host galaxies of Ca-rich transients.}
  \begin{tabular}{llcccccccc}
  \hline
SN name  & Host galaxy & \vrec{} & Telescope & Obs. date & Seeing       & Filter name & \Ha{} range & NCR index & SF detected \\
         &             & (\kms{})&           &           & (arcseconds) &            & (\kms{})    &           & in host?      \\      
\hline            
2000ds	 & NGC~2768      & 1373  & INT       & Jan 2012 & 1.6  & `Halpha'  &  $-$1865--2357      & 0.000  & No \\
2001co   & NGC~5559      & 5166  & INT       & Mar 2007 & 1.6  & `Ha 6657' &  2400--6100   & 0.357  & Yes \\
2003H    & NGC~2207      & 2741  & LT:IO     & Sep 2012 & 1.8  & `Ha\_6634'&  2080--4520   & 0.312  & Yes \\
2003dg   & UGC~6934      & 5501  & INT       & Jan 2012 & 1.3  & `Ha 6657' &  2400--6100   & 0.626  & Yes \\
2003dr   & NGC~5714      & 2237  & INT       & Jan 2012 & 1.6  & `Halpha'  &  $-$1865--2357      & 0.000  & Yes \\
2005E    & NGC~1032      & 2694  & LT:IO     & Jan 2013 & 1.3  & `Ha\_6634'&  2080--4520   & 0.000  & No \\
2005cz   & NGC~4589      & 1980  & INT       & Jan 2012 & 1.2  & `Halpha'  &  $-$1865--2357      & 0.000  & No \\
2007ke   & NGC~1129      & 5194  & INT       & Jan 2012 & 1.7  & `Ha 6657' &  2400--6100   & 0.000  & No \\
PTF09dav & Anon          & 11123 & LT:IO     & Dec 2012 & 1.6  & `Ha\_6822'&  10747--13097 & 0.000  & Yes \\
2010et & Uncertain       & --    & LT:IO     & Mar 2013 & 2.9  & `Ha\_6705'&  4900--7640   & 0.000  & --  \\
PTF11bij & IC~3956       & 10406 & LT:IO     & Jan 2013 & 3.0  & `Ha\_6822 &  10747--13097 & 0.000  & No  \\
2012hn   & NGC~2272      & 2130  & LT:IO     & Feb 2013 & 2.4  & `Ha\_6566'&  $-$1865--2164      & 0.000  & No \\
\hline
\end{tabular}
\label{tab:carich_obs}
\end{minipage} 
\end{table*}

\begin{table*}
 \centering
 \begin{minipage}{140mm}
  \caption{Properties of the SN2002cx-like transients and their host galaxies.}
  \begin{tabular}{lllccl}
  \hline
SN name & Host galaxy  & Host type    & \vrec{}    & Discovery Abs. mag &  IAU classn. \\
        &              &              & (\kms{})   & (unfiltered mag)   &            \\
\hline            
1991bj	& IC~344       & SBcd\footnote{Classified by PAJ based on our imaging}        &  5440       & $-15.46$  & Ia         \\
2004gw  & CGCG~283-003 & Sab\footnotemark[1]& 5102  & $-16.33$  & Ia         \\
2005P   & NGC~5468     & SABcd        &  2842       & $-15.14$  & ?          \\
2005cc  & NGC~5383     & SBb pec      & 2270        & $-15.18$  & ?          \\
2005hk  & UGC~272      & SABd         & 3895        & $-17.05$\footnote{$M_R$ at discovery taken from \citet{phil07}} & Ia-p  \\
2006hn  & UGC~6154     & SBa          &  5156       & $-18.69$  & Ia         \\
2007J   & UGC~1778     & Sdm          &  5034       & $-15.92$  & Ia         \\
2008A   & NGC~634      & Sa (edge-on) &  4925       & $-16.57$  & Iap        \\
2008ha  & UGC~12682    & Im           &  1393       & $-12.7$\footnote{\citet{puck08}}  & Ia?        \\
2009J   & IC~2160      & SBc pec      &  4739       & $-16.17$  & Ia-p       \\
2012Z   & NGC~1309     & SAbc         &  2136       & $-14.62$  & Ia-p       \\

\hline
\end{tabular}
\label{tab:2008ha_props}
\end{minipage} 
\end{table*}

\begin{table*}
 \centering
 \begin{minipage}{210mm}
  \caption{Observations of the host galaxies of SN2002cx-like transients.}
  \begin{tabular}{llcccccccc}
  \hline
SN name  & Host galaxy & \vrec{} & Telescope & Obs. date & Seeing       & Filter name & \Ha{} range & NCR index & SF detected\\
         &             & (\kms{})&           &           & (arcseconds) &            &(\kms{})      &           &  in host?  \\  
\hline            
1991bj  & IC~344       &  5440     & INT       & Jan 2012  & 1.9   & `Ha 6657' & 2400--6100 &  0.163     & Yes         \\
2004gw  & CGCG~283-003 &  5102     & INT       & Jan 2012  & 1.7   & `Ha 6657' & 2400--6100 &  0.000     & Yes         \\
2005P   & NGC~5468     &  2842     & INT       & Feb 2008  & 1.4   & `Ha 6657' & 2400--6100 &  0.055     & Yes         \\
2005cc  & NGC~5383     &  2270     & LT:RATCam & Dec 2005  & 1.8   & `H-alpha-100' & $-$2093--2478&  0.621     & Yes         \\
2005hk  & UGC~272      &  3895     & LT:IO     & Oct 2012  & 1.1   & `Ha 6634' & 2080--4520 &  0.000     & Yes         \\
2006hn  & UGC~6154     &  5156     & INT       & Jan 2012  & 1.3   & `Ha 6657' & 2400--6100 &  0.289     & Yes         \\
2007J   & UGC~1778     &  5034     & INT       & Jan 2012  & 1.1   & `Ha 6657' & 2400--6100 &  0.904     & Yes         \\
2008A   & NGC~634      &  4925     & INT       & Jan 2012  & 0.9   & `Ha 6657' & 2400--6100 &  0.000     & Yes         \\
2008ha  & UGC~12682    &  1393     & LT:IO     & Oct 2012  & 1.9   & `Ha\_6566'&  $-$1865--2164   &  0.407     & Yes         \\
2009J   & IC~2160      &  4739     & MPI2.2    & Feb 2010  & 1.9   & `665/12'  &  1616--6857&  0.000     & Yes         \\
2012Z   & NGC~1309     &  2136     & LT:RATCam & Aug 2009  & 1.5   & `H-alpha-100' & $-$2093--2478&  0.000     & Yes         \\  
\hline
\end{tabular}
\label{tab:2008ha_obs}
\end{minipage} 
\end{table*}

\linethickness{0.05cm}
\begin{figure*}
  \begin{center}
  \begin{overpic}[width=0.5\columnwidth]{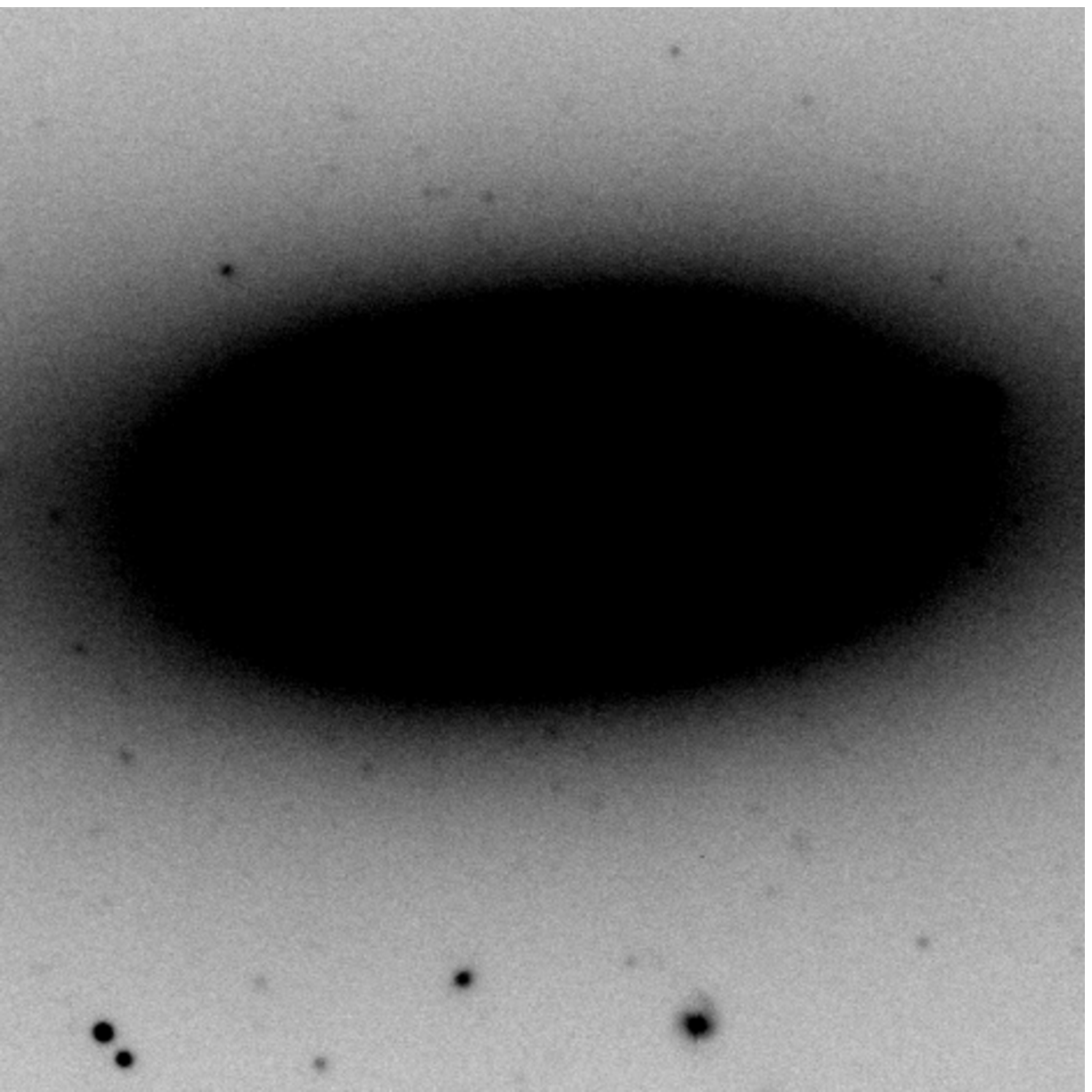}
  \put(10,10){\color{blue}\line(1,0){20}}
  \put(10,12){\color{blue}2.9}
  \end{overpic}
  \begin{overpic}[width=0.5\columnwidth]{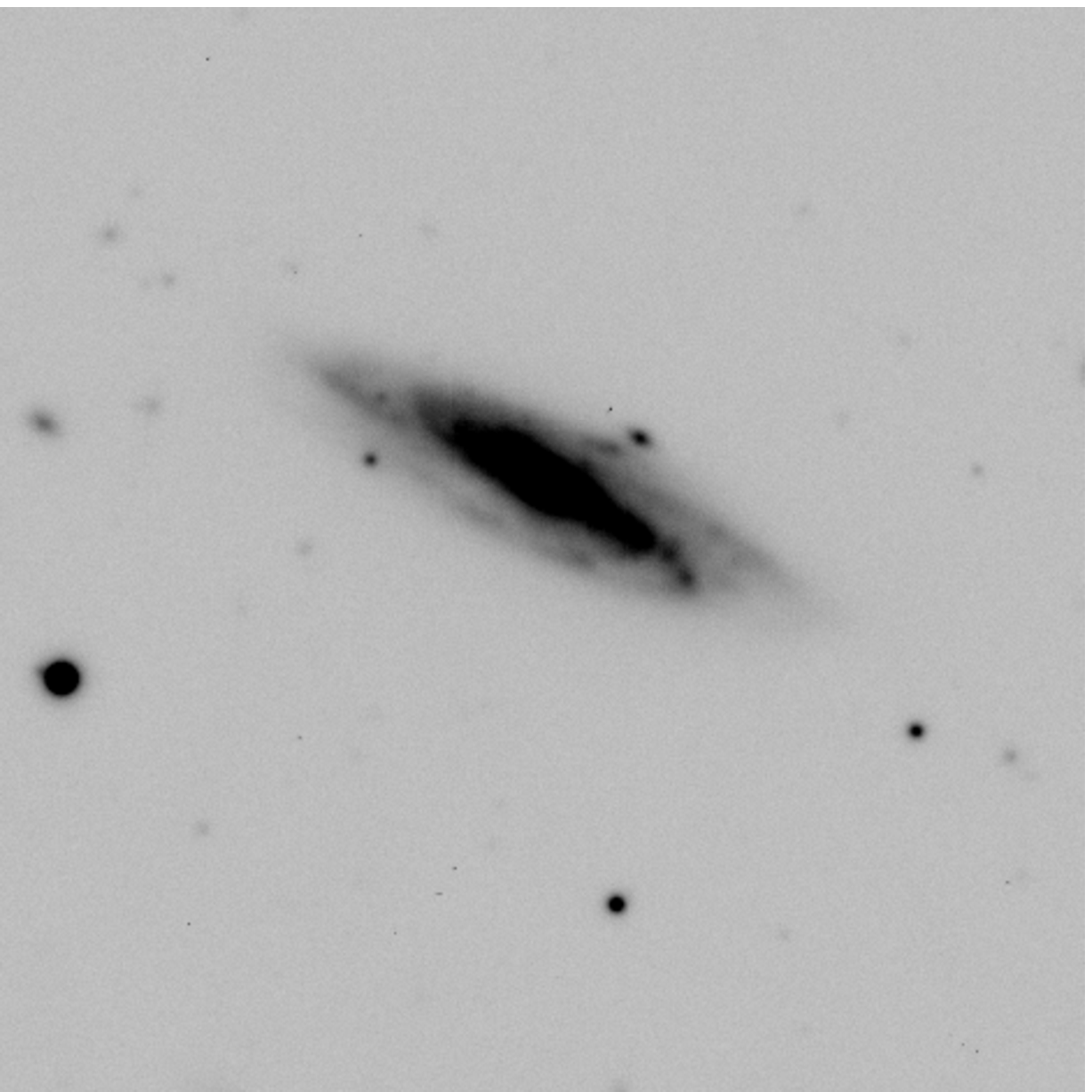}
  \put(10,10){\color{blue}\line(1,0){20}}
  \put(10,12){\color{blue}10}
  \end{overpic}
  \begin{overpic}[width=0.5\columnwidth]{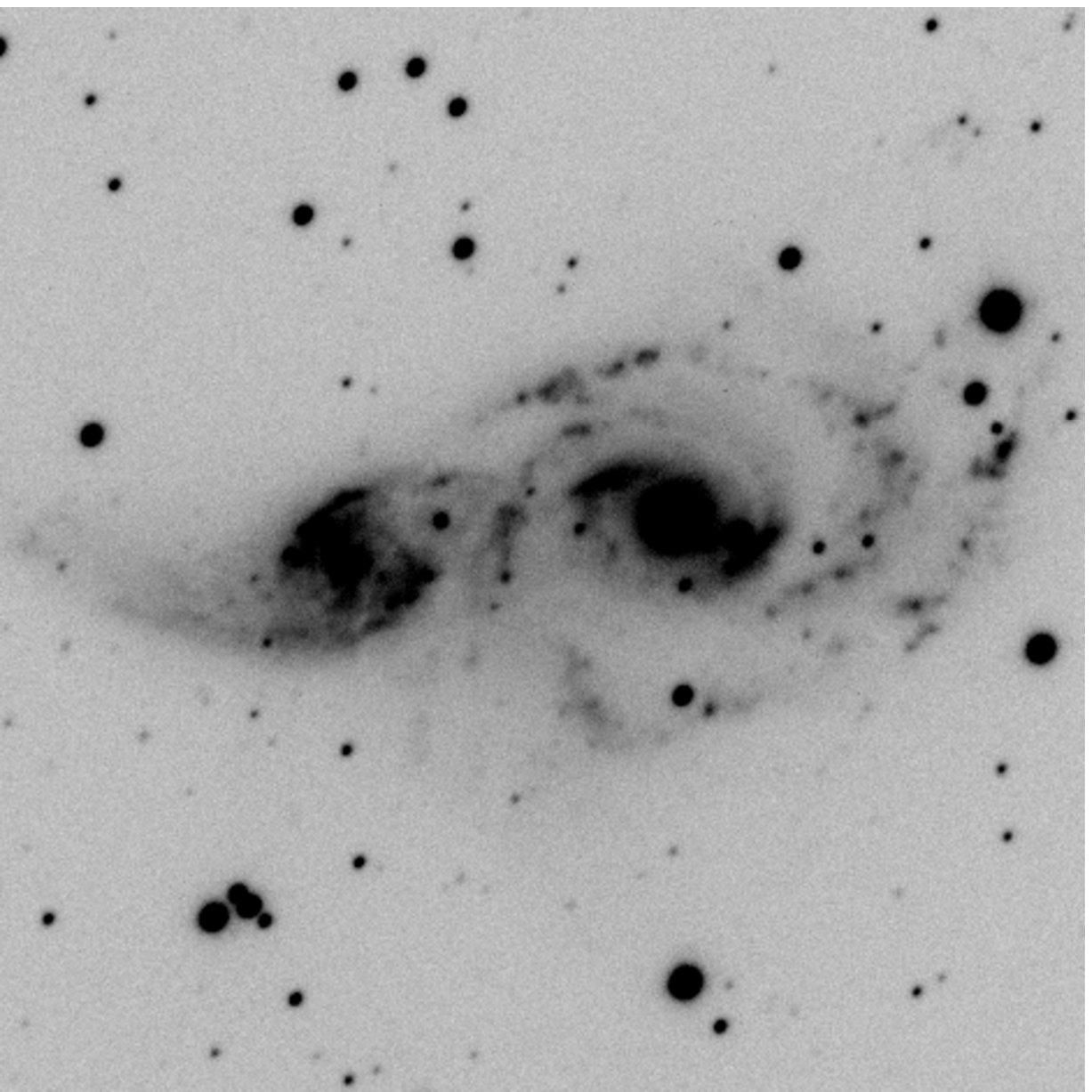}
  \put(10,10){\color{blue}\line(1,0){10}}
  \put(10,12){\color{blue}3.8}
  \end{overpic}\\
  \begin{overpic}[width=0.5\columnwidth]{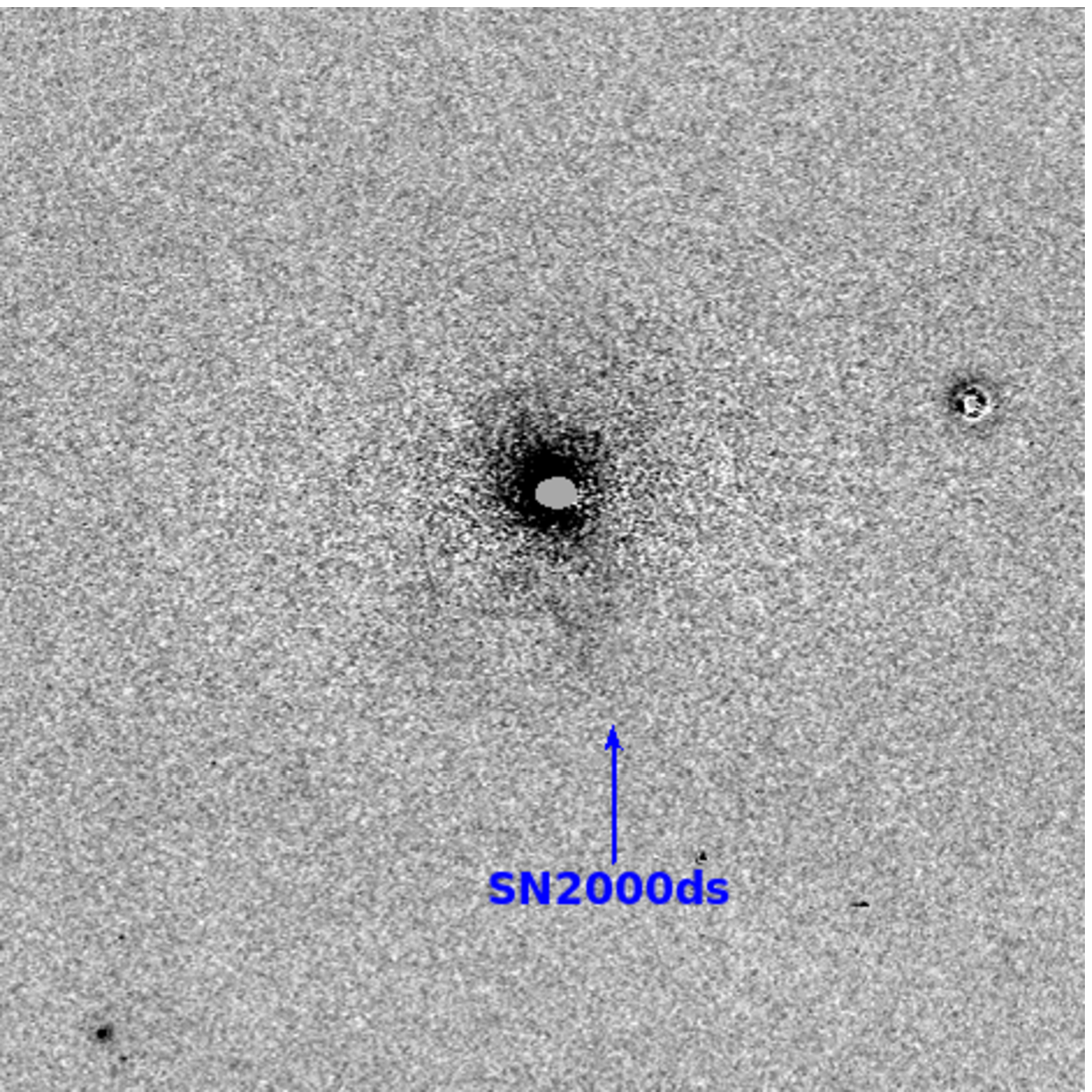}
  \end{overpic}
  \begin{overpic}[width=0.5\columnwidth]{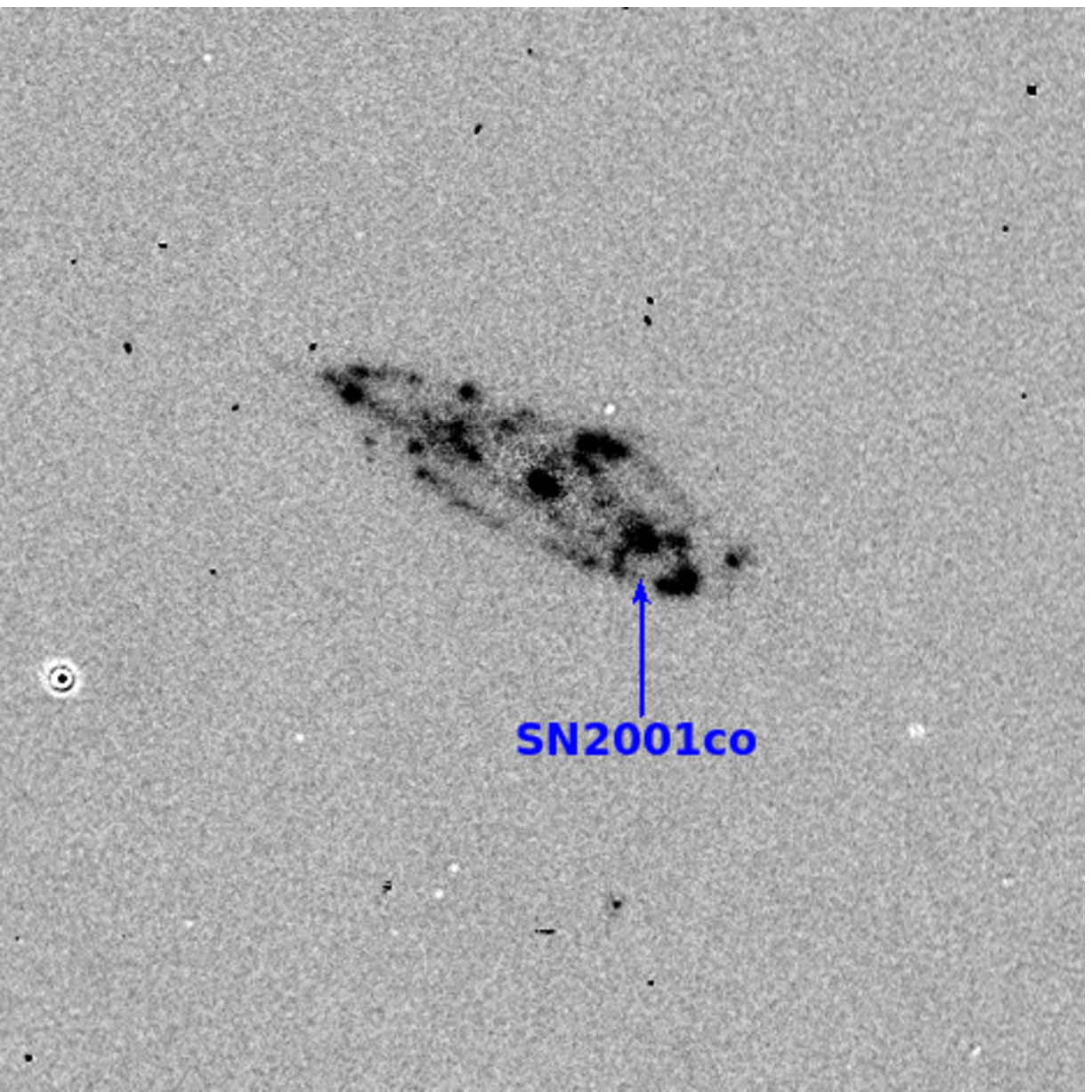}
  \end{overpic}
  \begin{overpic}[width=0.5\columnwidth]{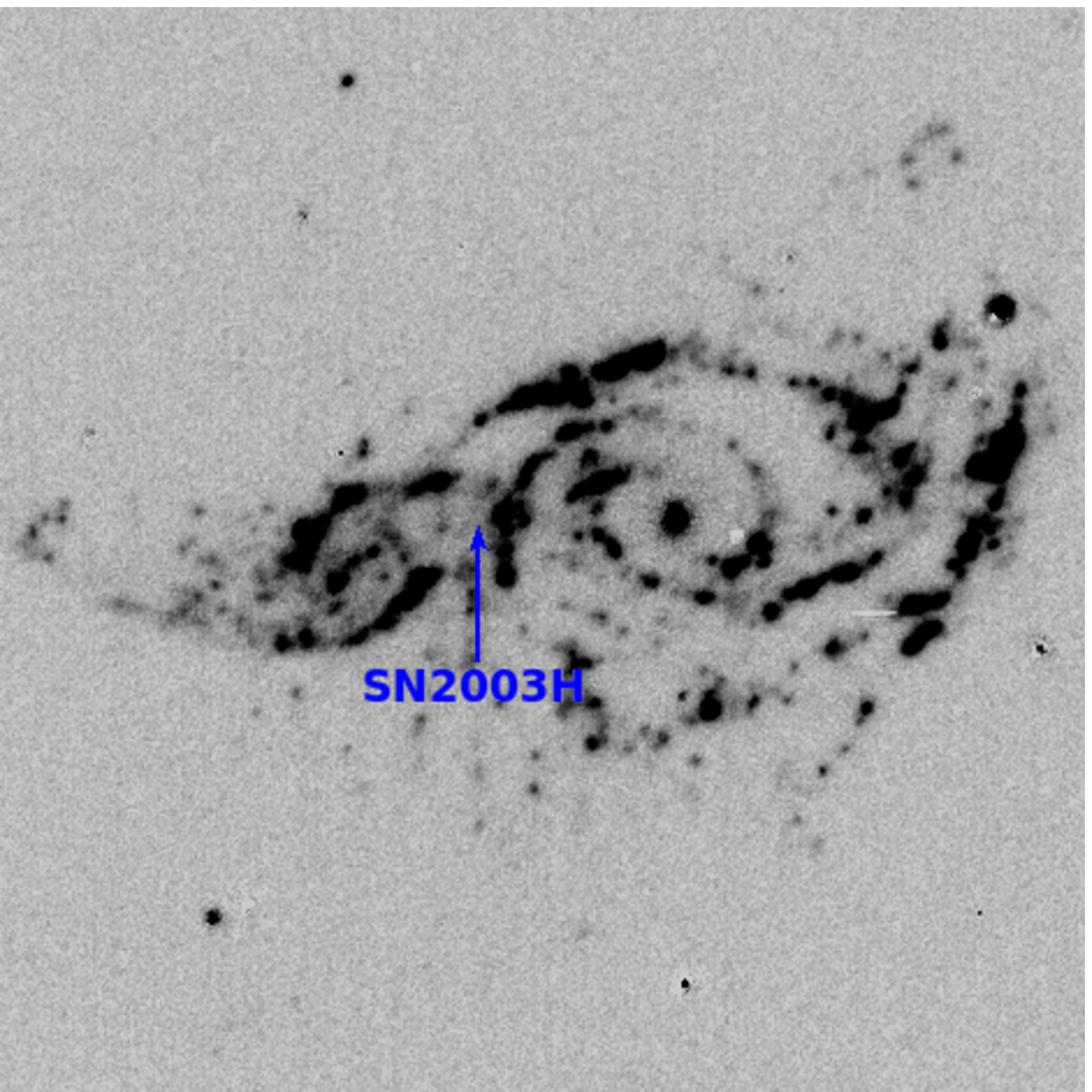}
  \end{overpic}\\
   \vspace{0.2cm}
  \begin{overpic}[width=0.5\columnwidth]{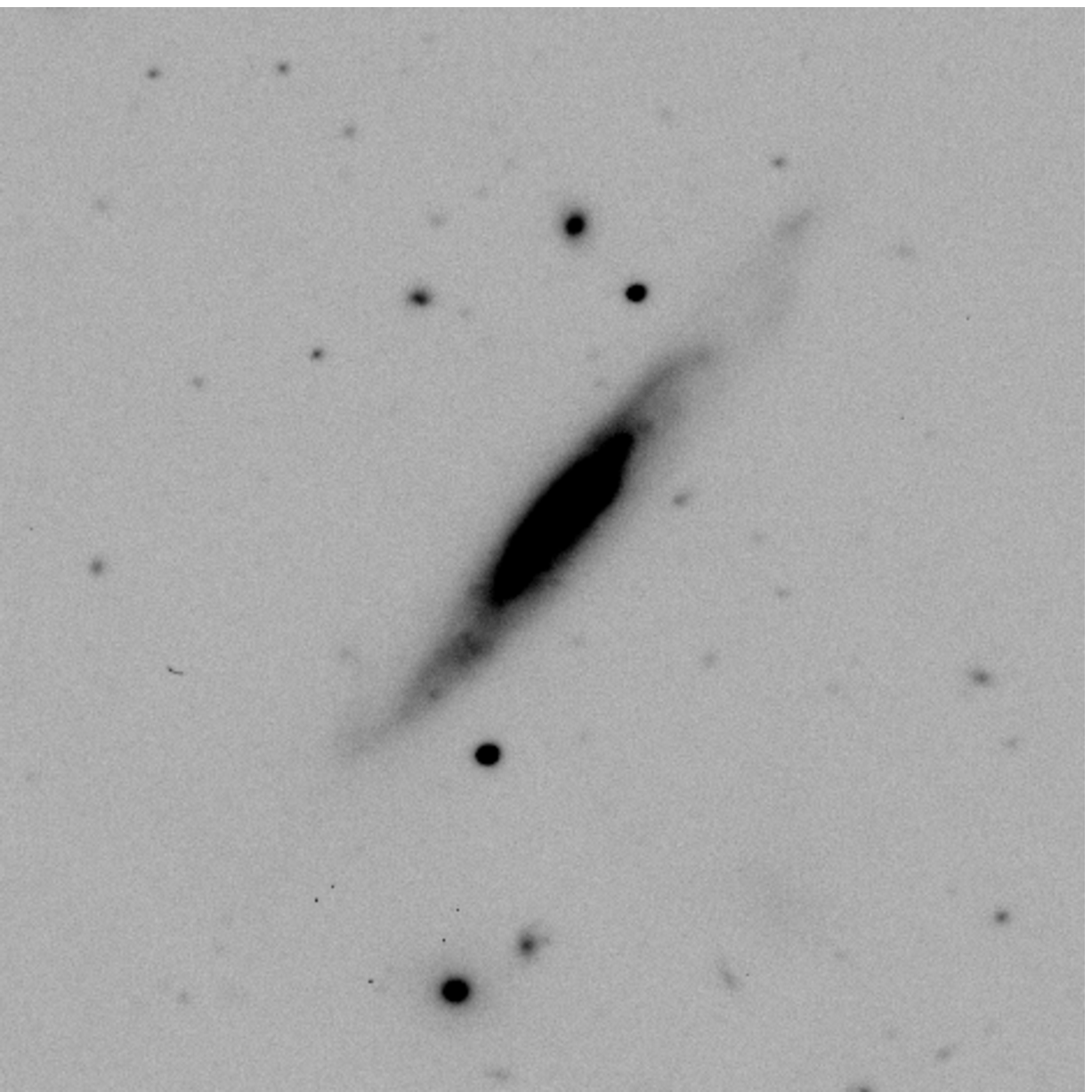}
  \put(10,10){\color{blue}\line(1,0){20}}
  \put(10,12){\color{blue}12.2}
  \end{overpic}
  \begin{overpic}[width=0.5\columnwidth]{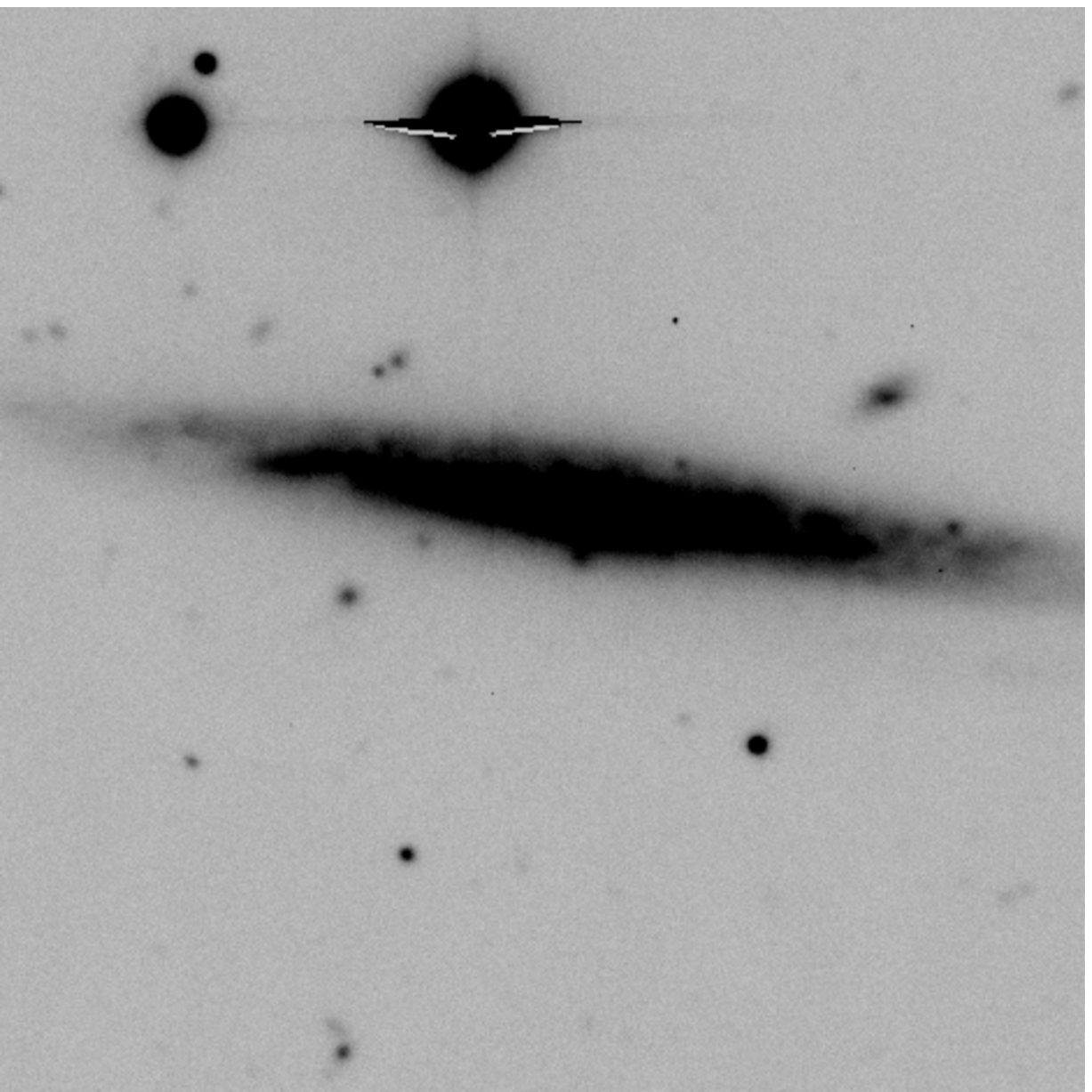}
  \put(10,10){\color{blue}\line(1,0){20}}
  \put(10,12){\color{blue}5.9}
  \end{overpic}
  \begin{overpic}[width=0.5\columnwidth]{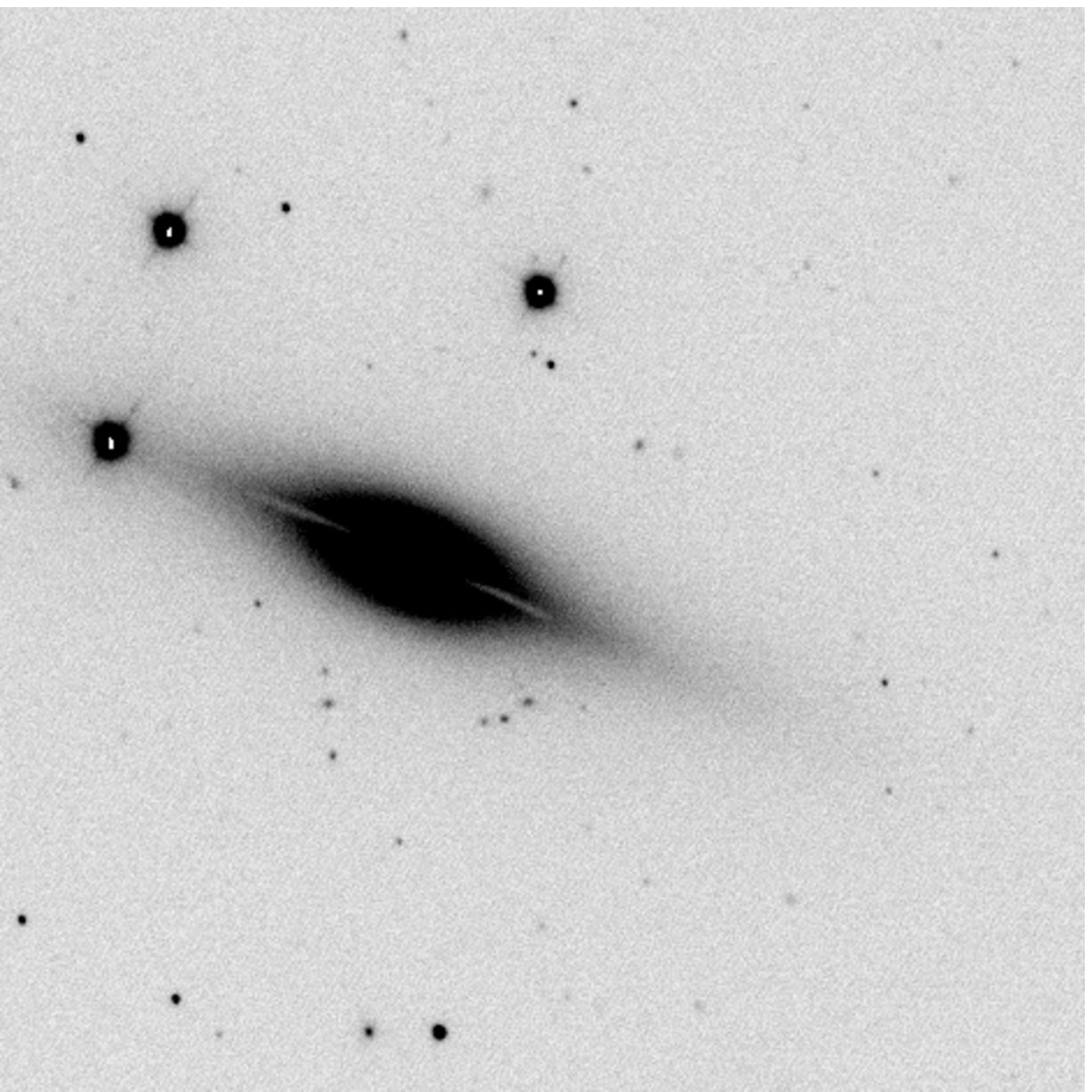}
  \put(10,10){\color{blue}\line(1,0){9}}
  \put(10,12){\color{blue}4.9}
  \end{overpic}\\
  \begin{overpic}[width=0.5\columnwidth]{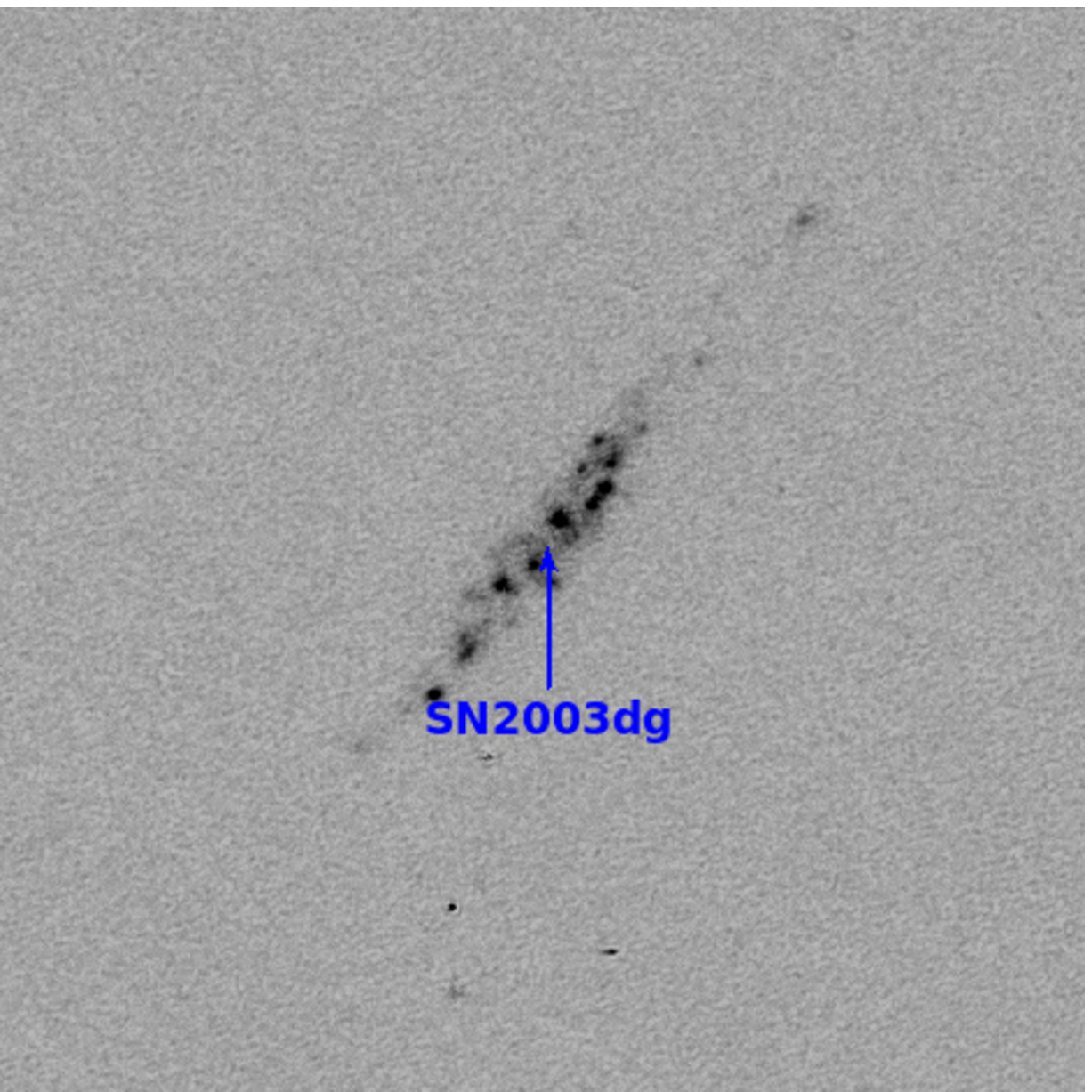}
  \end{overpic}
  \begin{overpic}[width=0.5\columnwidth]{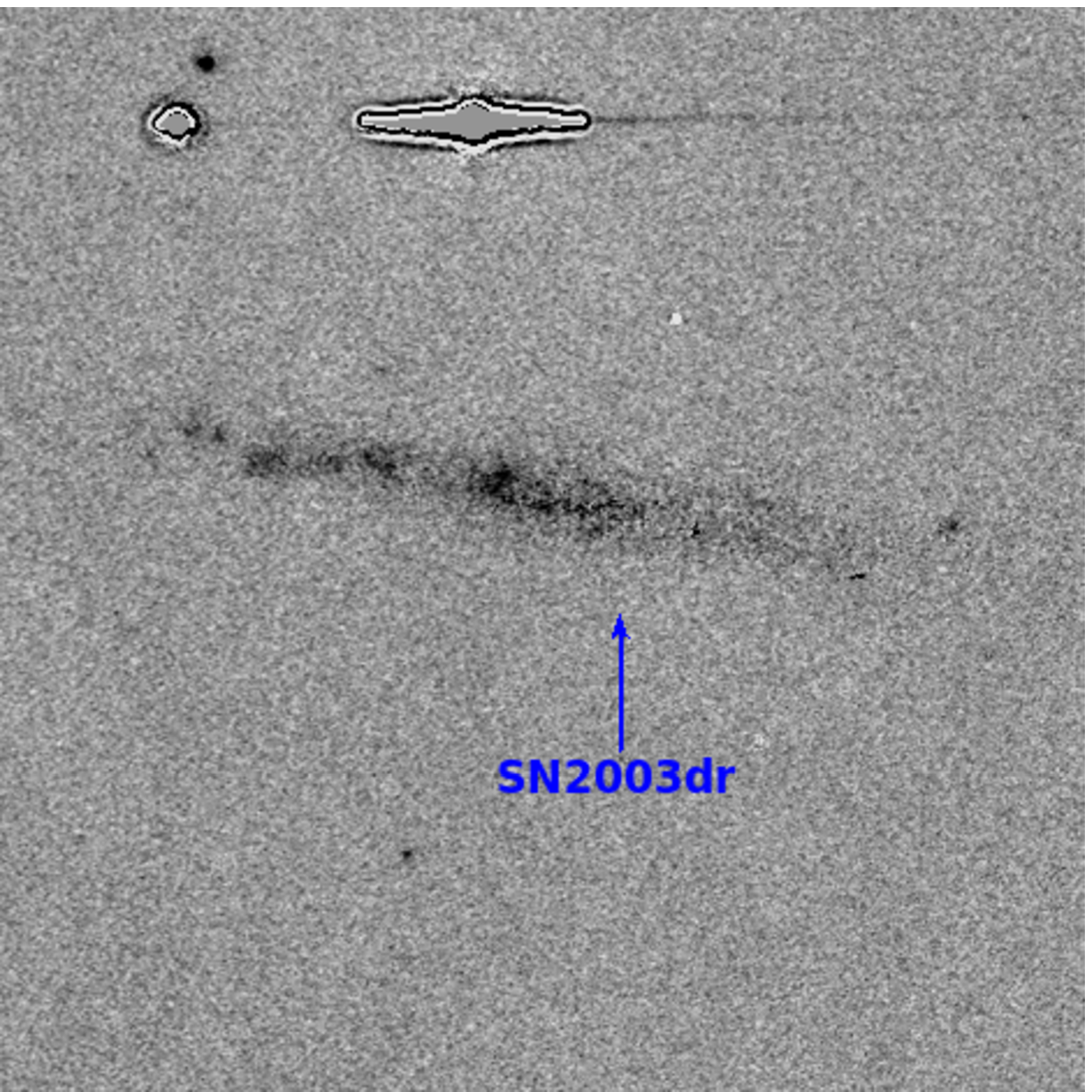}
  \end{overpic}
  \begin{overpic}[width=0.5\columnwidth]{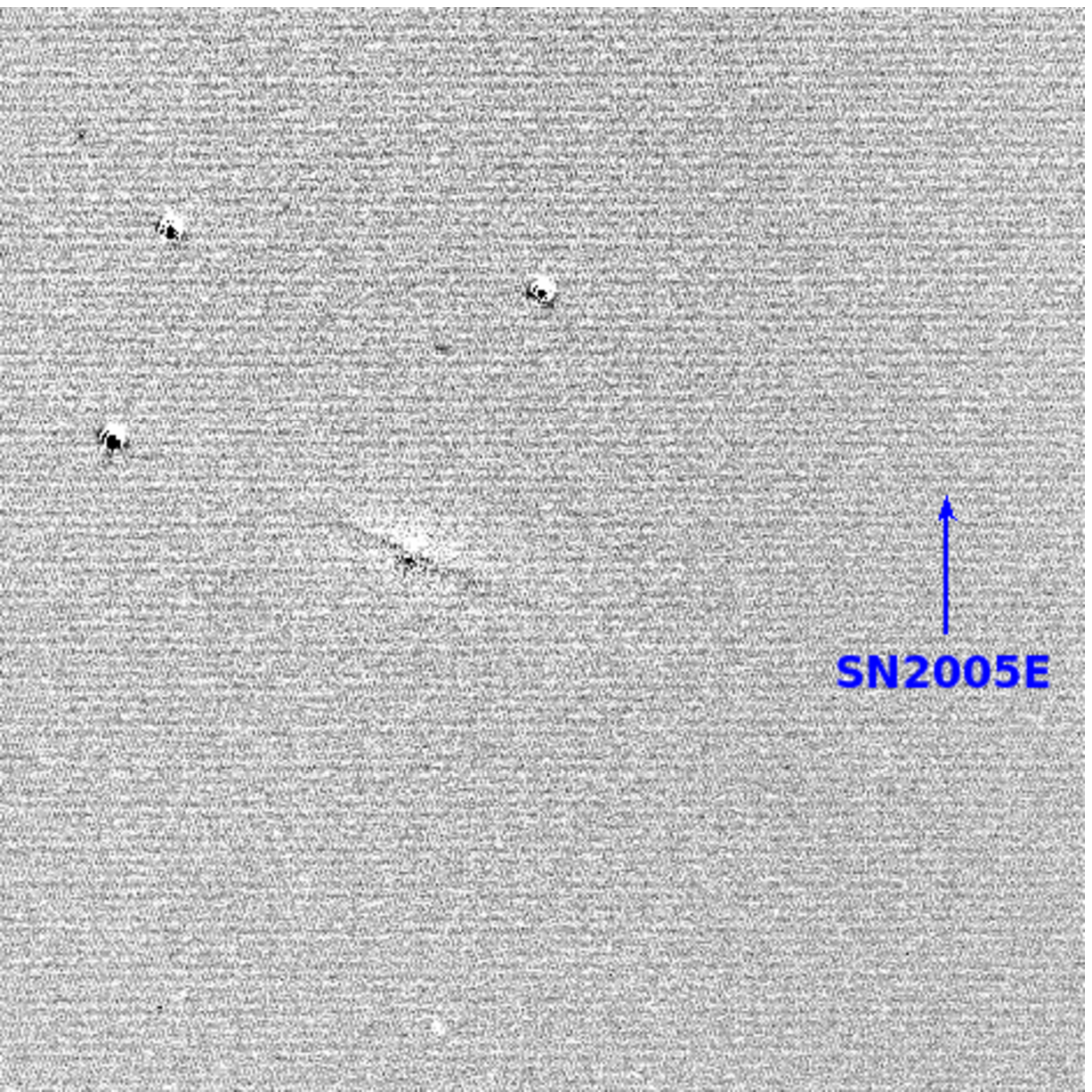}
  \end{overpic}\\
  \end{center}

  \caption{R band (top) and continuum-subtracted \Ha{} (bottom) images of Ca-rich transients. The location of the transient is marked in each case on the continuum-subtracted \Ha{} image. The bars in each R band image indicate 30$^{\prime\prime}$ and are labelled with the linear size at the distance of the host in kpc. For all images North is up, East is left.}
\label{fig:carich_imgs}
\end{figure*}

\addtocounter{figure}{-1}
\begin{figure*}
  \begin{center}
  \begin{overpic}[width=0.5\columnwidth]{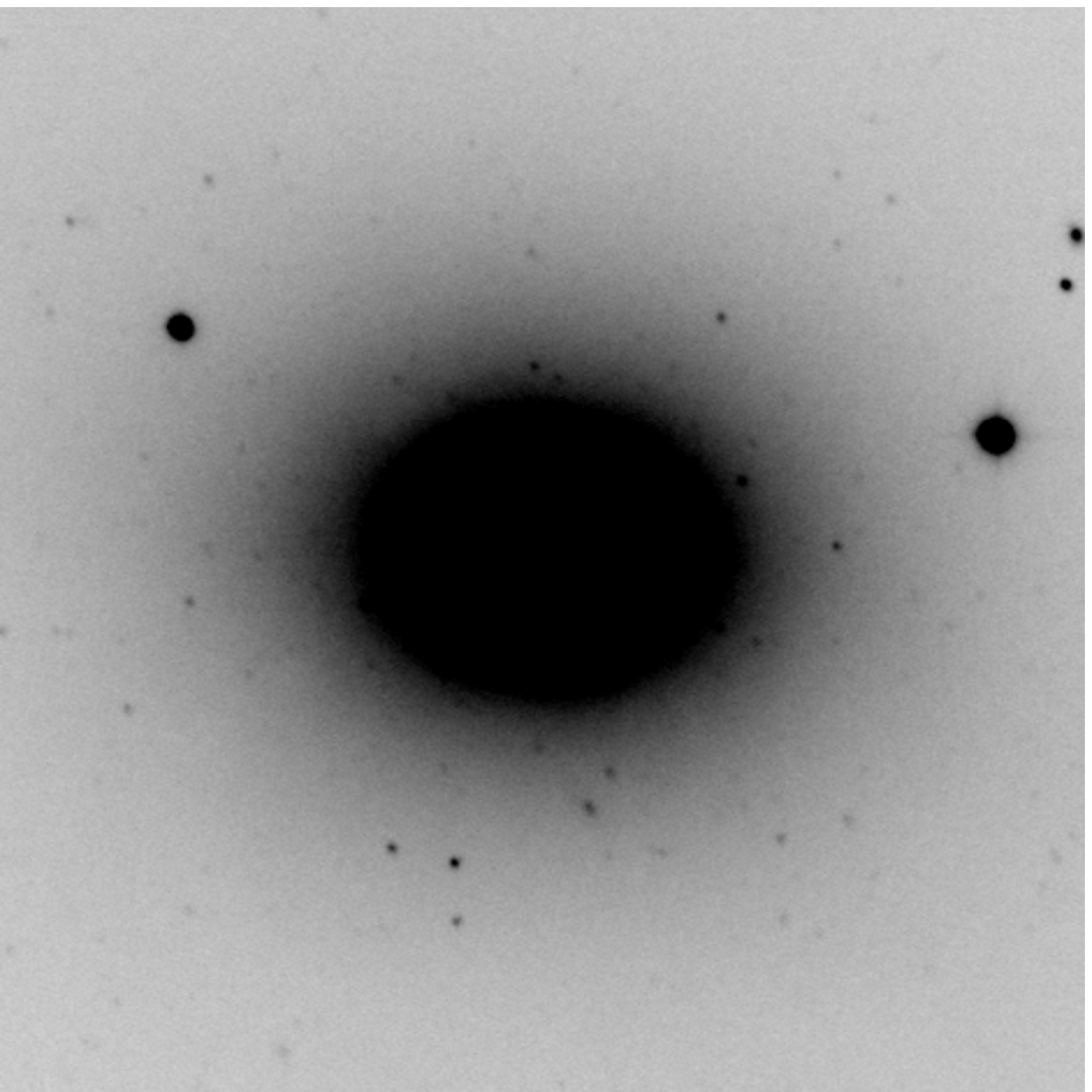}
  \put(10,10){\color{blue}\line(1,0){20}}
  \put(10,12){\color{blue}4.4}
  \end{overpic}
  \begin{overpic}[width=0.5\columnwidth]{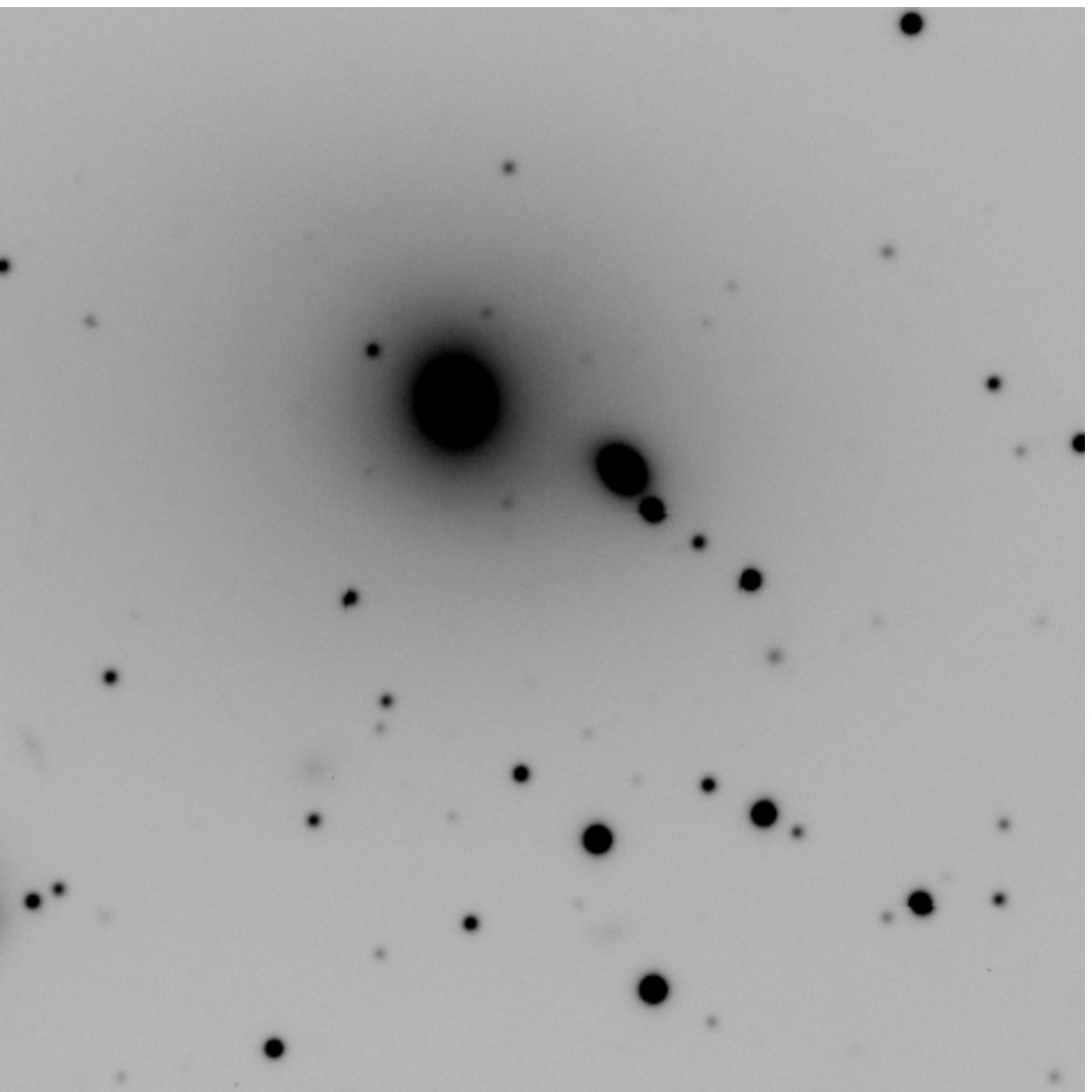}
  \put(10,10){\color{blue}\line(1,0){20}}
  \put(10,12){\color{blue}11.5}
  \end{overpic}
  \begin{overpic}[width=0.5\columnwidth]{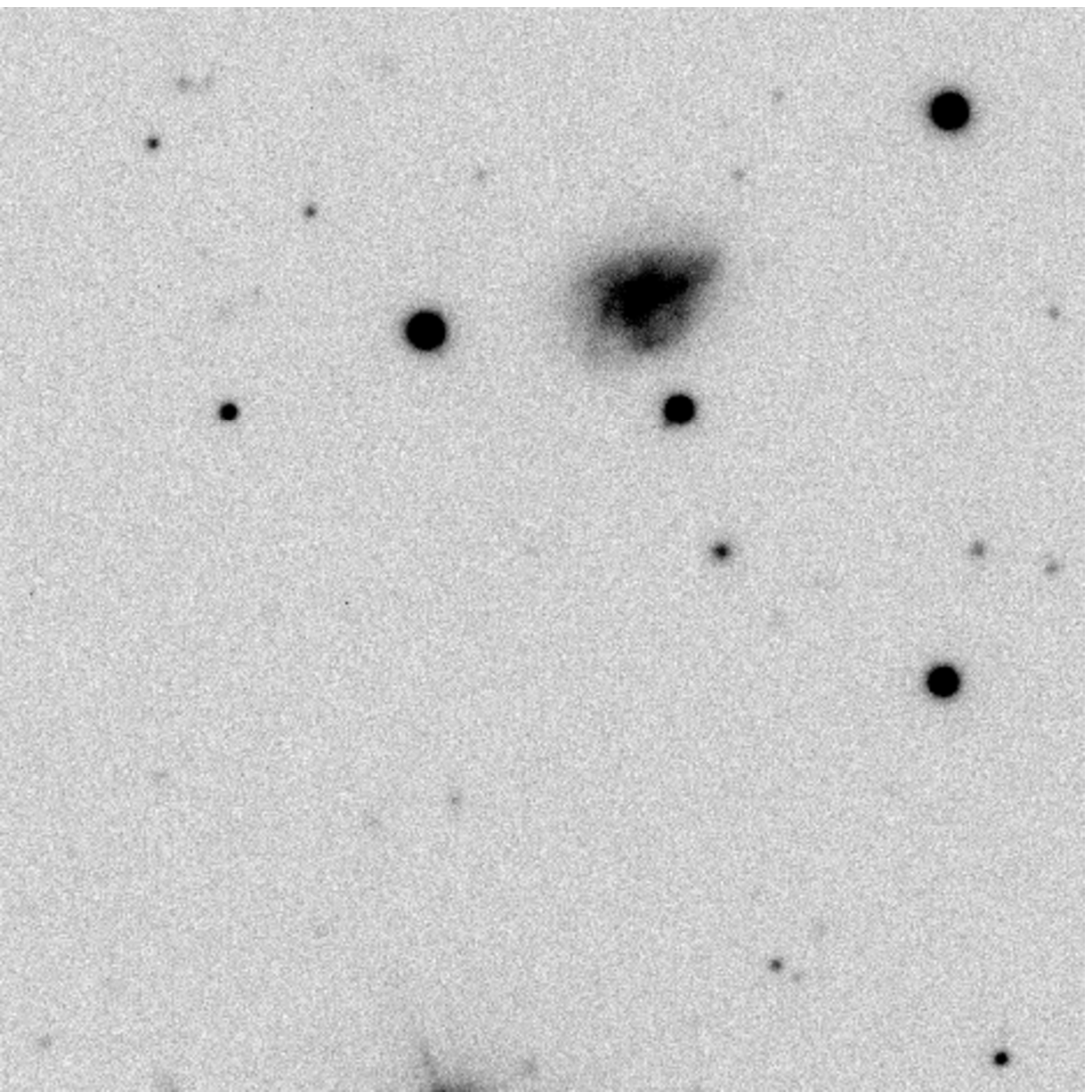}
  \put(10,10){\color{blue}\line(1,0){22}}
  \put(10,12){\color{blue}22.9}
  \end{overpic}\\
  \begin{overpic}[width=0.5\columnwidth]{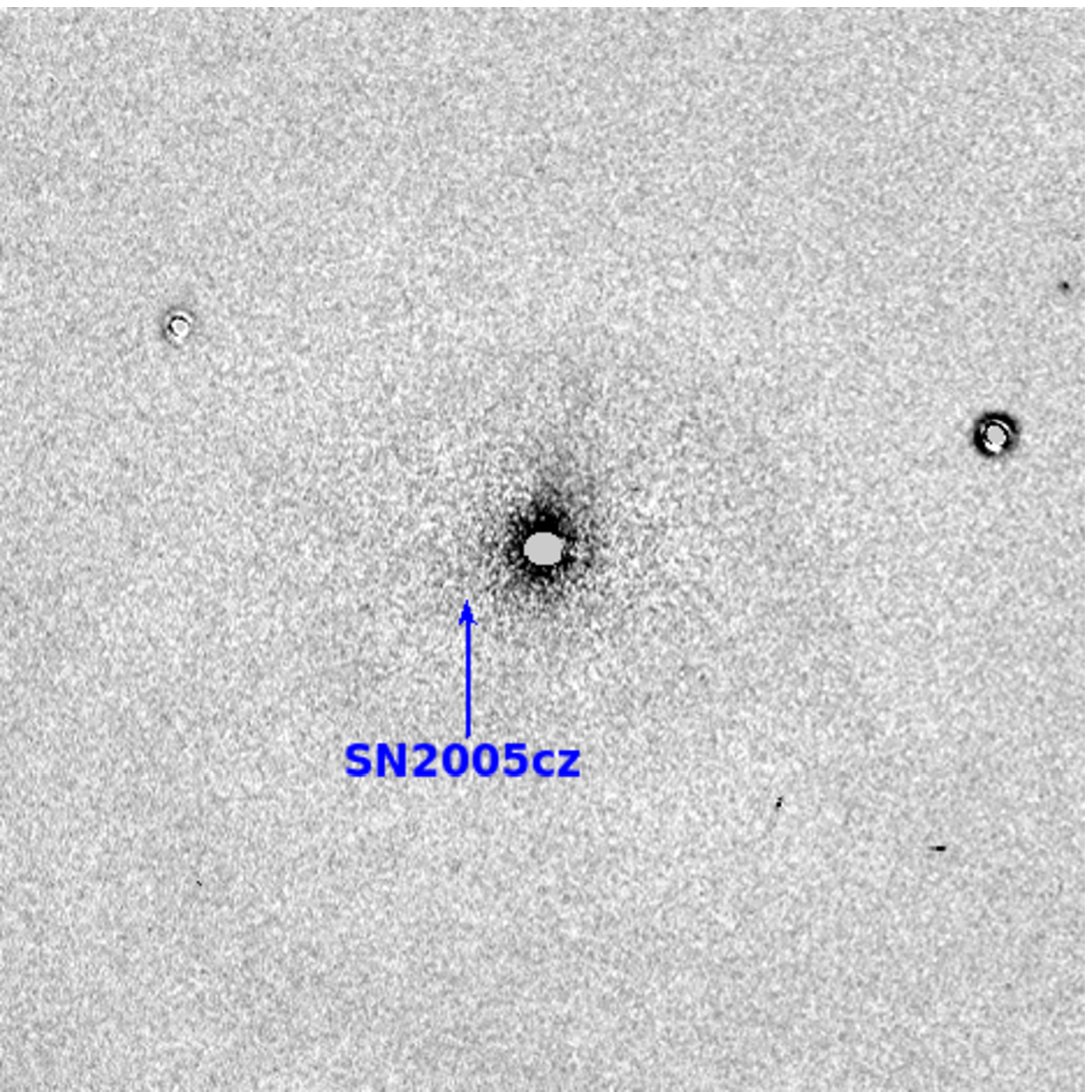}
  \end{overpic}
  \begin{overpic}[width=0.5\columnwidth]{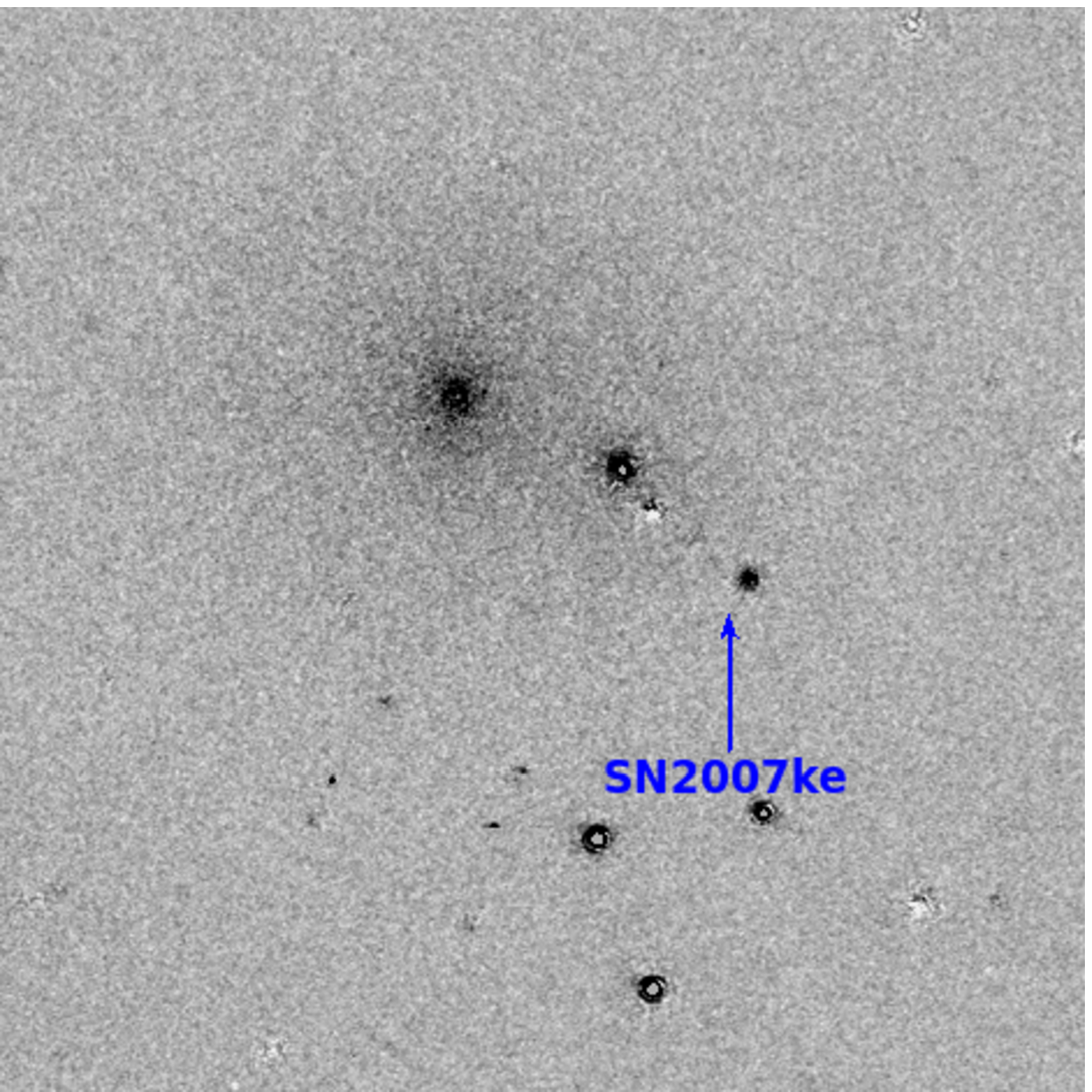}
  \end{overpic}
  \begin{overpic}[width=0.5\columnwidth]{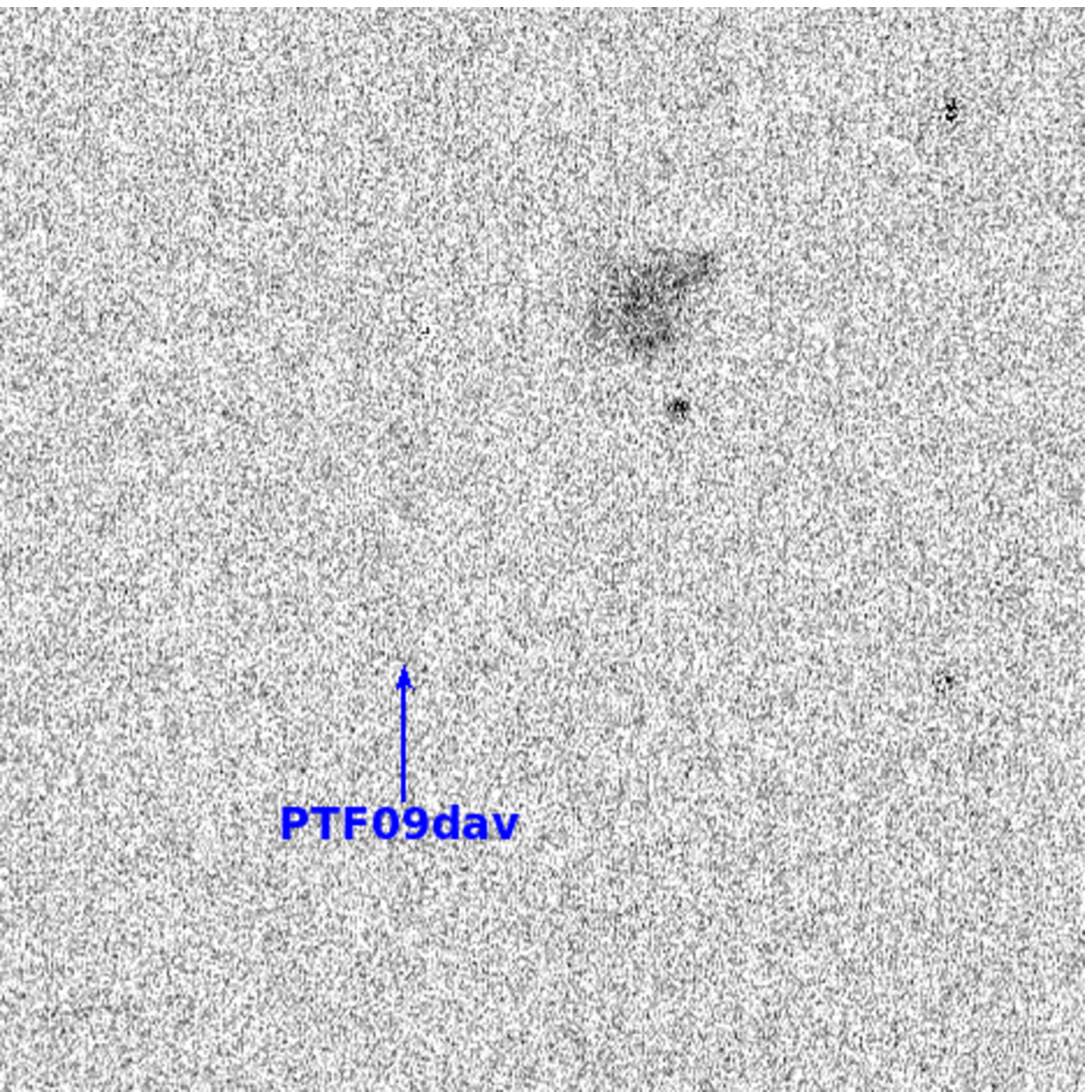}
  \end{overpic}\\
  \vspace{0.2cm}
  \begin{overpic}[width=0.5\columnwidth]{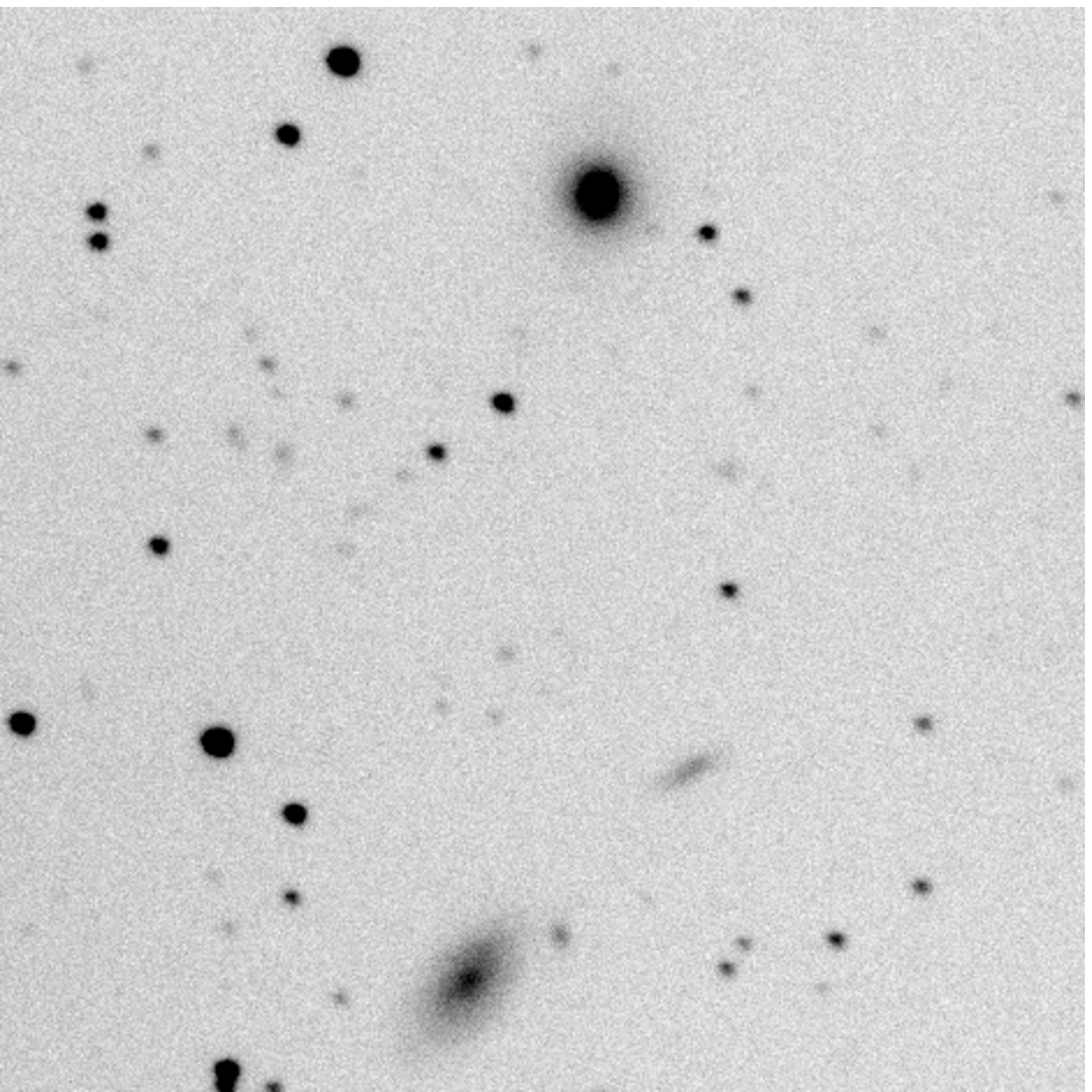}
  \put(10,10){\color{blue}\line(1,0){22}}
  \put(10,12){\color{blue}14.6}
  \end{overpic}
  \begin{overpic}[width=0.5\columnwidth]{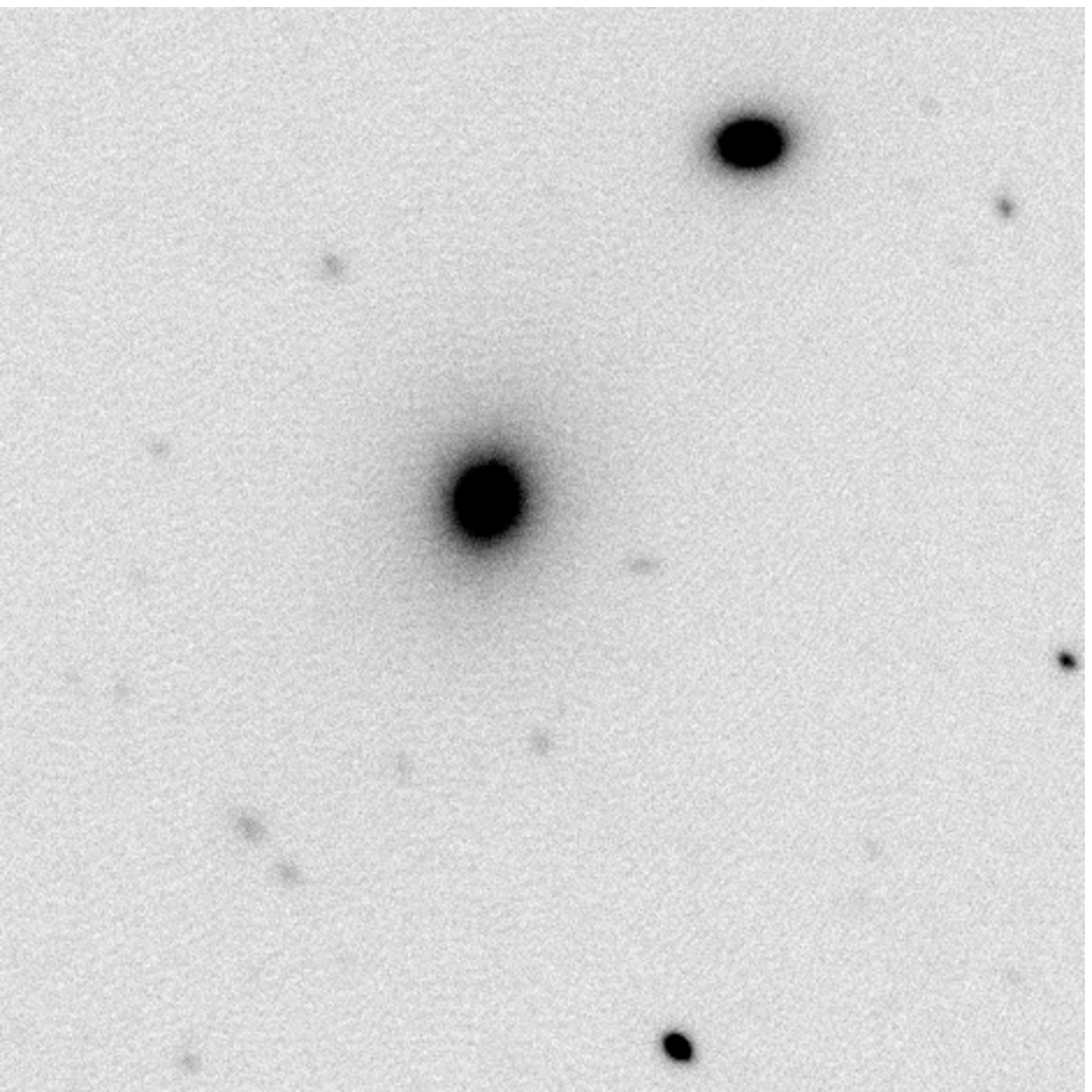}
  \put(10,10){\color{blue}\line(1,0){22}}
  \put(10,12){\color{blue}21.9}
  \end{overpic}
  \begin{overpic}[width=0.5\columnwidth]{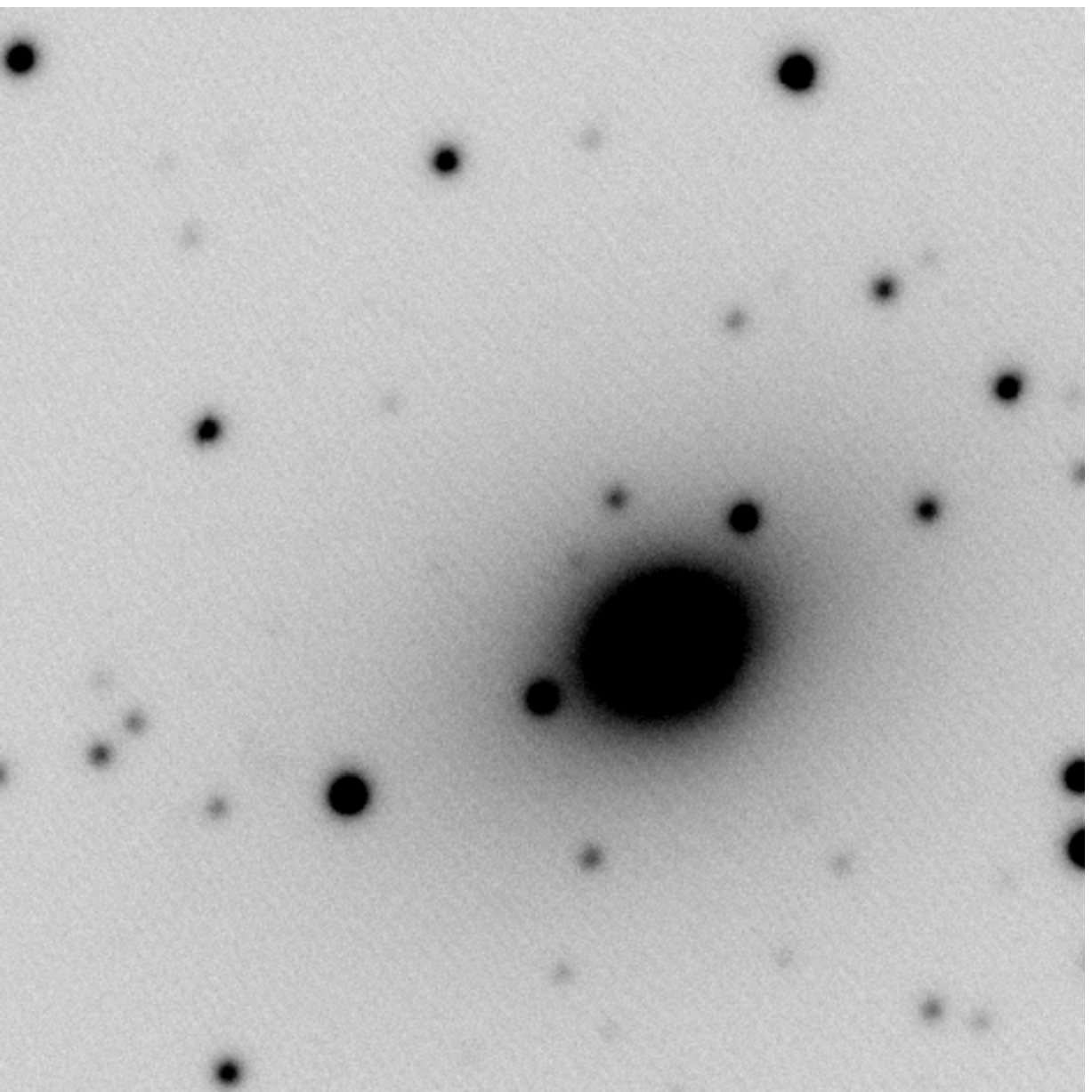}
  \put(10,10){\color{blue}\line(1,0){22}}
  \put(10,12){\color{blue}4.8}
  \end{overpic}\\
  \begin{overpic}[width=0.5\columnwidth]{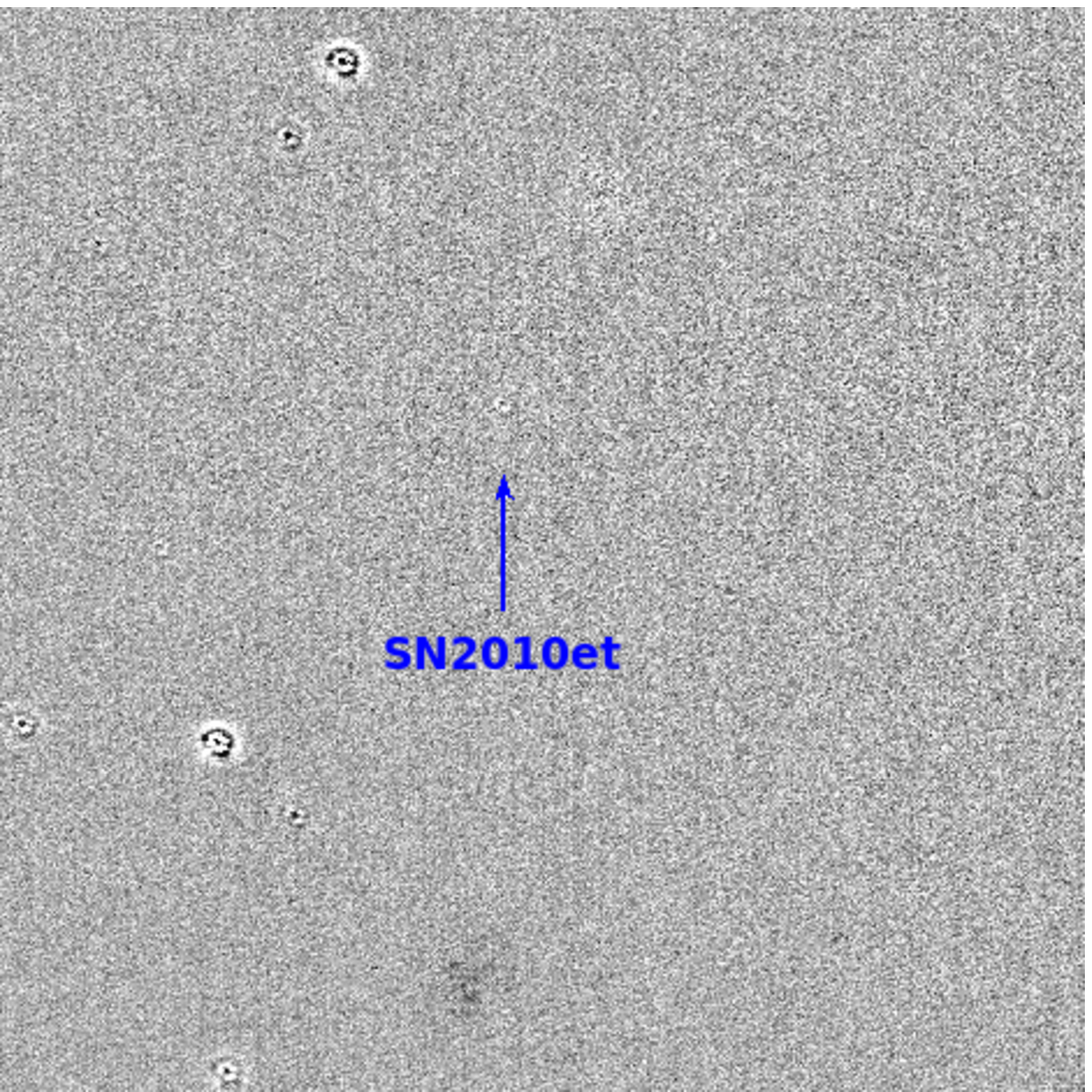}
  \end{overpic}
  \begin{overpic}[width=0.5\columnwidth]{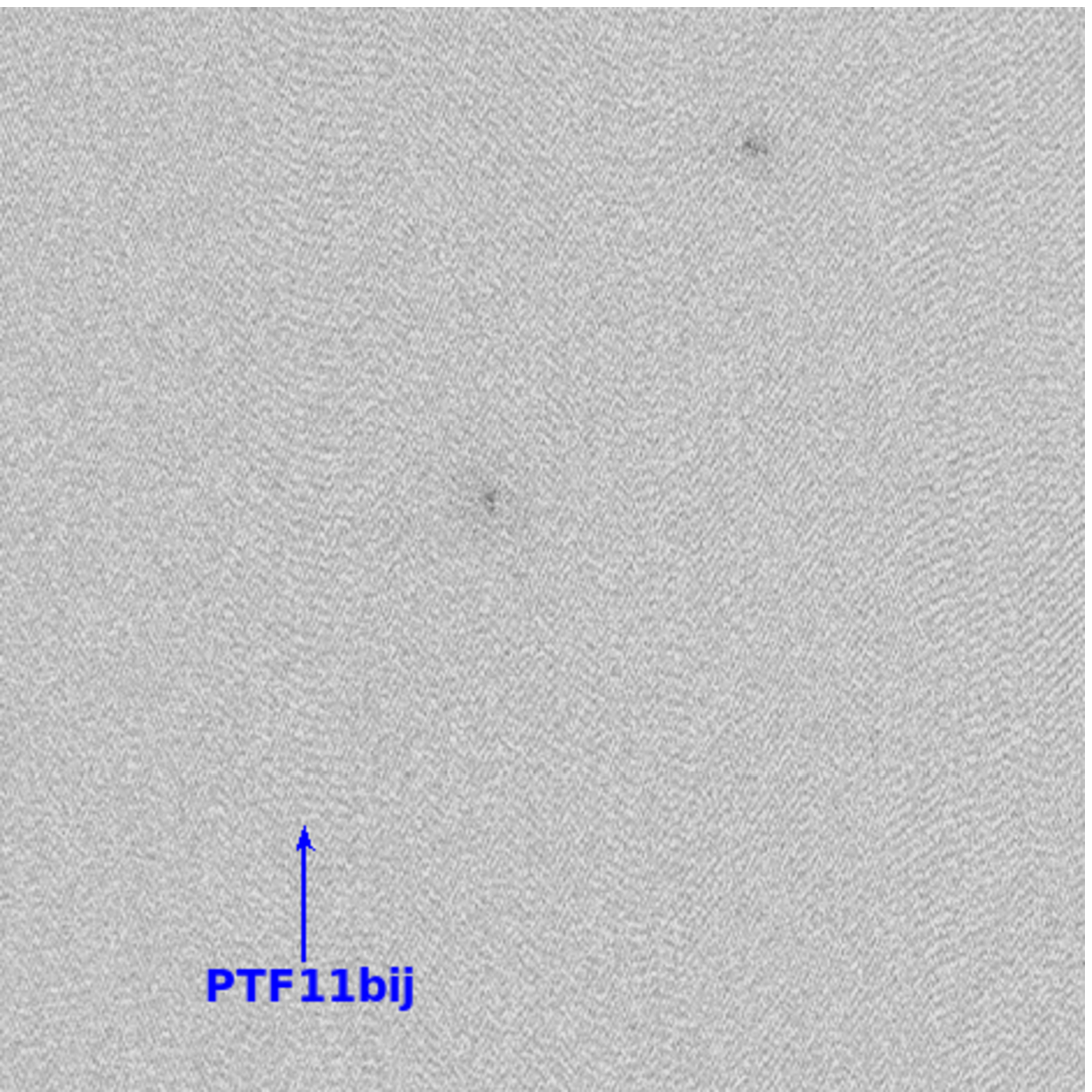}
  \end{overpic}
  \begin{overpic}[width=0.5\columnwidth]{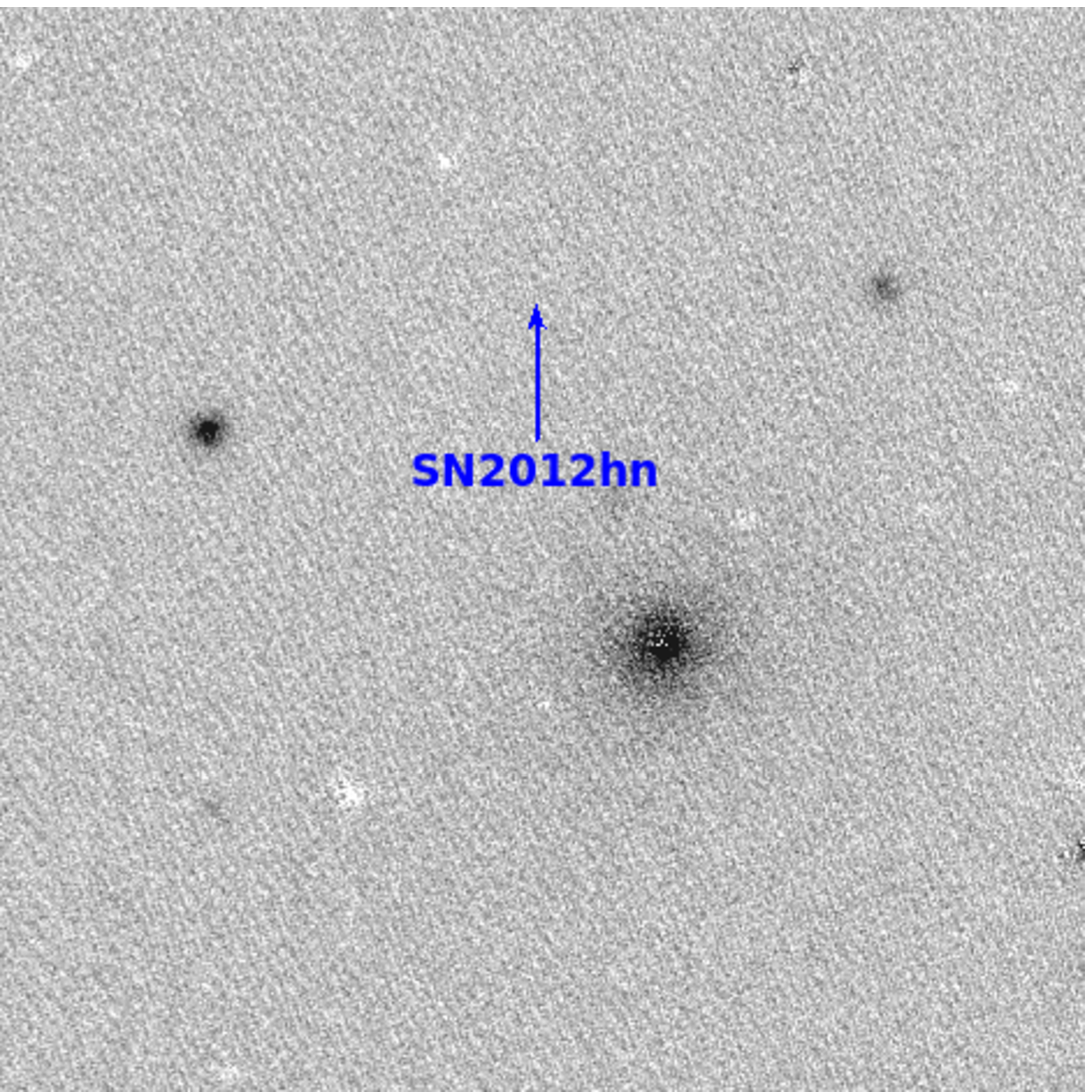}
  \end{overpic}\\
  \end{center}

  \caption{cont.}
\label{fig:carich_imgs2}
\end{figure*}

\begin{figure*}
  \begin{center}
  \begin{overpic}[width=0.5\columnwidth]{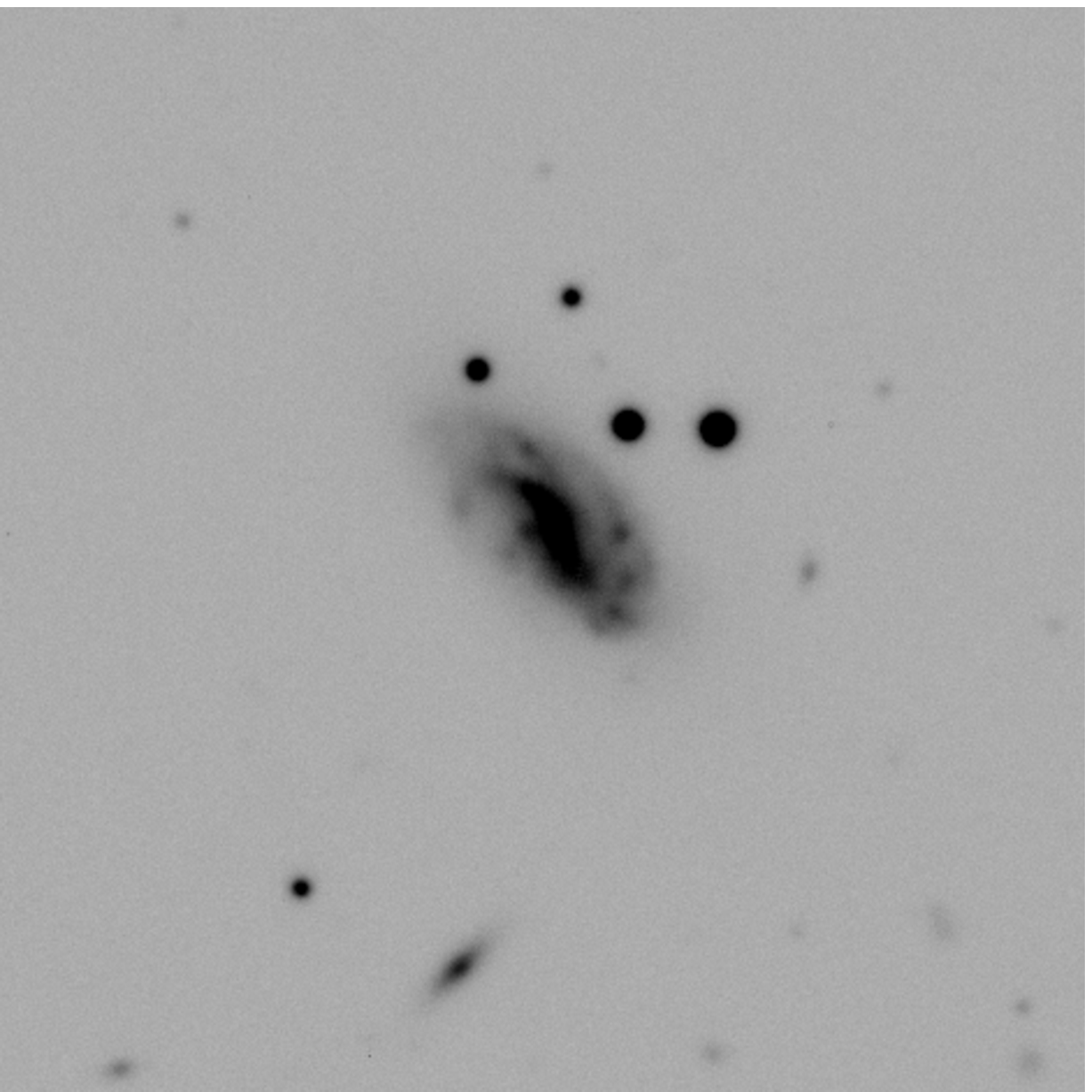}
  \put(10,10){\color{blue}\line(1,0){20}}
  \put(10,12){\color{blue}10.3}
  \end{overpic}
  \begin{overpic}[width=0.5\columnwidth]{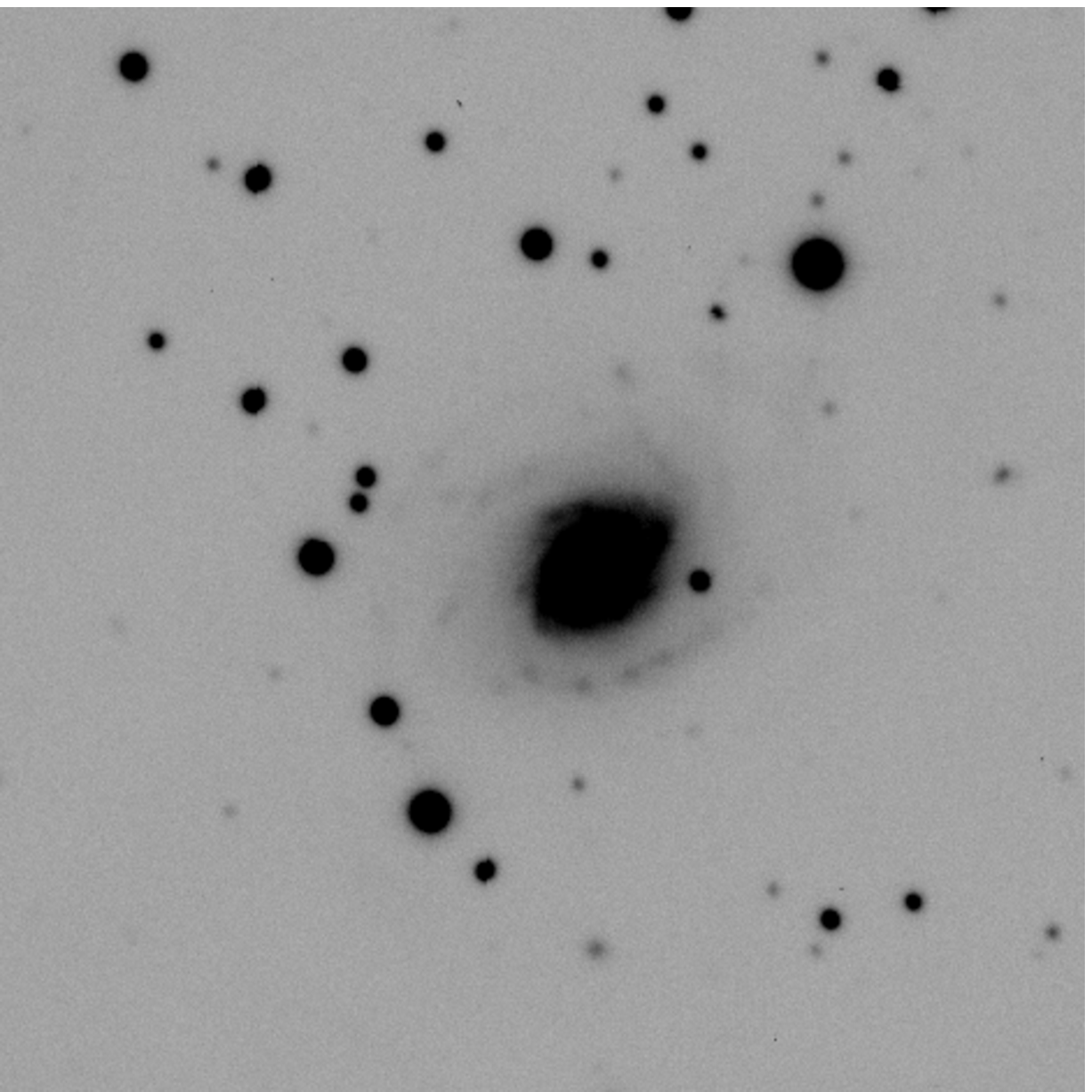}
  \put(10,10){\color{blue}\line(1,0){20}}
  \put(10,12){\color{blue}10.6}
  \end{overpic}
  \begin{overpic}[width=0.5\columnwidth]{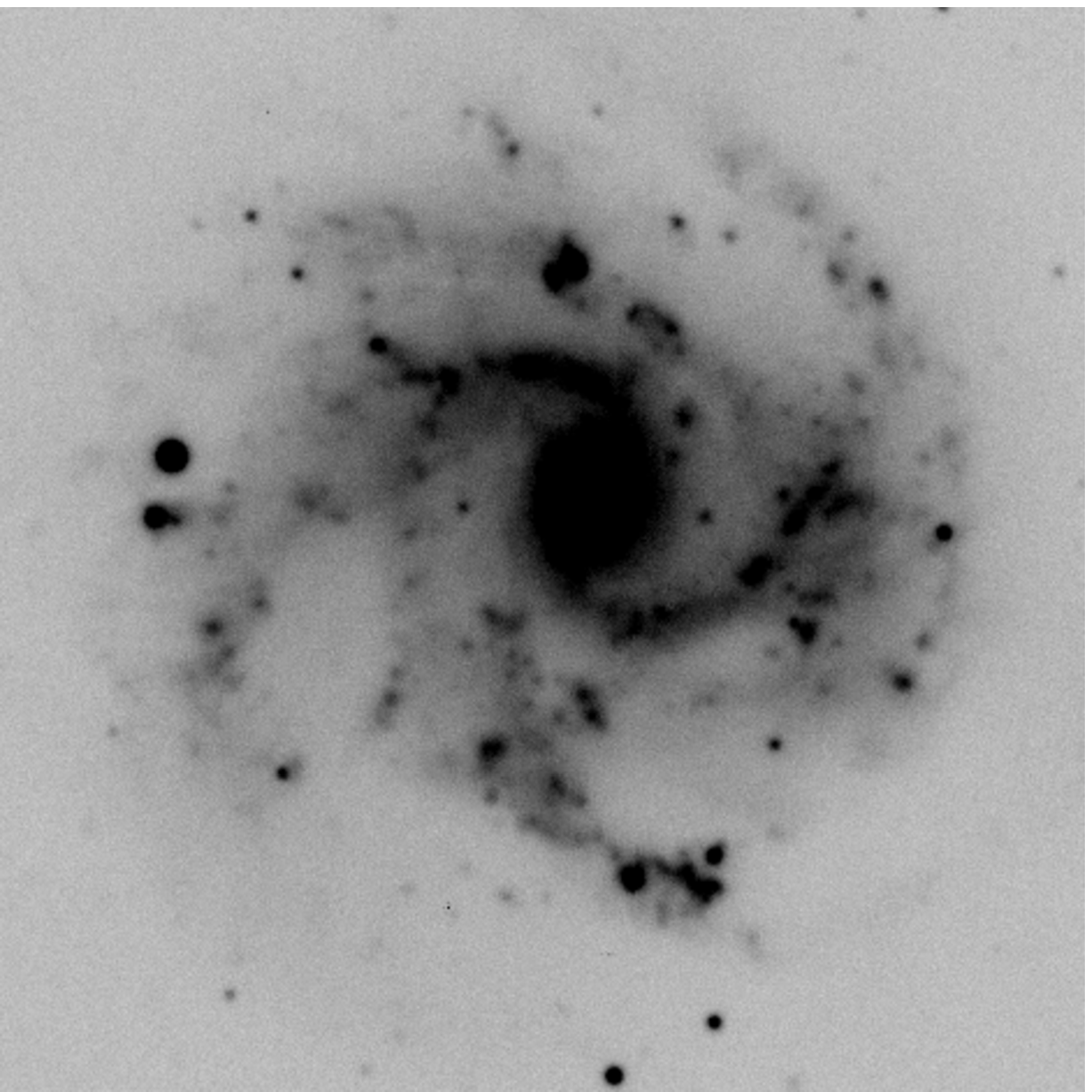}
  \put(10,10){\color{blue}\line(1,0){20}}
  \put(10,12){\color{blue}6.4}
  \end{overpic}\\
  \begin{overpic}[width=0.5\columnwidth]{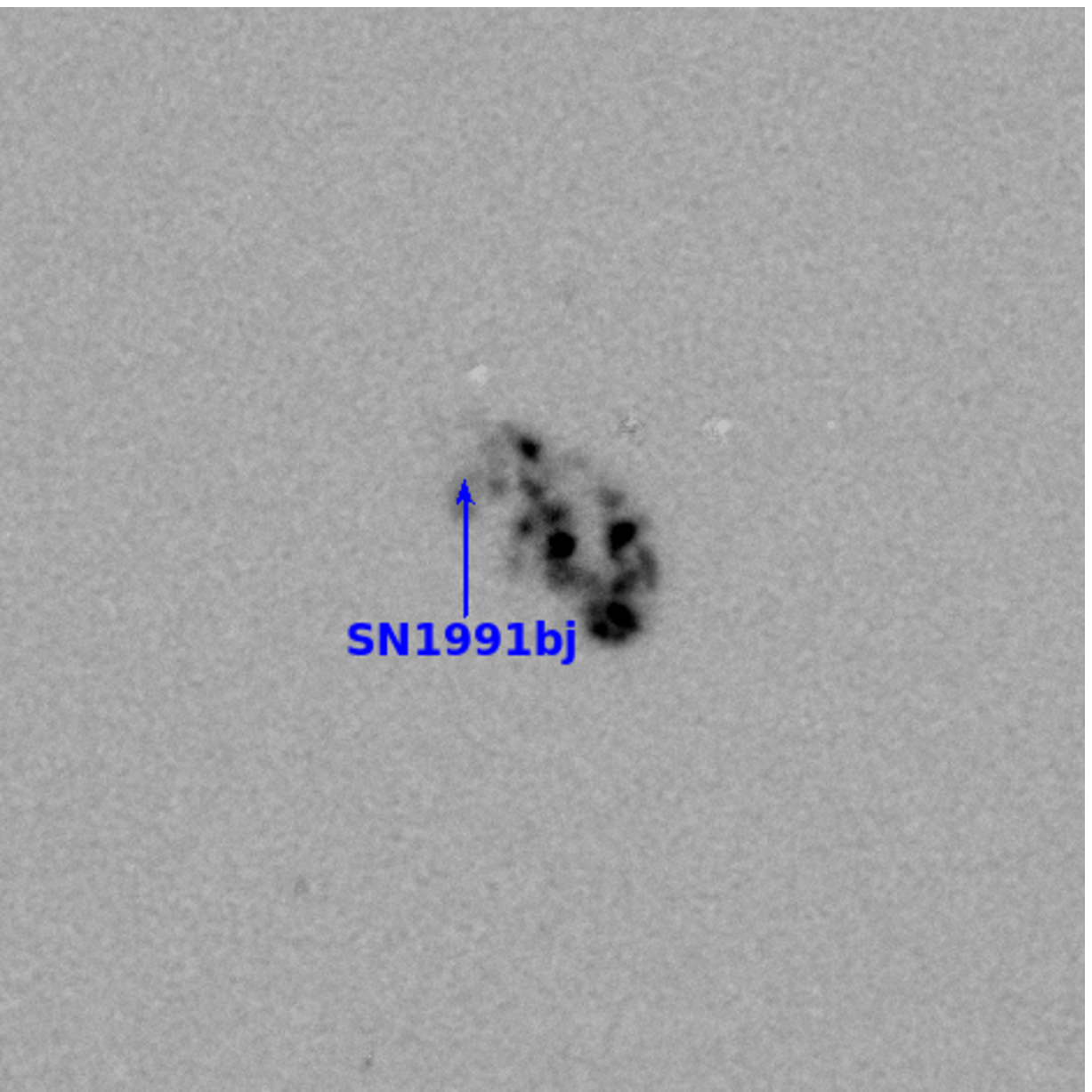}
  \end{overpic}
  \begin{overpic}[width=0.5\columnwidth]{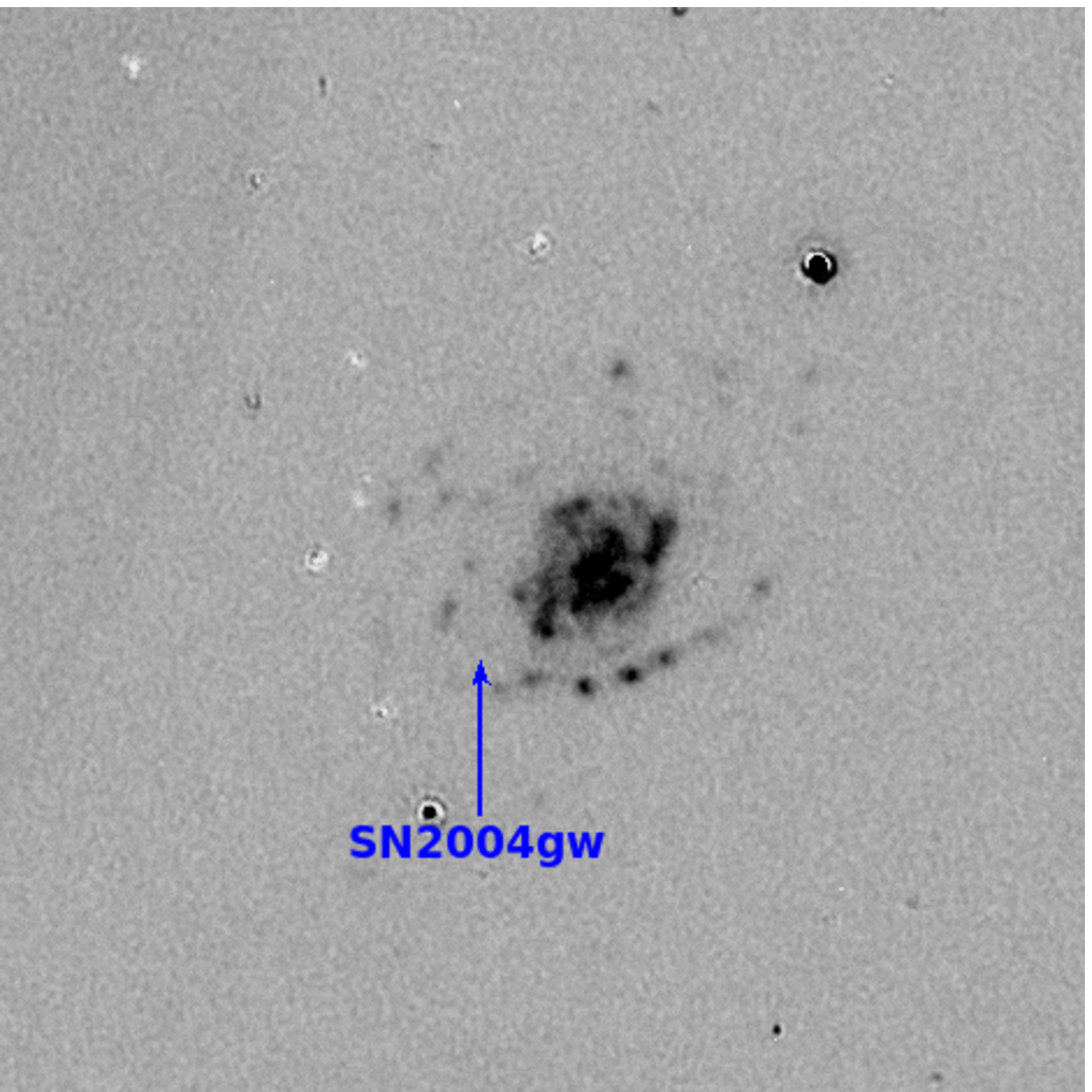}
  \end{overpic}
  \begin{overpic}[width=0.5\columnwidth]{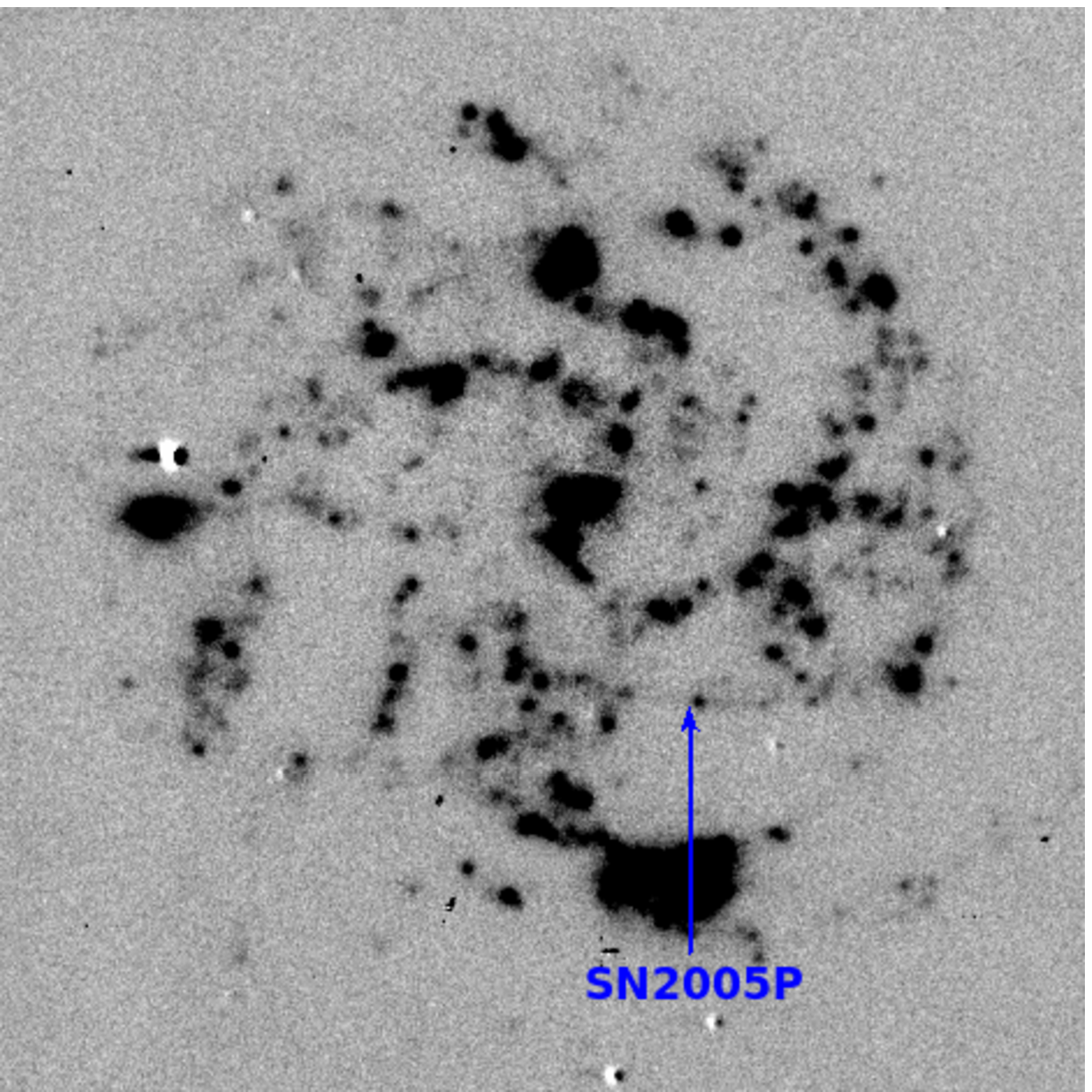}
  \end{overpic}\\
  \vspace{0.2cm}
  \begin{overpic}[width=0.5\columnwidth]{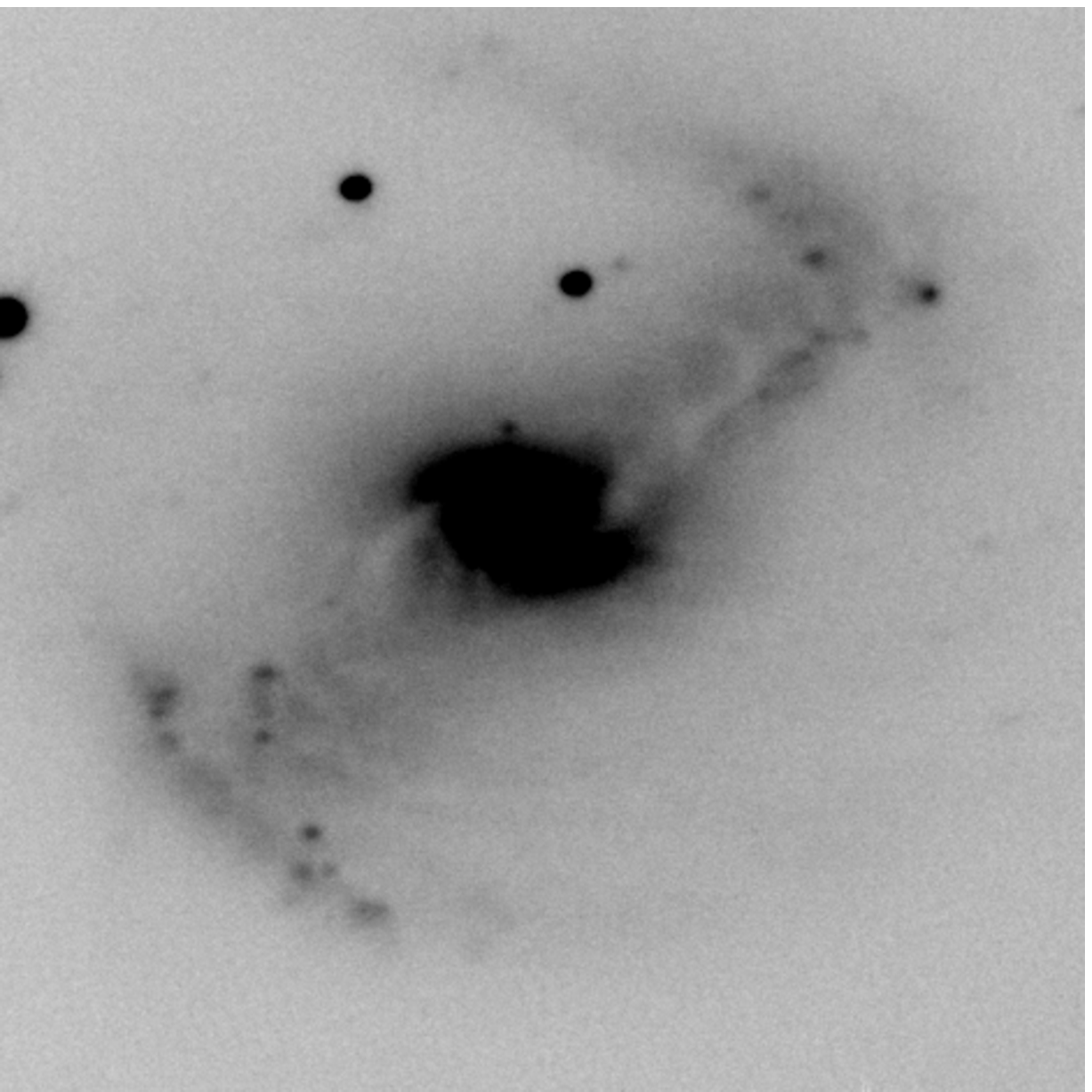}
  \put(10,10){\color{blue}\line(1,0){24}}
  \put(10,12){\color{blue}5.4}
  \end{overpic}
  \begin{overpic}[width=0.5\columnwidth]{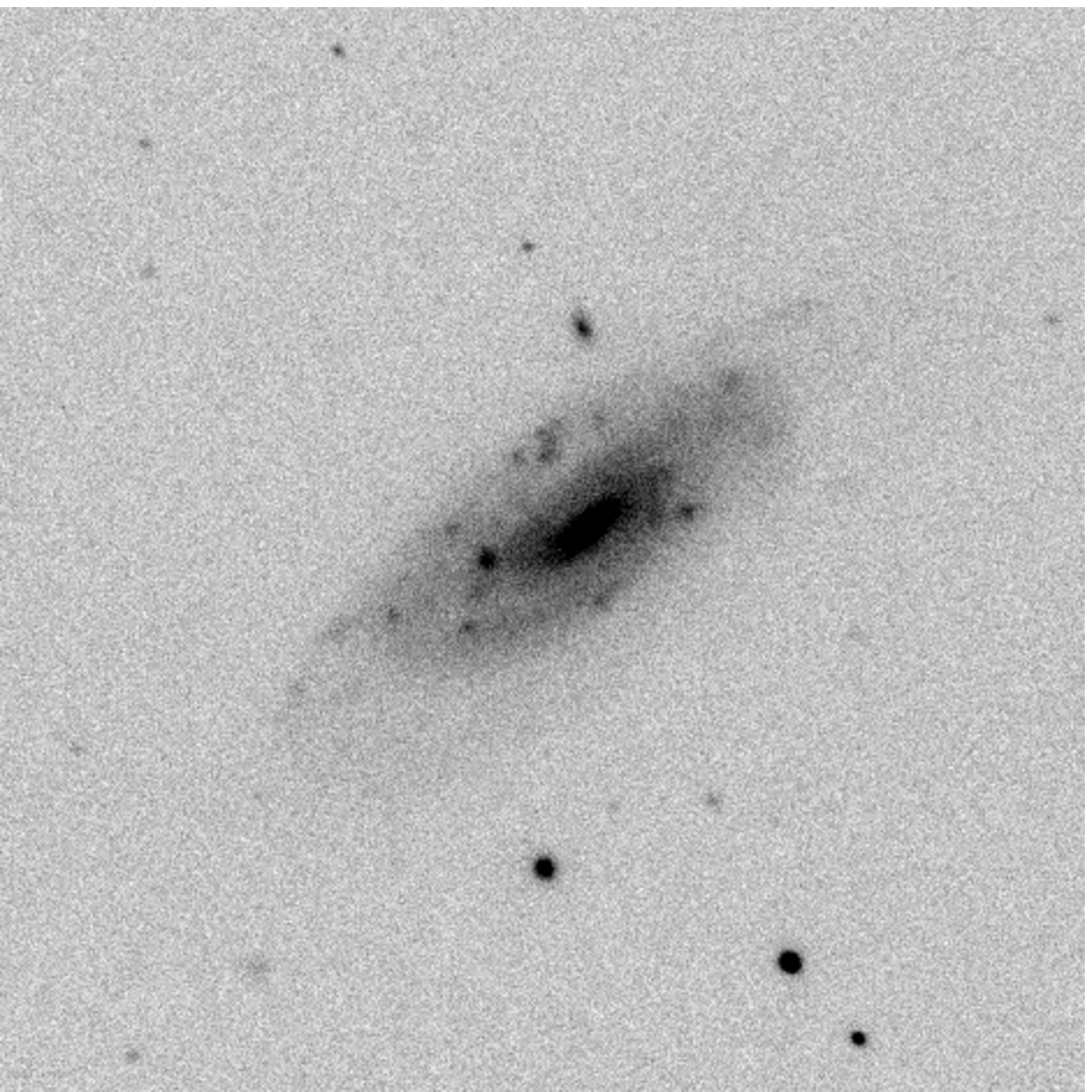}
  \put(10,10){\color{blue}\line(1,0){22}}
  \put(10,12){\color{blue}8.5}
  \end{overpic}
  \begin{overpic}[width=0.5\columnwidth]{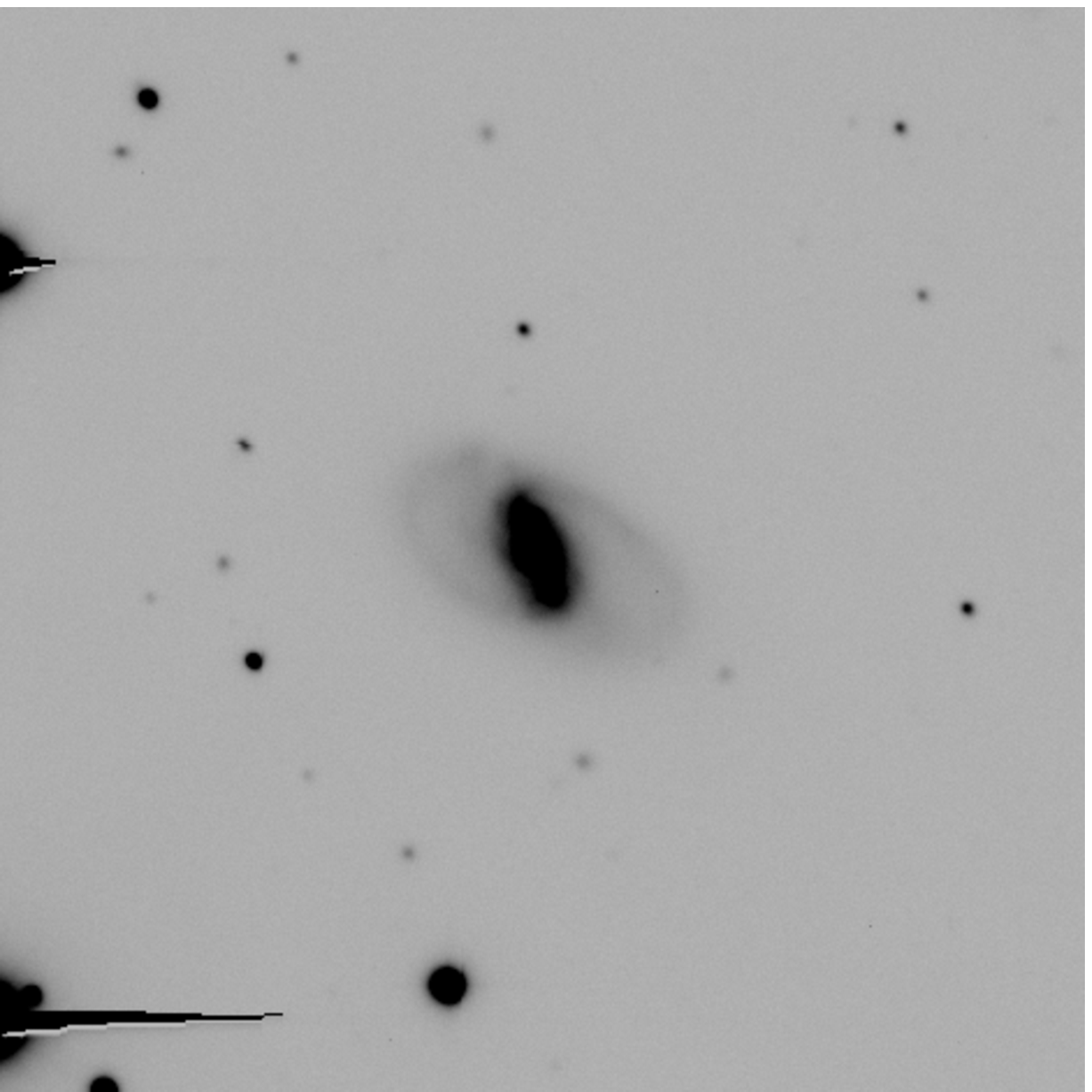}
  \put(10,10){\color{blue}\line(1,0){20}}
  \put(10,12){\color{blue}10.7}
  \end{overpic}\\
  \begin{overpic}[width=0.5\columnwidth]{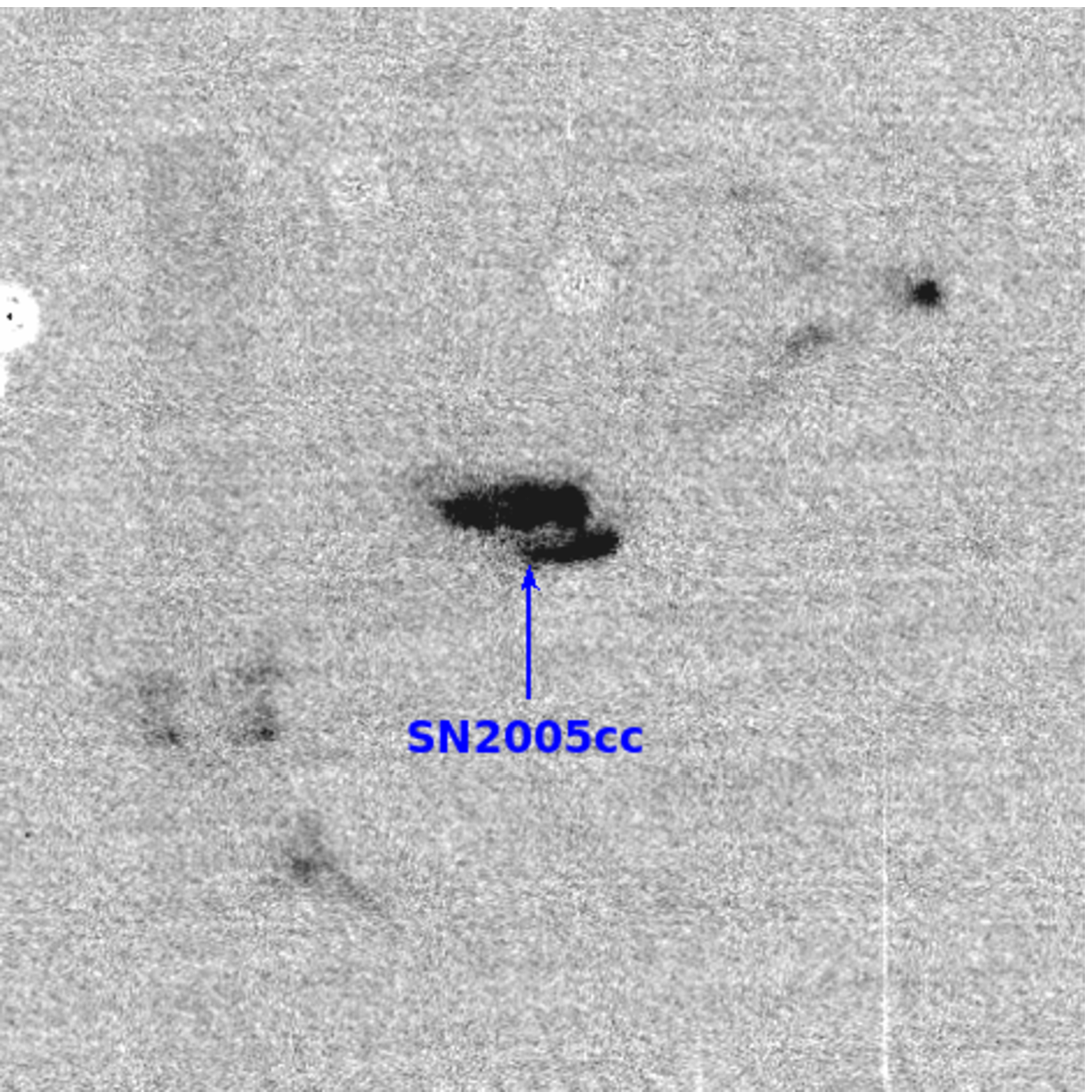}
  \end{overpic}
  \begin{overpic}[width=0.5\columnwidth]{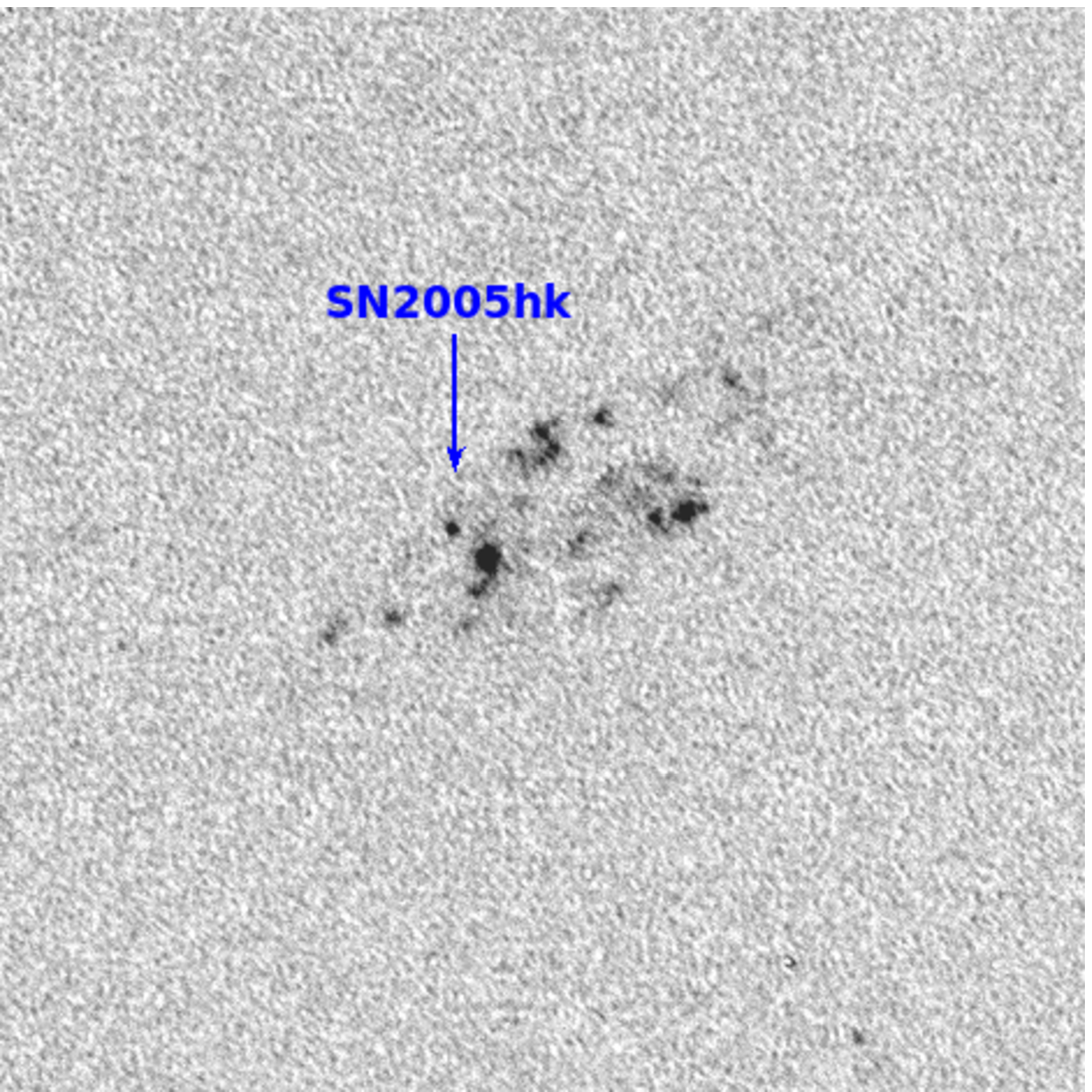}
  \end{overpic}
  \begin{overpic}[width=0.5\columnwidth]{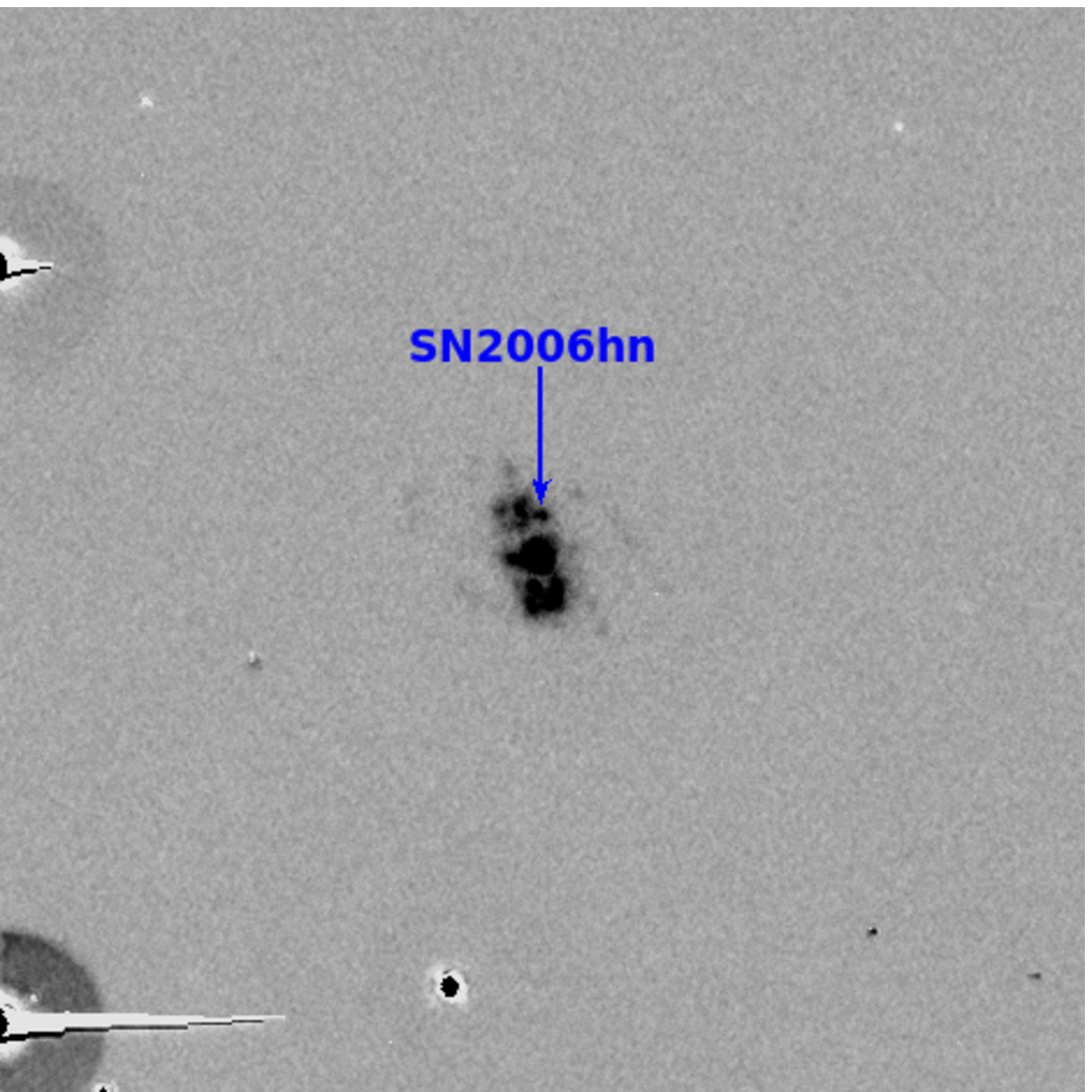}
  \end{overpic}\\
  \end{center}

  \caption{Same as Figure \ref{fig:carich_imgs} but for the SN2002cx-like transients.}
\label{fig:2008ha_imgs}
\end{figure*}

\addtocounter{figure}{-1}
\begin{figure*}
  \begin{center}
  \begin{overpic}[width=0.5\columnwidth]{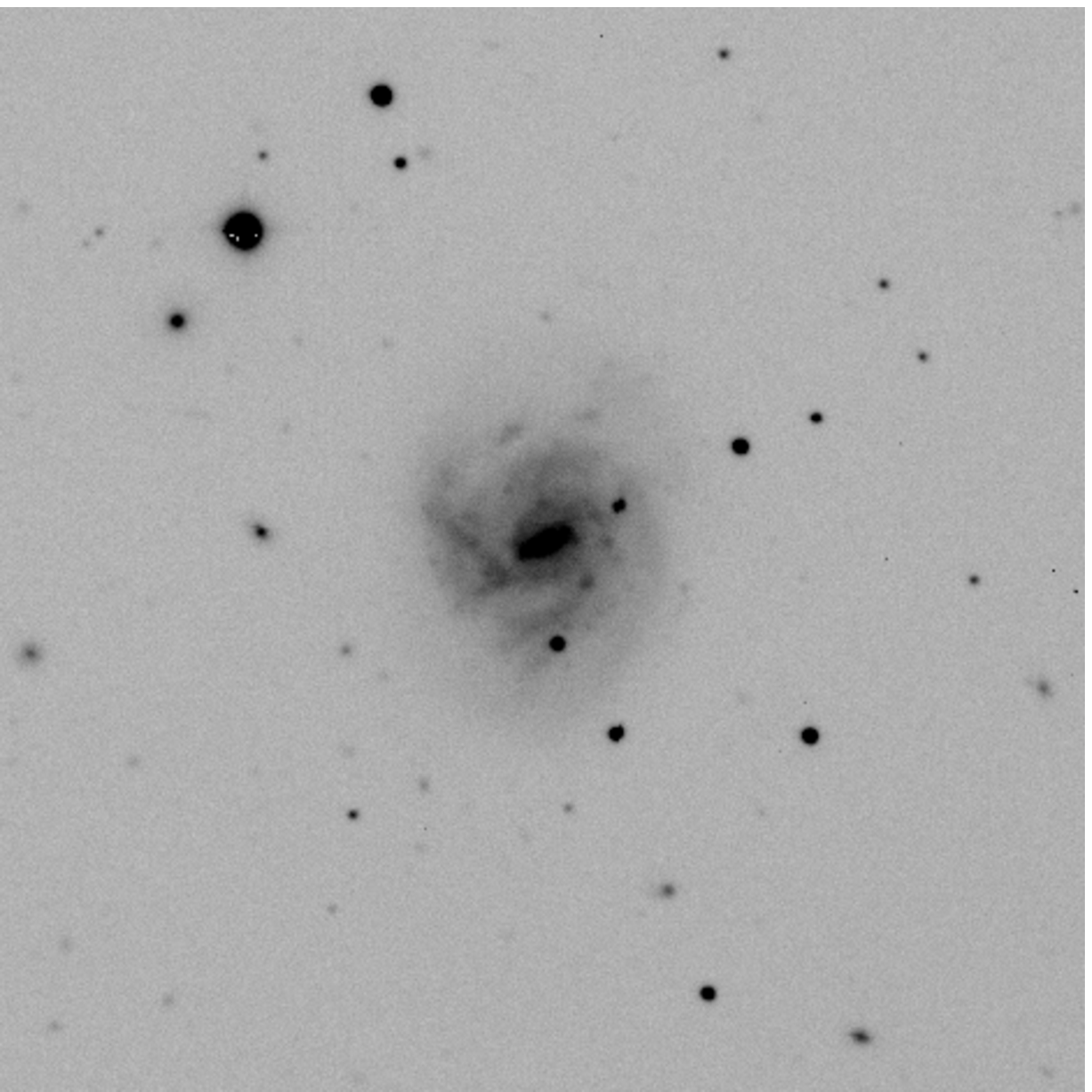}
  \put(10,10){\color{blue}\line(1,0){20}}
  \put(10,12){\color{blue}10.4}
  \end{overpic}
  \begin{overpic}[width=0.5\columnwidth]{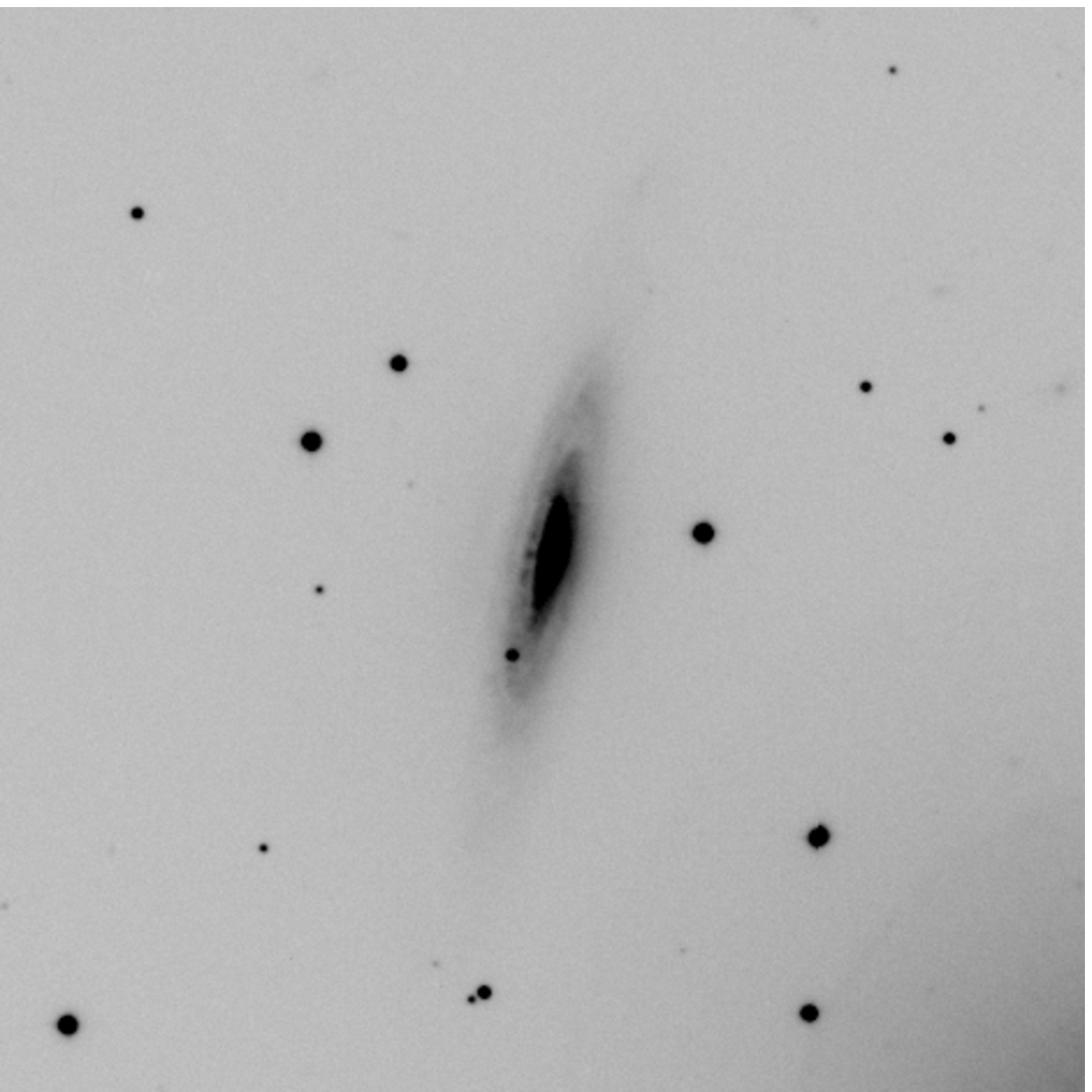}
  \put(10,10){\color{blue}\line(1,0){20}}
  \put(10,12){\color{blue}10.0}
  \end{overpic}
  \begin{overpic}[width=0.5\columnwidth]{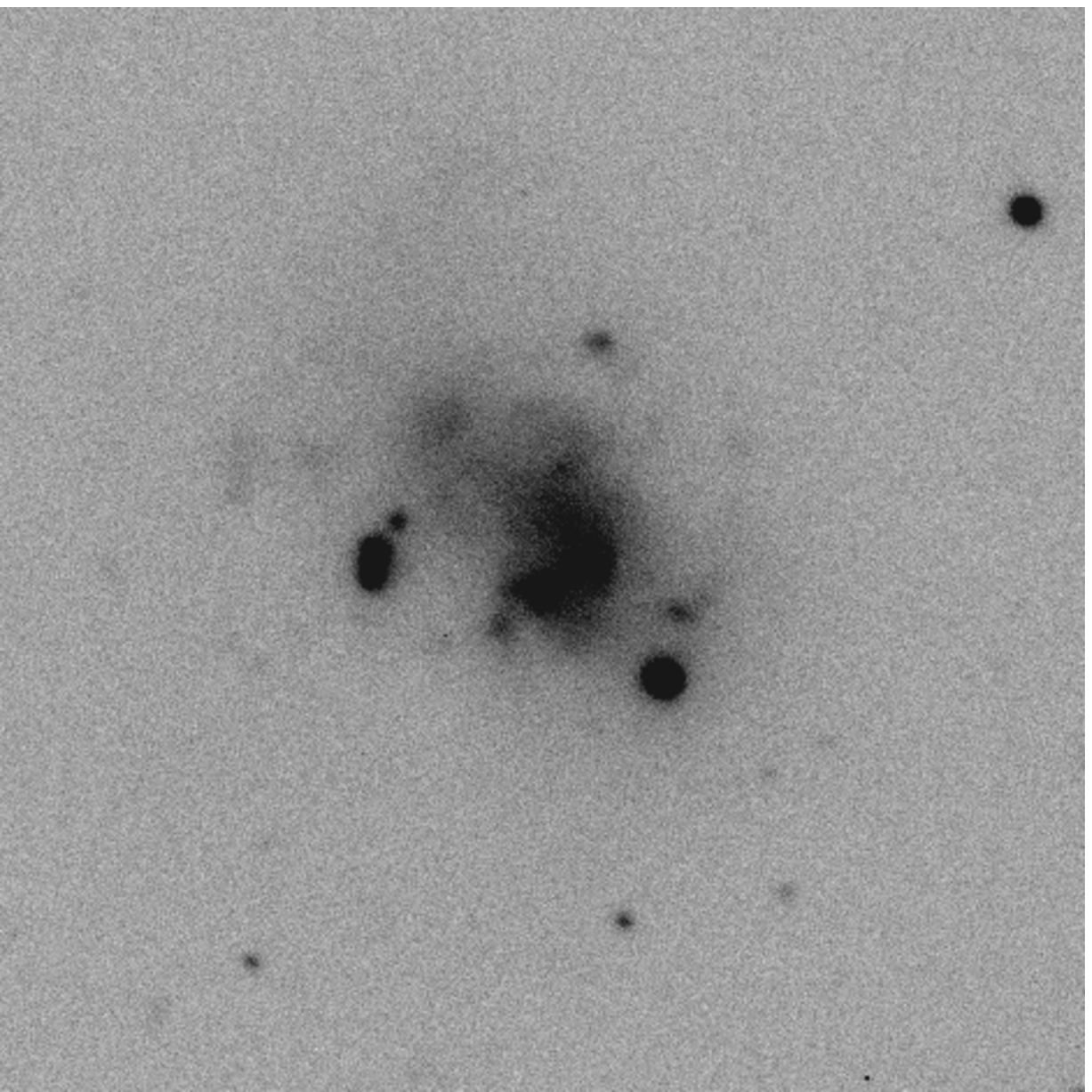}
  \put(10,10){\color{blue}\line(1,0){22}}
  \put(10,12){\color{blue}10.7}
  \end{overpic}\\
  \begin{overpic}[width=0.5\columnwidth]{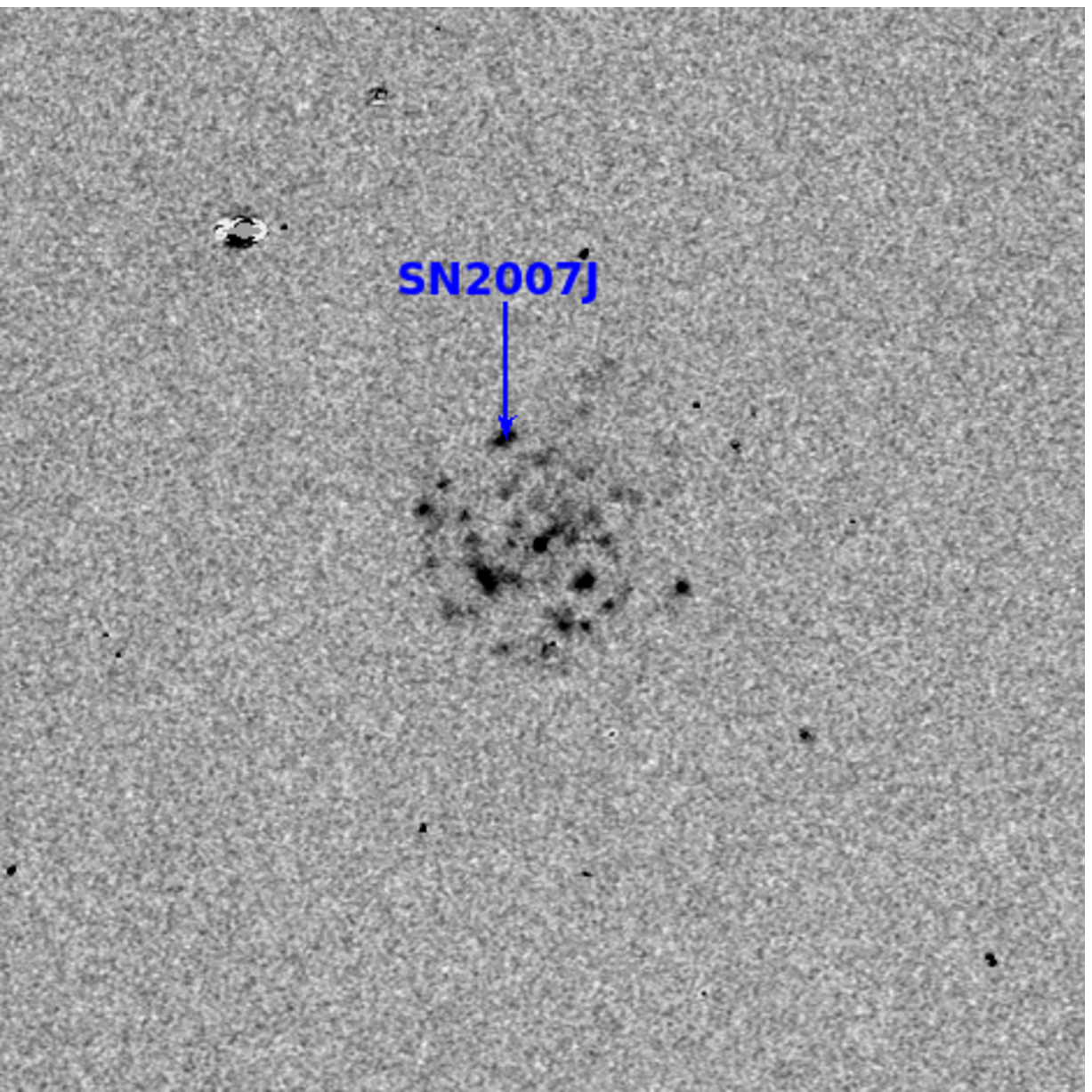}
  \end{overpic}
  \begin{overpic}[width=0.5\columnwidth]{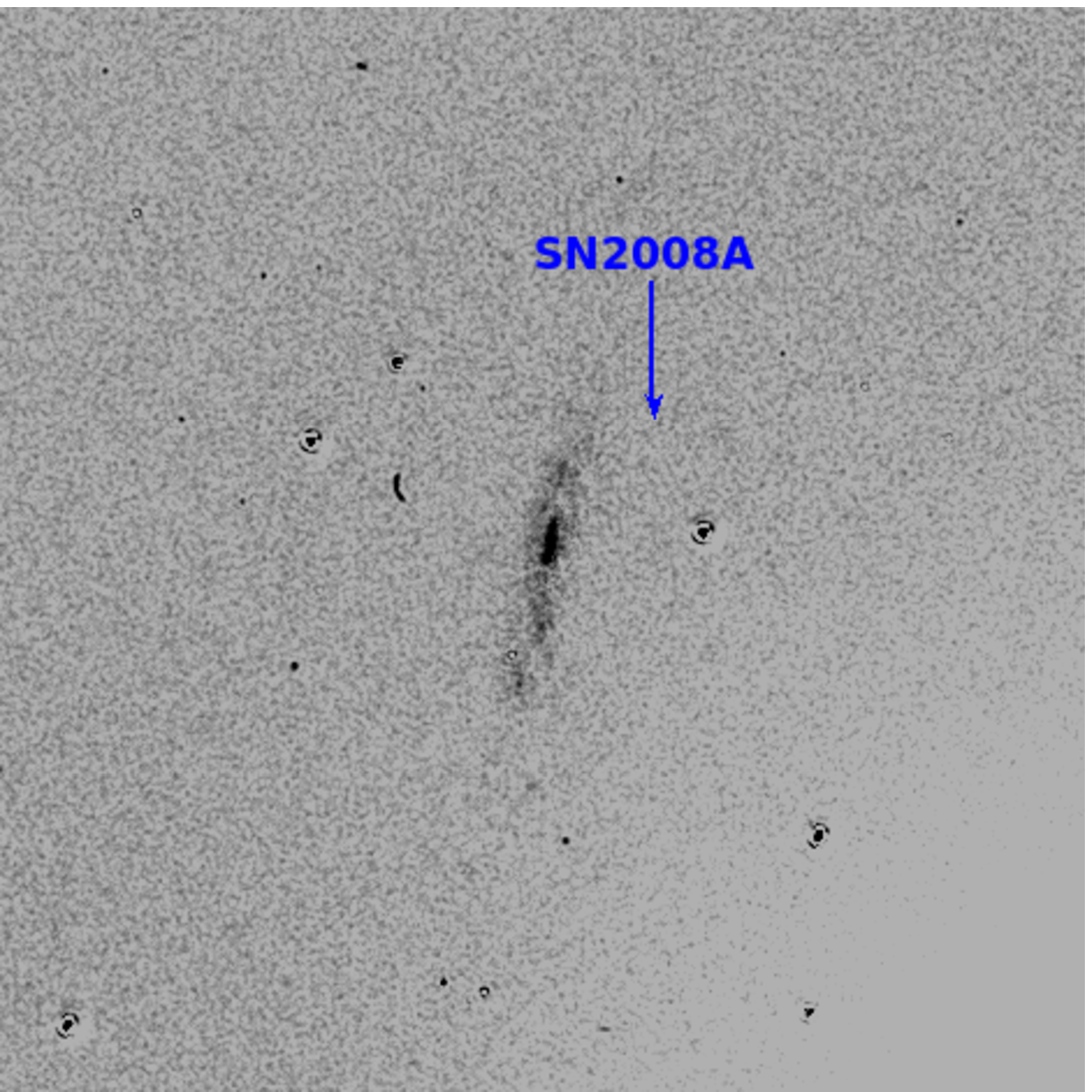}
  \end{overpic}
  \begin{overpic}[width=0.5\columnwidth]{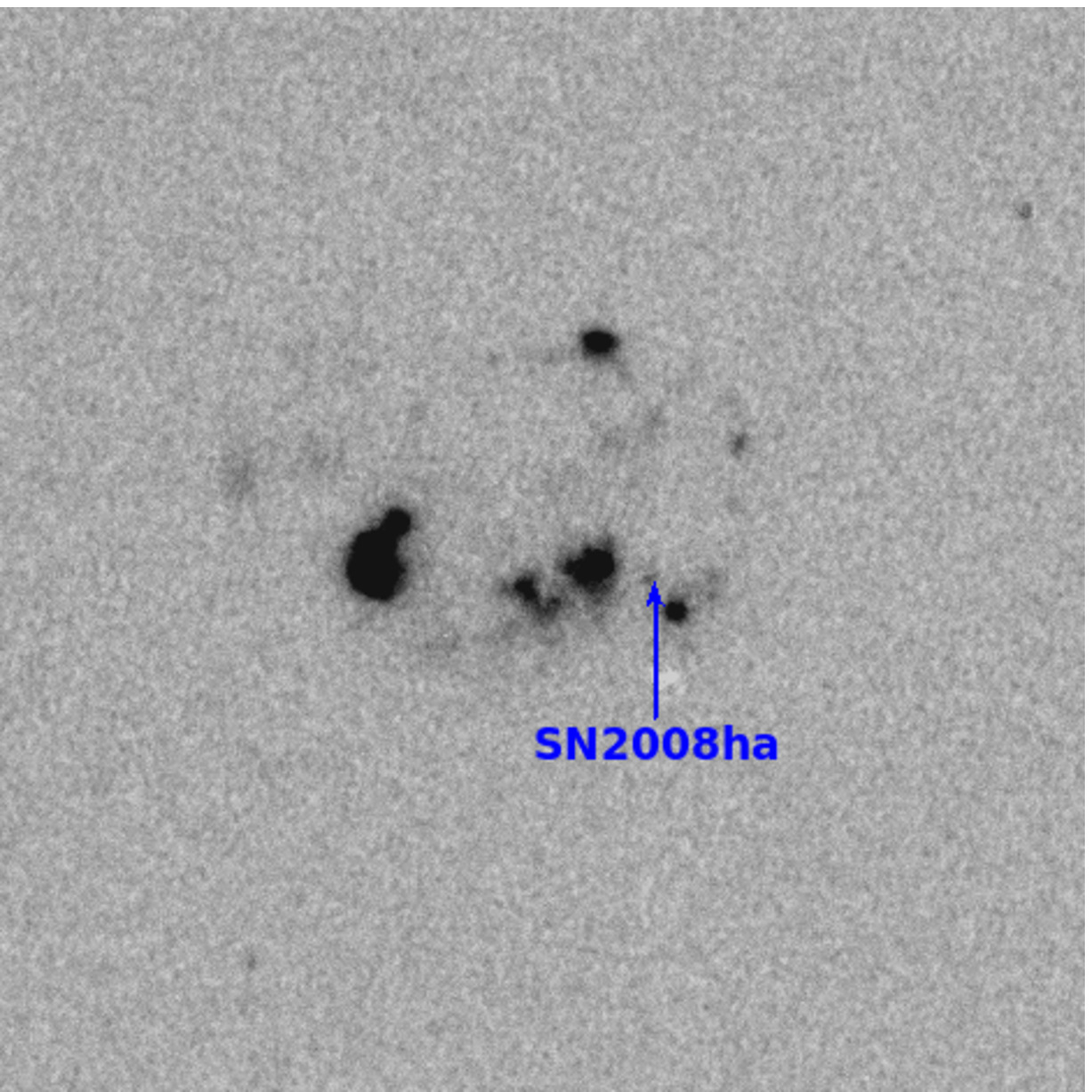}
  \end{overpic}\\
  \vspace{0.2cm}
  \begin{overpic}[width=0.5\columnwidth]{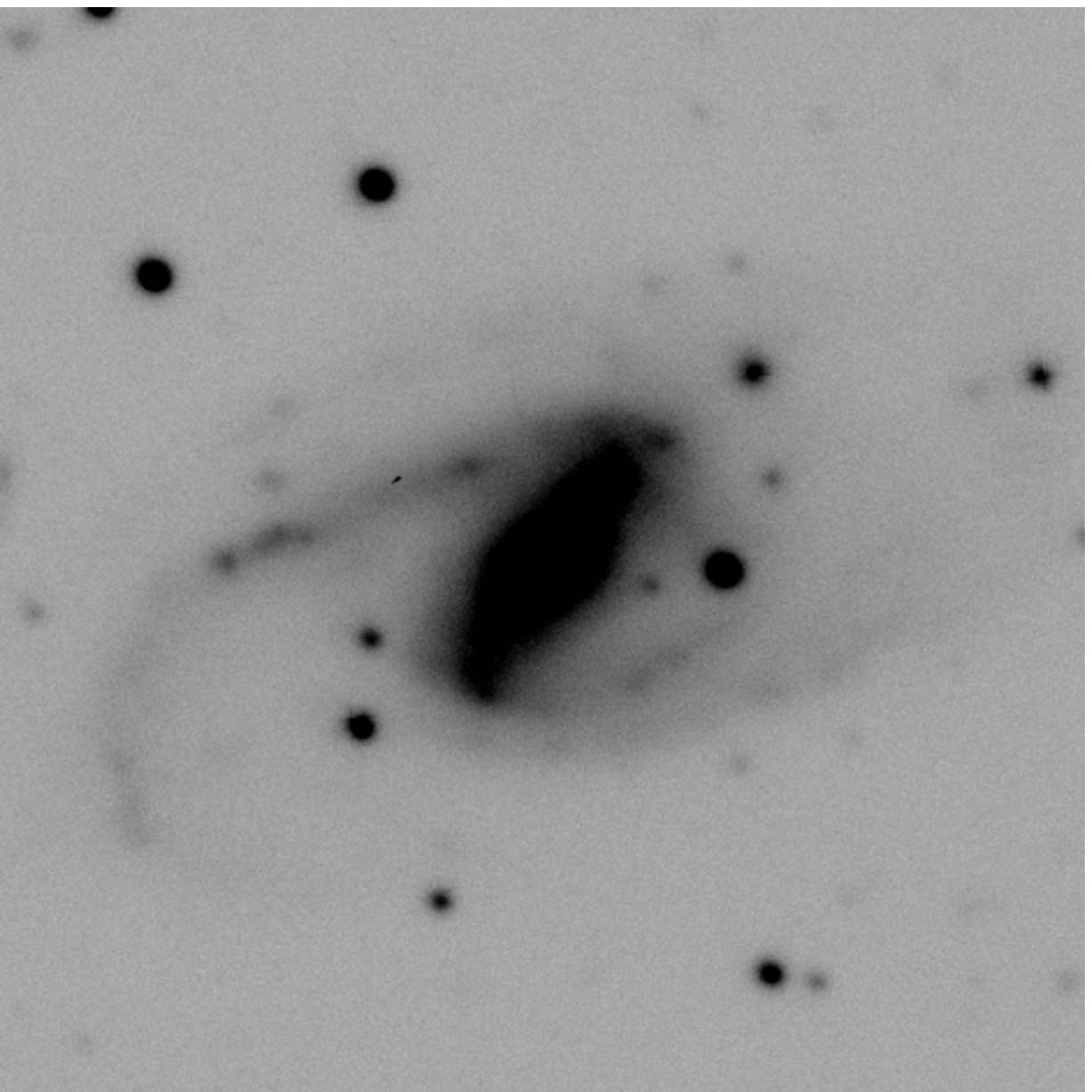}
  \put(10,10){\color{blue}\line(1,0){25}}
  \put(10,12){\color{blue}8.3}
  \end{overpic}
  \begin{overpic}[width=0.5\columnwidth]{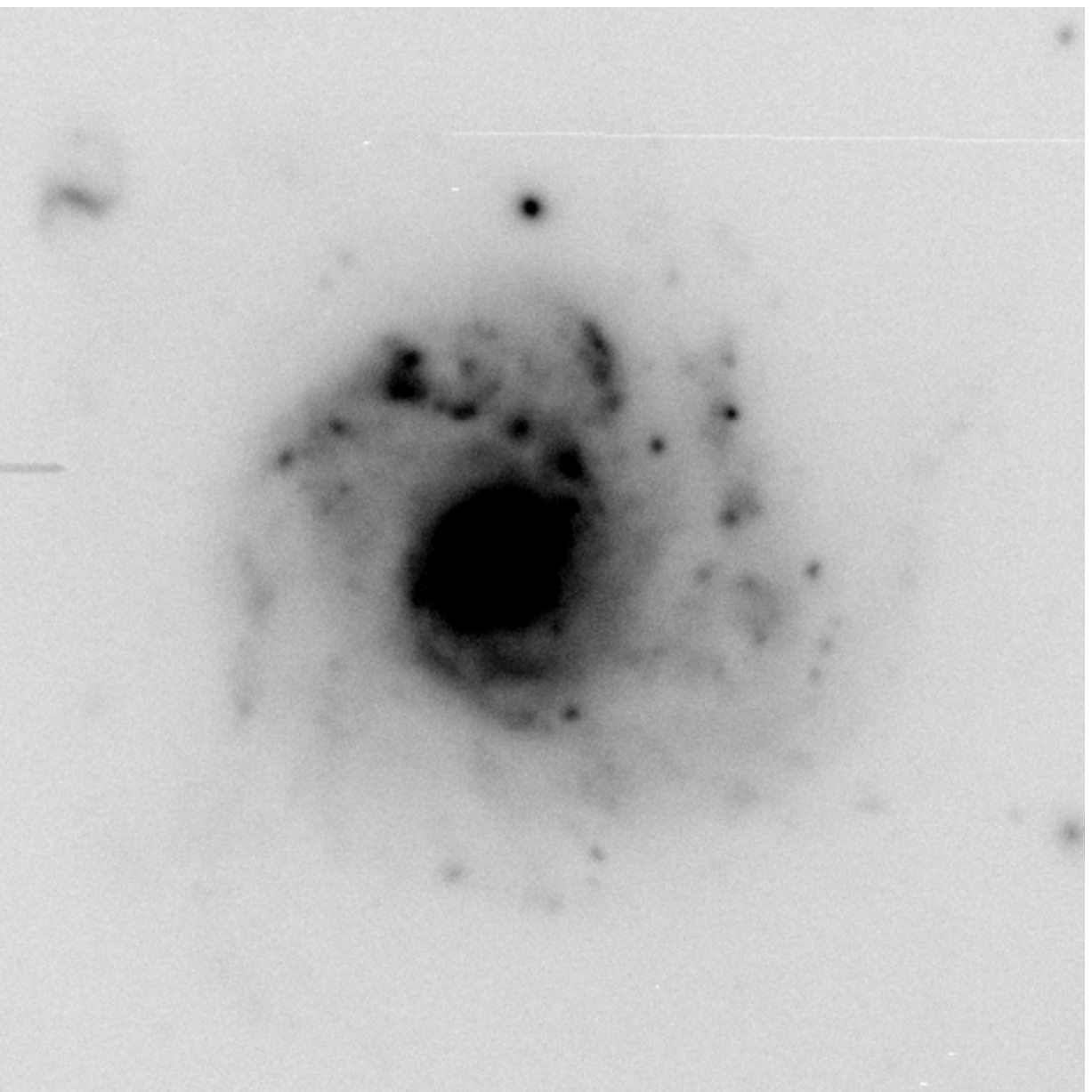}
  \put(10,10){\color{blue}\line(1,0){24}}
  \put(10,12){\color{blue}4.0}
  \end{overpic}
  \hspace{0.495\columnwidth}\hbox{}\\
  \begin{overpic}[width=0.5\columnwidth]{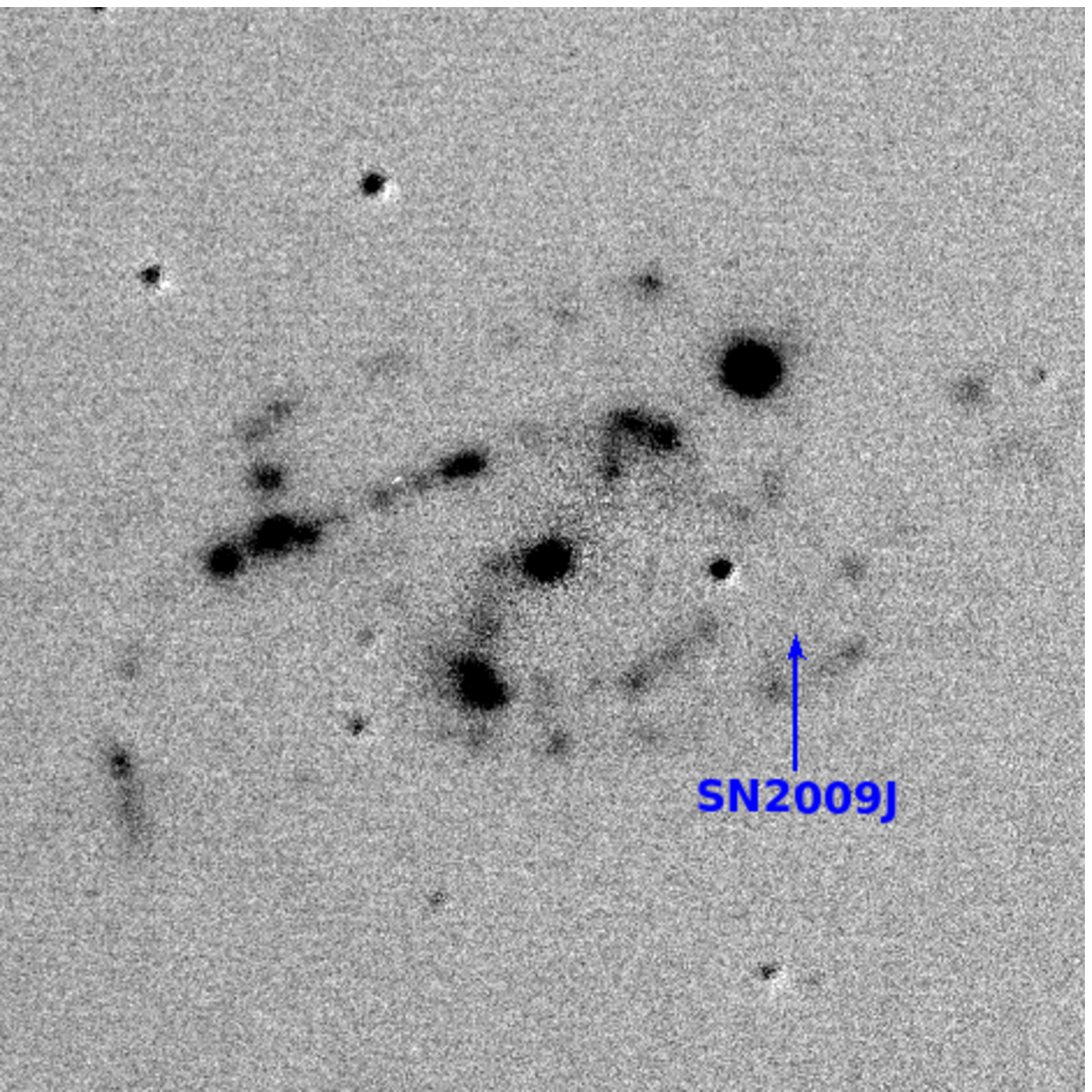}
  \end{overpic}
  \begin{overpic}[width=0.5\columnwidth]{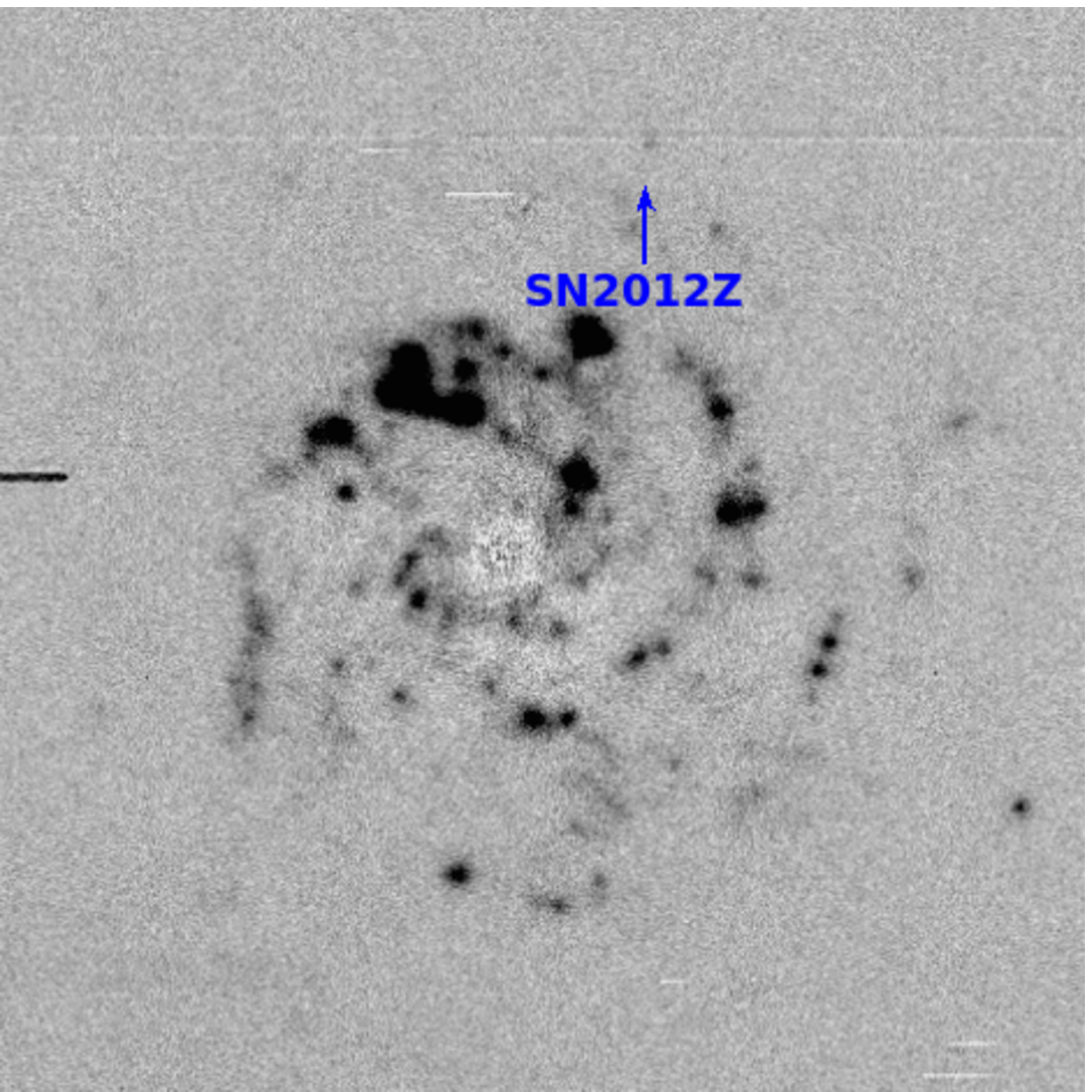}
  \end{overpic}
  \hspace{0.495\columnwidth}\hbox{}\\
  \end{center}
  
  \caption{cont.}
\label{fig:2008ha_imgs2}
\end{figure*}

\begin{figure*}
  \begin{center}
  \includegraphics[width=80mm,angle=0]{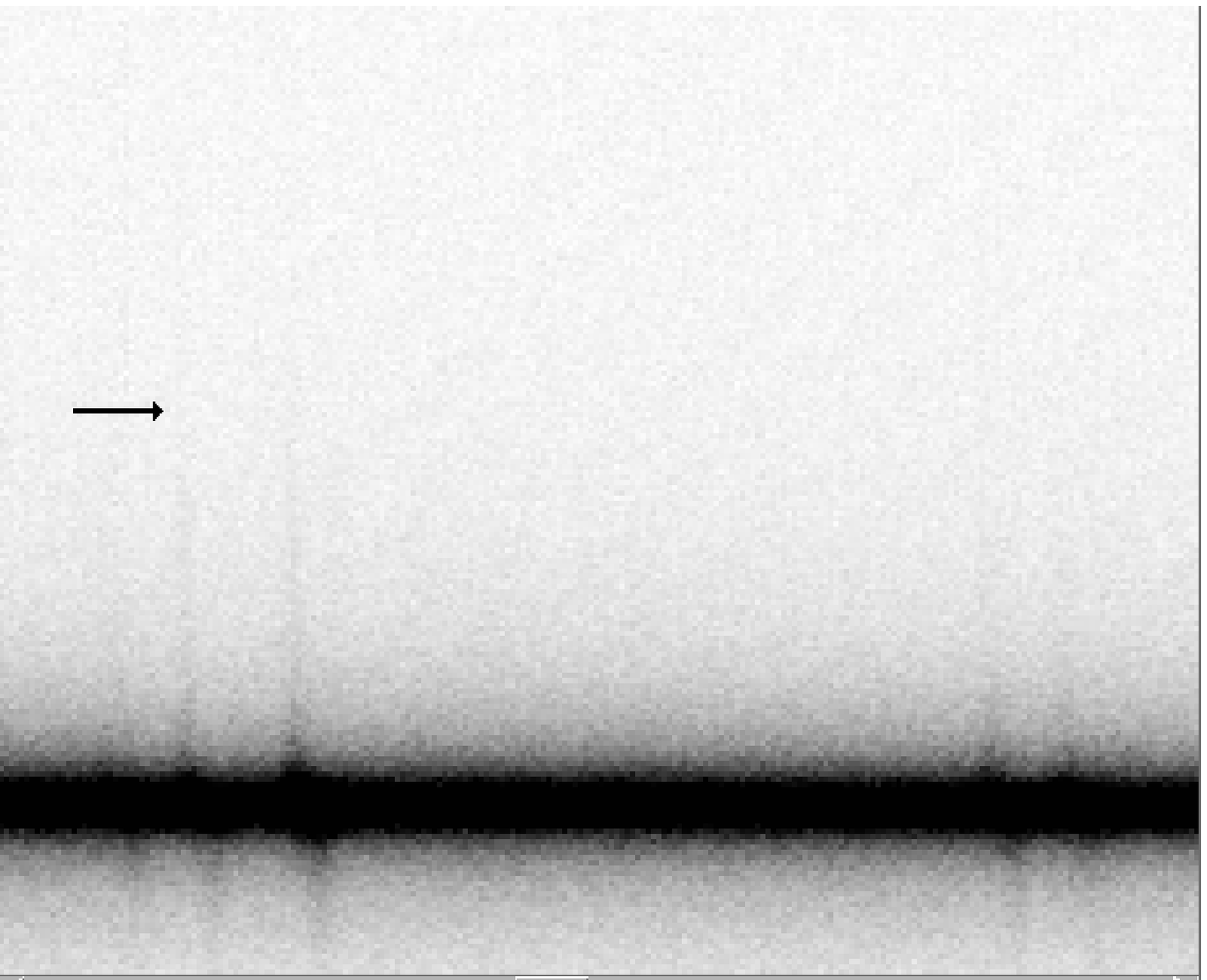}
  \includegraphics[width=80mm,angle=0]{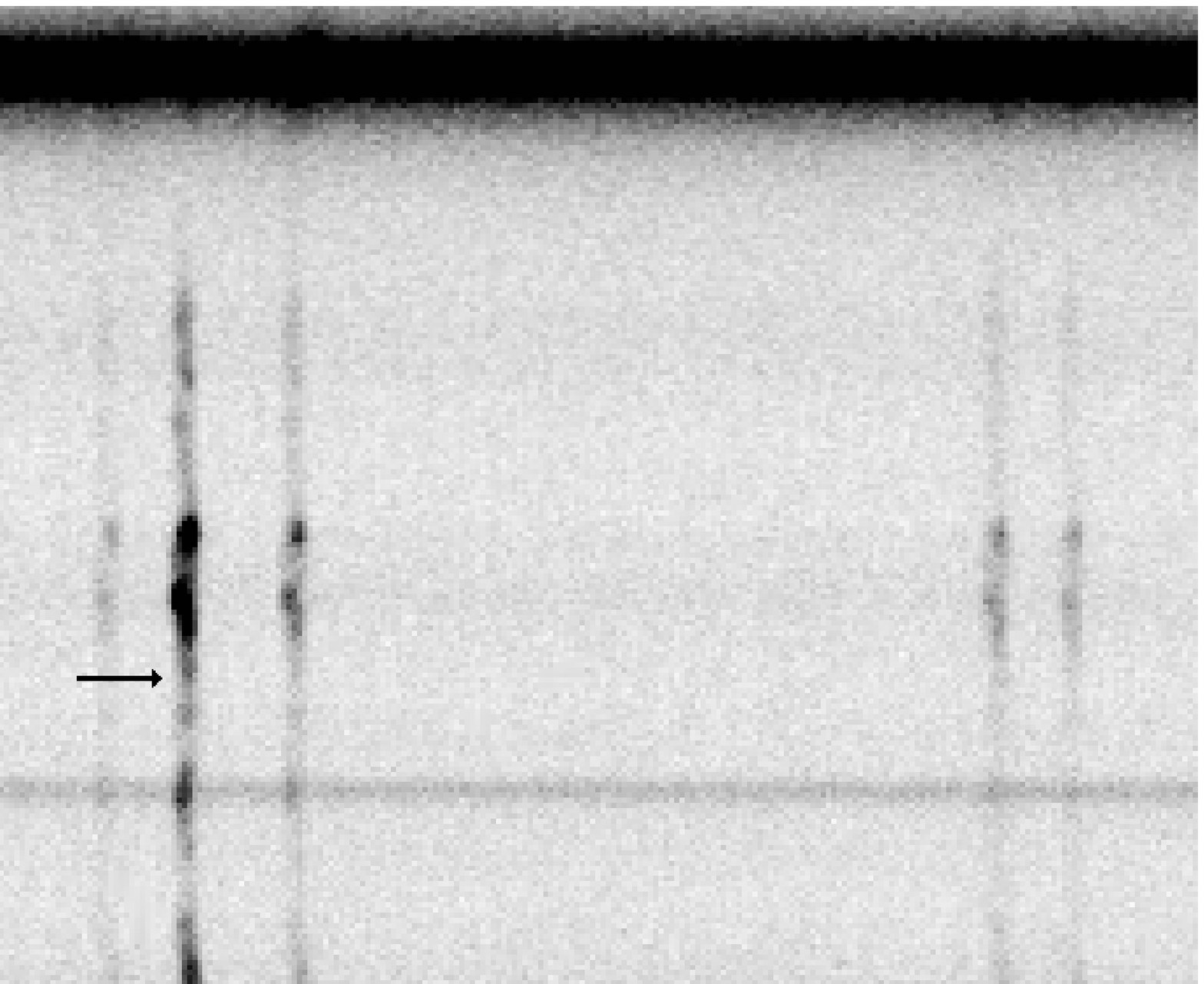}\\
  \includegraphics[trim = 22mm 0mm 10.2mm 140mm, clip, width=80mm,angle=0]{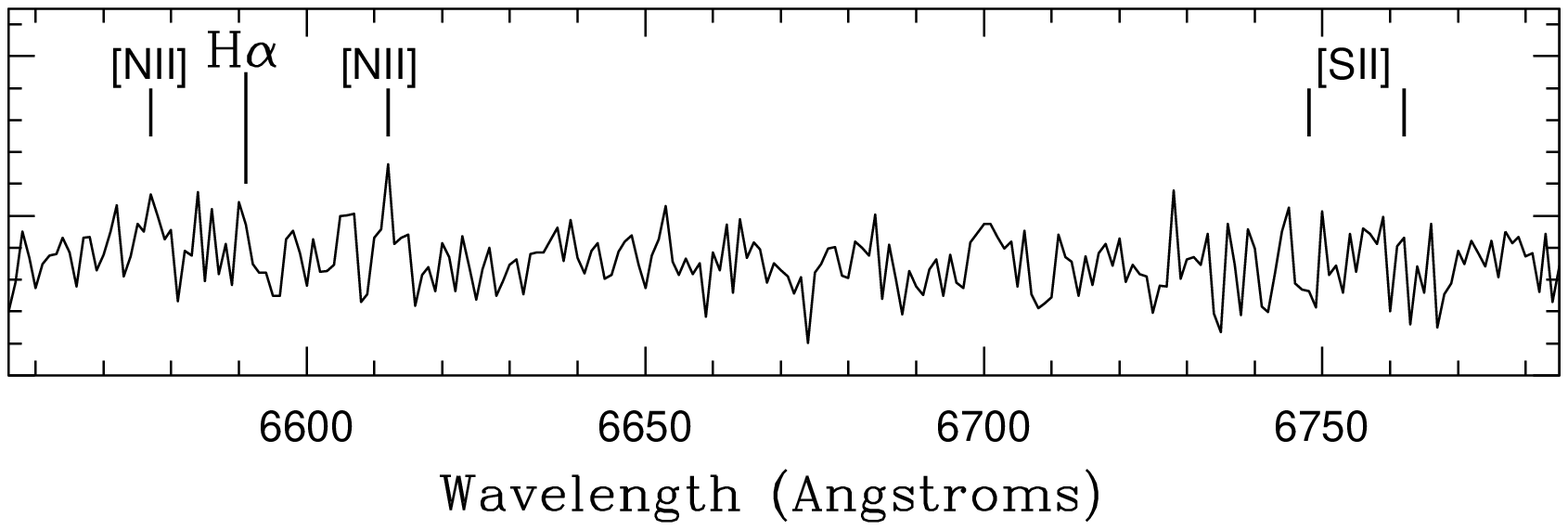}
  \includegraphics[trim = 22mm 0mm 10.25mm 140mm, clip, width=80mm,angle=0]{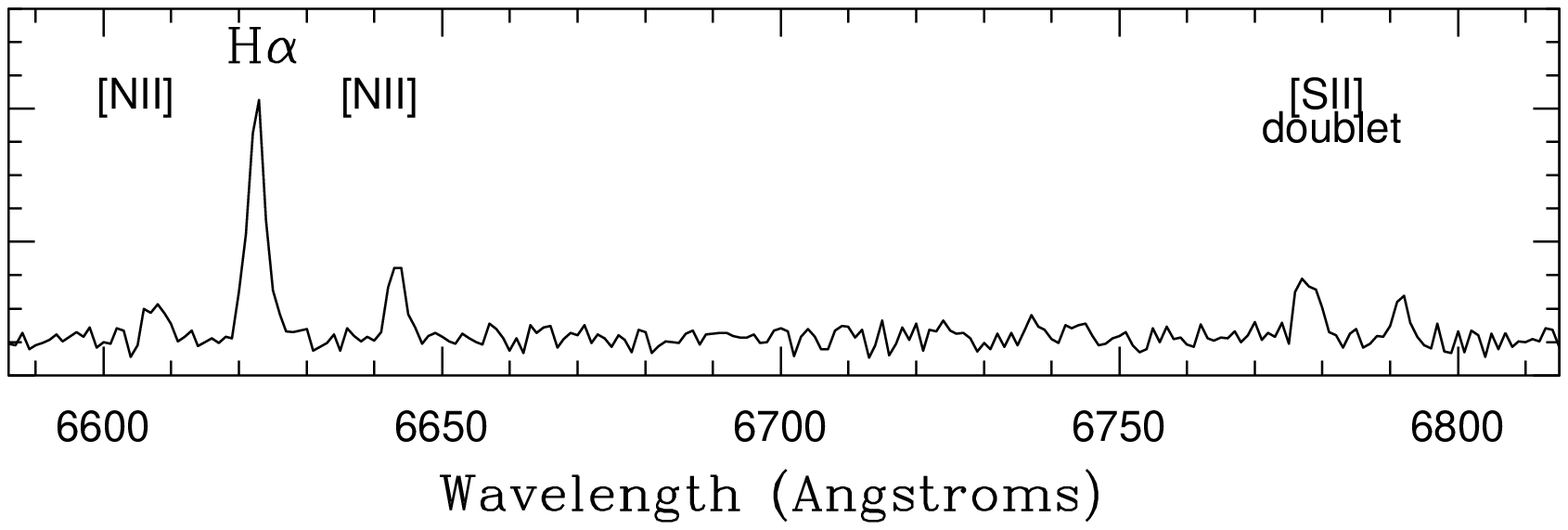}
  \end{center}
  \caption{Spectra showing the H$\alpha$ region of the long slit spectra of SN2000ds in NGC~2768 (left) and SN2003H in NGC~2207 (right).  The arrows indicate positions corresponding to H$\alpha$ emission from the location where each transient occurred. A 1D spectrum is extracted for each at the position of the transient and shown below in relative flux. SN2000ds and SN2003H are located 33.4$^{\prime\prime}$ and 51.4$^{\prime\prime}$ away from their respective hosts' nuclei.}
\label{fig:2000ds2003H_spec}
\end{figure*}

\section{Individual properties of the transients and their environments}

\subsection{Ca-rich transients}

\noindent
{\bf SN2000ds in NGC~2768.}
NGC~2768 is classified as an E6 galaxy in NED, and in the Third Reference Catalog \citep{deva91}. This classification is discussed by \citet{hako08}, who ultimately prefer a classification of S0. As expected, we find no \Ha{} in our obervations (apart from the region affected by subtraction artefacts at the very centre) indicating a lack of recent star formation at the transient location, or indeed anywhere within this host galaxy.  Our INT+IDS long-slit spectrum crossing the nucleus of NGC~2768 and the location of SN2000ds is shown in Fig.~\ref{fig:2000ds2003H_spec}, confirming the lack of any line emission close to the location of the SN.  There is weak, diffuse line emission in H$\alpha$ and [NII] in the central regions of the galaxy, far from the SN location, that is probably related to the known LINER nucleus of this galaxy.

\noindent
{\bf SN2001co in NGC~5559.}
An inclined spiral galaxy, NGC~5559 displays prominent star formation throughout the disc. SN2001co is located near the edge of the disc and is coincident with some diffuse star formation.

\noindent
{\bf SN2003H in NGC~2207.}
NGC~2207 is a close interaction with Sc galaxy IC~2163 at 2765~\kms; SN2003H lies immediately between the bulges of the two galaxies on an area of intermediate-level \Ha. For the purposes of the NCR analysis, the pixels used included those from both galaxies since we cannot cleanly distinguish them as separate systems. As such, SN2003H's NCR value is relative to the interacting system as a whole. A long-slit spectrum crossing the nucleus of NGC~2207 and the location of SN2003H is shown in Fig.~\ref{fig:2000ds2003H_spec}, showing that there is clearly detectable star formation at the location of SN2003H, although it appears to lie in the outer regions of a star-formation complex. The interacting system of NGC~2207 and IC~2163 has also hosted SNe 1975A (Ia), 1999ec (Ib) and 2010jp (IIn).

\noindent
{\bf SN2003dg in UGC~6934.}
The host displays strong HII regions along its highly inclined disc. SN2003dg appears to be somewhere in the plane of the disc, but due to line of sight effects it cannot be determined where in the disc it lies. This means the NCR value may not be accurate (see Sect. \ref{sect:ncrdiscuss}). From the projected view, SN2003dg is coincident with some fairly bright \Ha{} emission.

\noindent
{\bf SN2003dr in NGC~5714.}
SN2003dr occurred in another galaxy that is viewed almost exactly edge-on, but in this case the transient location lies well outside the plane of the disc. Thus we can be more confident to say that it is in a region of no recent star formation. The only apparent star formation in NGC~5714 is diffuse and concentrated along the plane.

\noindent
{\bf SN2005E in NGC~1032.}
NGC~1032 is an S0/a galaxy, and we find no \Ha{} along the plane of the disc lending weight to the argument this is a lenticular galaxy. The host is edge-on and the transient well separated from the disc plane with no \Ha{} evident at its location.

\noindent
{\bf SN2005cz in NGC~4589.}
NGC~4589 is classified as an E2 elliptical galaxy in NED, and in the Third Reference Catalog \citep{deva91}.  \citet{moel89} find unusual central kinematics, and a minor axis dust lane, which they interpret as the result of merging activity.  However, they conclude from the regular shape, and a smooth light profile that follows the classic $R^{\frac{1}{4}}$ profile characteristic of elliptical galaxies \citep{deva48}, that `the merging already is in an advanced state'. As with NGC~2768, in the very bright central region we observe a saturated core with subtraction residuals that accounts for the apparent \Ha{} emission seen in the continuum-subtracted image. No other detected star formation is seen from the host, as is expected if we accept its E2 morphology. The transient is located fairly close to the centre of the galaxy, although it is still outside the region of subtraction residuals.

\noindent
{\bf SN2007ke in NGC~1129.}
The central excess in the continuum-subtracted frame may again be due to saturation effects, although it is less clear in this case. However, SN2007ke is very distant from the centre of this halo on a location of no detected \Ha{}. (Note that the bright spot nearest SN2007ke in the subtracted frame is a foreground star residual and was masked prior to NCR analysis.) Although NGC~1129 is the proposed host, clearly seen between this galaxy and the transient is another galaxy, MCG+07-07-003, at \vrec{}~$= 4967$~\kms. Due to the similarity of the velocities of the two galaxies, the chosen narrowband filter would have detected any \Ha{} from both these galaxies, so we can be confident that we are not missing potential star formation from MCG+07-07-003 (the location of MCG+07-07-003 was included in the NCR analysis since it lies between the putative host and the transient). MCG+07-07-003 appears to be an elliptical galaxy from our imaging, possibly of the compact cE type, and so it is immaterial whether this galaxy or NGC~1129 is adopted as the host for the discussion of the statistics of host types in Sect. \ref{sect:ncrdiscuss}. 

\noindent
{\bf PTF09dav.}
The most distant transient in our sample, this could prove a problem for the NCR method when trying to compare consistently with the other, much nearer examples where the resolving distance at the host will be much smaller. However the extreme separation of the transient from the host negates this problem and we detect no \Ha{} anywhere near the transient, though there is clear star formation in the disc of the putative host galaxy $\sim$40~kpc away. \citet{kasl12} present a limiting magnitude of $M_{R}\sim-10$ for any underlying dwarf host at the location of the transient.

\noindent
{\bf SN2010et.}
SN2010et~$=$~PTF10iuv was discovered by PTF in a very isolated location, with no obvious host galaxy.  Our images, shown in Fig.~\ref{fig:carich_imgs2}, contain three galaxies which probably constitute a small galaxy group, since they have similar recession velocities.   These are 6997~\kms{} for the elliptical galaxy toward the right-hand edge of the images, 7132~\kms{} for the faint edge-on spiral galaxy, and 7407~\kms{} for the brighter spiral toward the left-hand edge of the images. The brighter spiral is the only galaxy in the frame to show evidnce for \Ha{} emission, and hence for ongoing star formation, but this galaxy is very remote from the location of SN2010et. The elliptical galaxy is marginally the most likely host, given its luminosity and somewhat lower (but still subtantial) projected distance. However, it would be very misleading to claim any strong preference for a host galaxy in this case, and so SN2010et is omitted from the analysis of host galaxy types presented later in this paper. The NCR index for the location of SN2010et is unsurprisingly 0.000, i.e. consistent with an empty `sky' location. The limiting magnitude for an underlying dwarf galaxy is estimated as $M_R\sim-12$  by \citet{kasl12}.

\noindent
{\bf PTF11bij in IC~3956.}
Another relatively distant example, the transient is located 33~kpc from IC~3956, an elliptical galaxy that displays no definite \Ha{} emission in the continuum-subtracted image. The recession velocity, unfortunately for this study, lies in the overlap region of the transmission curves of two \Ha{} filters, where both have transmissions of about half of their peak values. The filter with the slightly better transmission at the \vrec{} of IC~3956 was chosen, however we must attach a strong caveat to the analysis of this transient as there is a possibility that our observations miss potential \Ha{} emission. Regardless of this problem, the remote location around an early type galaxy would indicate an unlikely place for significant star formation and hence \Ha{} emission. \citet{kasl12} present a limiting magnitude of $M_{R}\sim-12.5$ for any underlying dwarf host.

\noindent
{\bf SN2012hn in NGC~2272.}
The host galaxy is an early type (SAB0 in NED, E/S0 in HyperLeda\footnote{\url{http://leda.univ-lyon1.fr/}}) with no detected star formation in our imaging, although it should be noted that the recession velocity of the host galaxy would put any H$\alpha$ emission only just within the half-peak transmission limit of the filter used.  The transient location lies well away from the nucleus, and no emission is seen close to its location in either our broad or narrow-band imaging.

\subsection{SN2002cx-like transients}

\noindent
{\bf SN1991bj in IC~344.}
IC~344 is a spiral galaxy showing clumpy star formation in strong HII regions. SN1991bj lies on a region of weak \Ha{} emission within the disc.


\noindent
{\bf SN2004gw in CGCG~283-003.}
Weak star formation is displayed throughout the disc, apart from the southerly arm, which displays several bright areas of \Ha{} emission. The bulk of the star formation is centrally located. The transient location is close to regions of very diffuse \Ha{} emission, but is not coincident with any.

\noindent
{\bf SN2005P in NGC~5468.}
Clumpy \Ha{} structure with some regions of extremely intense star formation are seen in this face-on spiral. SN2005P is located on the edge of a fairly bright HII region, although the low NCR value of 0.055 is warranted by the other, intensely bright regions in the host.  NGC~5468 also hosted SN1999cp (type Ia), SN2002cr (type Ia) and SN2002ed (type II-P).

\noindent
{\bf SN2005cc in NGC~5383.}
A strongly barred galaxy, NGC~5383 displays strong \Ha{} emission in the centre of the bar including an intense star burst region. Lower-level, diffuse emission occurs near the ends of the bar and the base of the spiral arms. SN2005cc is located on a bright region on the southern edge of the bulge.

\noindent
{\bf SN2005hk in UGC~272.}
The transient is located towards the outer edge of the host's disc, which displays several regions of strong \Ha{} emission. SN2005hk is located close to some very faint emission but is coincident with an area devoid of any detected flux and thus has NCR~$= 0$.

\noindent
{\bf SN2006hn in UGC~6154.}
Star formation is concentrated around the bar region in this spiral with little elsewhere in the disc. The transient is located on the cusp of a moderately bright HII region.

\noindent
{\bf SN2007J in UGC~1778.}
Star formation is clumpy, spread evenly across nearly all of the disc. SN2007J lies towards the outer edge of the disc coincident with one of the brightest HII regions.

\noindent
{\bf SN2008A in NGC~634.}
NGC~634 is a highly inclined spiral galaxy that shows reasonably strong \Ha{} emission in the central region with weaker emission coming from the disc plane. Line-of-sight effects mean the NCR can potentially be erroneously high, but for SN2008A, NCR~$= 0$, meaning it is likely to indeed be in a region of no star formation. NGC~634 also hosted SN2006Q (type given in the IAU list as `II?').

\noindent
{\bf SN2008ha in UGC~12682.} 
An irregular galaxy, UGC~12682 displays several regions of strong star formation. SN2008ha is located directly on top of a region of moderate \Ha{} emission.

\noindent
{\bf SN2009J in IC~2160.}
A strongly barred spiral showing some clumpy star formation. The transient, SN2009J is near very low level star formation but coincident with a region of no detected \Ha{} emission. IC~2160 also hosted SN2009iw (type Ia).

\noindent
{\bf SN2012Z in NGC~1309.}
Intense regions of \Ha{} emission are observed in the arms of the face-on spiral galaxy NGC~1309. The transient is located far out in the disc of the host, on a region devoid of \Ha{} emission.

\section{Strength of association of transients with ongoing star formation}

\subsection{Host galaxy classifications}
\label{sect:hostdiscuss}
The first indications of the association of these transients with ongoing star formation come from the Hubble types of the host galaxies. For the Ca-rich transients, the most important observation is that six of the eleven for which we have host types arise from early type galaxies (four ellipticals and two lenticulars) which, as expected, are shown by our observations to have no detectable star formation as revealed by \Ha{} emission.  The other five Ca-rich transient hosts are all nominally spiral galaxies; these are bright, star-forming galaxies with types in the range Sb--Scd, which are the types that dominate the overall star-formation rate in the local Universe \citep{jame08}.  One of these five is PTF09dav, which lies at a projected distance of 40.6~kpc from a bright, disturbed star-forming galaxy that was identified as the probable host by \citet{sull11}, and which we have classified as being of type Sb from our imaging. However, the identification of this galaxy as the host is far from certain, given the very substantial projected offset.

The host galaxies of the eleven SN2002cx-like transients are all clearly of star-forming types, ten being spiral galaxies and one a Magellanic-type irregular.  Two are classified as Sa, one as Sab, one as Sb, one as Sbc, one as Sc, two as Scd, one as Sd, one as Sdm and one as Im.

These distributions of host galaxy types can be compared with the expectations for the typical host environments of low- and high-mass stars, picked at random from across the ensemble of all galaxy types, with the important caveat that both sets of transients are likely to be subject to substantial selection biases.  For low-mass stars, a reasonable comparison is with the distribution of total stellar mass across the population of galaxies in the local Universe, which has been estimated by  \citet{driv07a, driv07b}. 
\citet{driv07a} gives the following fractions of stellar mass in different galaxy components: discs 58$\pm$6 per cent; elliptical galaxies 13$\pm$4 per cent; bulges 26$\pm$4 per cent; 3 per cent other.  While the significance is far from compelling, this indicates that the Ca-rich transients are if anything more strongly weighted towards elliptical galaxies than would be expected if they accurately traced the low-mass stellar population. The expectation might be for one to lie in an elliptical host, whereas four are actually found.  For the SN2002cx-like transients, the reverse is true; none of the hosts is an elliptical galaxy. From an inspection of Fig.~\ref{fig:2008ha_imgs}, all occurred within the star-forming disc components of their host galaxies, with the debatable exception of SN2008A, which would be a surprising finding if they follow the distribution of old stellar mass.

\begin{figure}
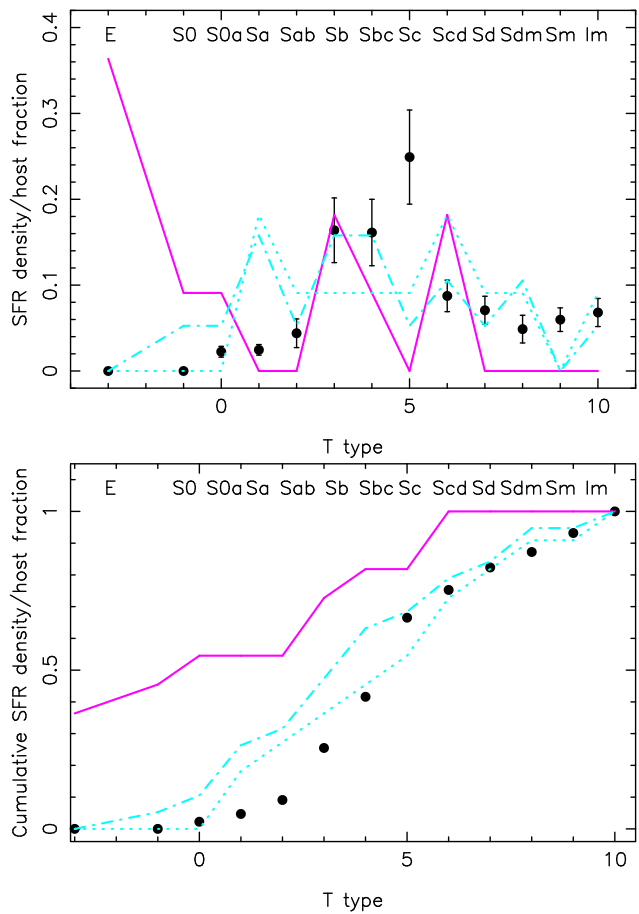

\includegraphics[width=60mm,angle=270]{tsfr_carichplot.ps}
\includegraphics[width=60mm,angle=270]{tsfrc_carichplot.ps}
\caption{The star-formation density of the local Universe as a function
of galaxy T-type (solid circles), compared with the distribution of
types of the host galaxies of Ca-rich (solid lines) and SN2002cx-like (dotted lines) transients. Also shown are the host types of the full sample SN2002cx-like events for which a good host classification can be made (see text; dot-dashed lines).
Both frames show the same data, but quantities plotted in the lower frame are
cumulative values along the sequence from early- to late-type galaxies.}
\label{fig:tsfr_carich}
\end{figure}

To determine the expectations for high-mass stars picked at random from local galaxies, we make use of the estimates of the contributions made by galaxies of different types to the star formation density of the local Universe, in \citet{jame08}.  This comparison is made in Fig.~\ref{fig:tsfr_carich}, where the filled circles in both frames represent the contributions made to the local star-formation rate density by galaxies of the different types.  Thus, Sc galaxies (T-type$=$5) make the largest single contribution, and host about 25 per cent of the current star formation in the local Universe.  The differences between the distribution of Ca-rich host types and those contributing to the star-formation rate density are striking.  The lower frame of Fig.~\ref{fig:tsfr_carich}, where all the quantities are plotted as cumulative, normalized distributions, confirms this  discrepancy for the Ca-rich transients (solid line), which show a large excess of early-type hosts. However, this cumulative distribution comparison shows that the SN2002cx-like transients (dashed line) much more closely match the expectations for a population that traces star formation and hence high-mass progenitors.

The numbers of transients involved in this study are small, and the sample is potentially subject to significant selection effects.  Noting these important caveats, it is still interesting to test the statistical significance of the difference between the distributions shown in Fig.~\ref{fig:tsfr_carich}.  A one-sample Kolmogorov--Smirnov (KS) test gives a critical $D$ value of 0.468 for a sample of eleven objects and a probability $P$ of 0.01; the observed maximum $D$ between the Ca-rich hosts and the star-formation density distribution summed over types is larger than the critical value, at 0.523.  Thus the Ca-rich hosts are significantly different from the expectations for a population that traces cosmic star formation. A two-sample KS test shows that Ca-rich and SN2002cx-like host galaxies differ significantly, with a maximum $D$ value of 0.55 and a probability of only 0.047 that the two sets of host galaxies could be  drawn from the same parent distribution.

It is useful, in the interests of increasing sample size, to look at the host types of other SN2002cx-like events, even those for which we have no \Ha{} imaging. Some events prohibit a confident host classification to be made due to the distance to and/or a lack of high resolution imaging of the host. Additional transients that we include are SN2002bp (UGC~6332, SBa), SN2003gq (NGC~7407, Sbc), SN2004cs (UGC 11001, Sdm), SN2008ge (NGC~1527, SAB0), SN2010ae (ESO~162-17, Sb), SN2010el (NGC~1566, SABbc), SN2011ay (NGC~2315,   S0/a) and SN2011ce (NGC~6708   Sb); see \citet{fole13} for a discussion of each event. As can be seen in Fig.~\ref{fig:tsfr_carich}, this enlarged sample of 19 transients broadly follows the host distribution of our SN2002cx-like sample, as expected, although one transient has an early-type host classification --- SN2008ge in NGC~1527. This may argue against a young age for the progenitor; indeed, \citet{fole10a} find an absence of evidence for recent star formation at the transient's location and conclude that the progenitor is likely to be a WD. SN2008ge is further discussed in Sect. \ref{sect:discuss}

\subsection{Transient locations and ongoing star formation}
\label{sect:ncrdiscuss}
To further quantify the apparent association of the two transient populations with current sites of star formation, we make use of the NCR statistic applied to a pixel-by-pixel analysis of our continuum-subtracted H$\alpha$ images, which was introduced in Sect.~\ref{sect:methods}.   

For the five Ca-rich transients in galaxies with detectable star formation, the mean NCR value is 0.259 (standard error on mean 0.119), range 0.000-0.626. Including the seven for which we have \Ha{} imaging which reveals no star formation anywhere in the galaxy (including SN2010et which has no obvious host), and which all thus have NCR values of 0.000, this value falls to 0.108 (0.060).

The Ca-rich transient with the strongest apparent association with an HII region is SN2003dg. This occurred in the disc plane of UGC~6934, an edge-on Scd spiral galaxy.  This galaxy orientation makes the interpretation of the NCR index highly ambiguous, with a greatly increased probability of line-of-sight projection effects resulting in spurious apparent correlations, and large, poorly-constrained extinction effects.  Thus, edge-on galaxies were excluded from the NCR analysis in all of our previous papers \citep{jame06, ande08, ande12}, with a limiting criterion of a major- to minor-axis ratio of 4.0.  For UGC~6934, this ratio is 7.7, so it would have been excluded from our earlier studies.  Removing SN2003dg from the sample, the mean NCR value falls to 0.061 (0.041).

Two more of the Ca-rich transients, SN2003dr and SN2005E, occurred in disc galaxies that are very close to being exactly edge-on.  In these cases, the transient occurred far from the disc plane, so the line-of-sight projection argument does not apply.  However, for the sake of consistency we recalculate the average NCR value with all three edge-on hosts removed, giving an average for the remaining nine of 0.074 (0.049). 

For the eleven SN2002cx-like transients, the mean NCR value is 0.222 (0.092), range 0.000 - 0.904. Removing SN2008A, which occurred in the edge-on Sa galaxy NGC~634, this mean is 0.244 (0.099).

\begin{figure}
\includegraphics[width=90mm,angle=0]{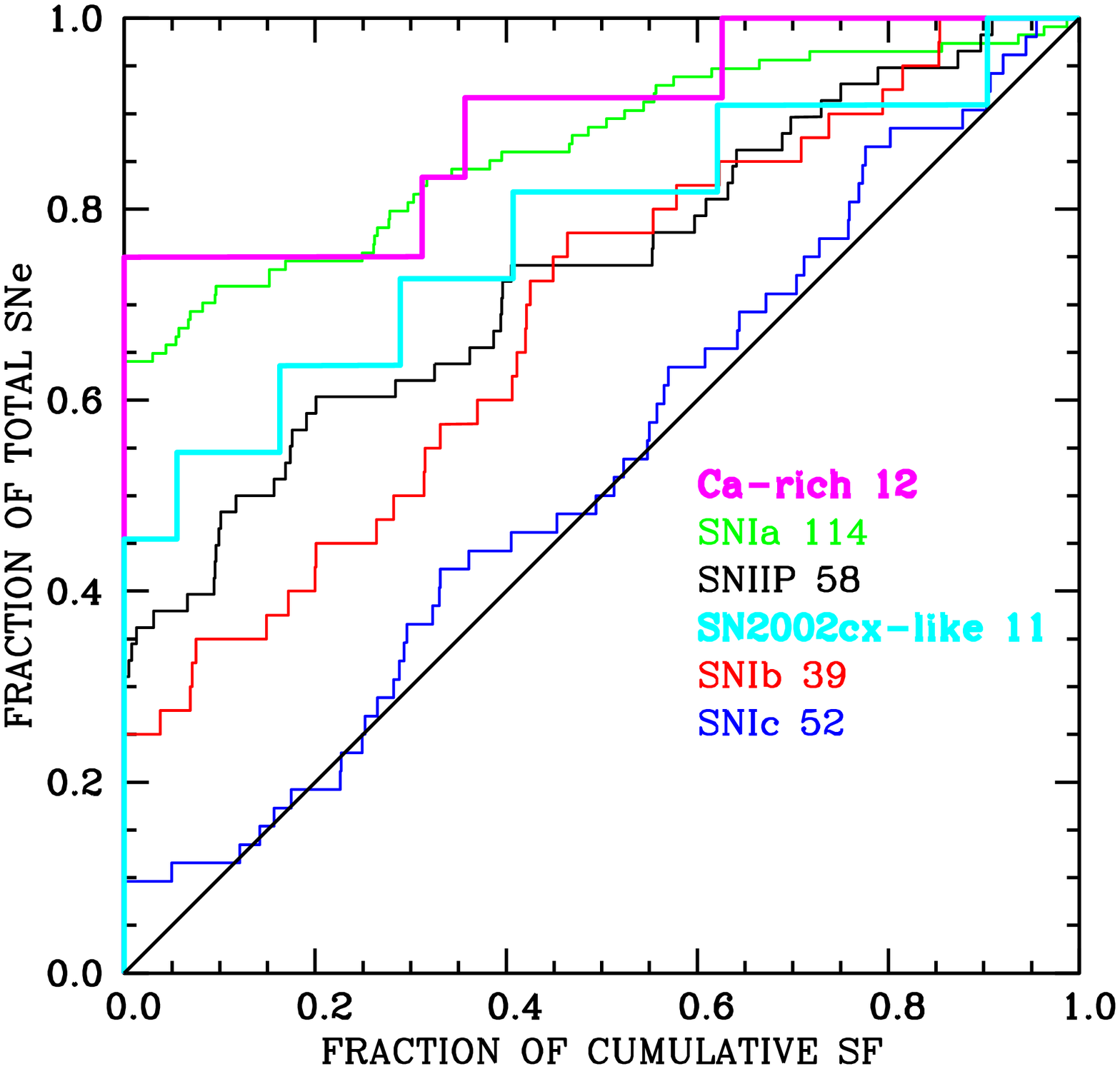}
\caption{Cumulative distribution of the Ca-rich and SN2002cx-like samples as compared to other SN types. Data for other SNe are based on \citet{ande12}, with a correction factor applied to the SNIa sample to account for the fact that only star-forming galaxies were included in that study (see text).}
\label{fig:ncrplot}
\end{figure}

Figure \ref{fig:ncrplot} shows the distributions of NCR values for our samples and of other well-known SN classes. The values for Ia, II-P, Ib and Ic types are taken from \citet{ande12}. The study of \citet{ande12} was of only star-forming galaxies however, whereas we have six early type hosts in the Ca-rich transient class that display no detected star formation. In order to compensate for this fact, we use the rate of SNeIa going off in early type galaxies from the Lick Observatory Supernova Search \citep{li11}, found to be $\sim$27 per cent. With no star formation, any location in the host will have NCR~$= 0$, so by adding 27 per cent to the SNeIa sample as NCR~$= 0$ events, this will give an expected SNeIa distribution across all galaxy types. The number of CCSNe that have been observed in non-star-forming early type hosts is $\sim$0, and as a result no correction was applied to the SN II, Ib or Ic distributions. 

We have applied the KS test to the distributions of NCR values shown in Fig.~\ref{fig:ncrplot}.  These confirm the extreme nature of the environments of the Ca-rich transients; even with a small sample (12 objects), the test conclusively shows that the values are not consistent with a distribution that perfectly traces the star-formation activity in the host galaxies (the black diagonal line in  Fig.~\ref{fig:ncrplot}), with a probability of these distributions being consistent of $<0.1$~per cent.  More importantly, the Ca-rich transient NCR values are inconsistent with the distributions for SNe of type II (including all subtypes and unclassified type II; see \citealp{ande12}) and Ib, with probabilities less than 2.5 per cent of being drawn from the distributions of either type; the consistency with the \mbox{II-P} sample shown in Fig.~\ref{fig:ncrplot} is $\sim$5~per cent. Notwithstanding the small sample size, a clear distinction exists between Ca-rich transients and SNeII/Ib.

The Ca-rich transient distribution is completely consistent with that of SNeIa, and by eye the two distributions overlay very closely, given the constraints of small number statistics. For the SN2002cx-like transients, the situation is less clear; formally the NCR values of these eleven transients could have been drawn from any of the other distributions shown in Fig.~\ref{fig:ncrplot}.  However, the distribution most closely approximates that of the type II-P SNe, and indeed, again considering the small number statistics, reproduces it well. The distribution of SN2002cx-like transients shows a stronger association to star formation than that of SNeIa, with the mean NCR being larger even in the case of considering only star-forming hosts of SNeIa (see Table \ref{tab:NCR_means}).

\begin{table*}
 \centering
 \begin{minipage}{140mm}
  \caption{Mean NCR values for the locations of different SN types.}
  \begin{tabular}{lrll}
  \hline
SN Type  & No. & Mean   & Std. err. \\
\hline            
Ca-rich &  9  & 0.074\footnote{All edge-on hosts excluded.} & 0.049 \\
Ca-rich & 12  & 0.108 & 0.060 \\
Ia      & 98  & 0.114\footnote{Corrected for an assumed early-type fraction of 27 per cent; mean in star-forming hosts is 0.157.} & 0.019 \\
SN2002cx-like  & 11  & 0.222 & 0.092 \\
SN2002cx-like  & 10  & 0.244\footnote{Edge-on host excluded.}  & 0.099 \\
II-P    & 58  & 0.264 & 0.039 \\
Ib      & 39  & 0.318 & 0.045 \\
Ic      & 52  & 0.469 & 0.040 \\
\hline
\end{tabular}
\label{tab:NCR_means}
\end{minipage} 
\end{table*}

\section{Discussion}
\label{sect:discuss}
Though the total rate of Ca-rich and SN2002cx-like transients might be significant compared to type Ia SNe, the actual number of observed events is still small, mostly due to their low luminousity which limits their detection at large distances. For this reason the sample sizes in this study are limited. Nevertheless, a clear picture emerges from our results, pointing to significant differences between the host environments of these two transient types, which, in turn, implies different types of progenitor systems. The clear distinction between the two classes is strengthened by the possible contamination from misclassified transients in each sample. Such contamination would serve to dilute any distinct behaviour between the two samples.

The first indications for such difference come from the host galaxy populations analysis. All SN2002cx-like transients have host galaxies that display strong, recent star formation activity. The progenitor systems are therefore likely associated with a young stellar population, quite similar to that of CCSNe. Conversely, six of the eleven Ca-rich hosts (diregarding SN2010et, where the host is not certain) are early-type galaxies with no detected star formation, and therefore point to an old stellar population lacking any young, massive stars. 

Our host galaxy distributions provide strong support to the suggestion of \citet{pere10} of an old progenitor system for Ca-rich transients. Their original analysis of a smaller sample of events, showed the host galaxy distribution of various SN types compared with the Ca-rich events. The distribution of Ca-rich transient hosts displays similarities with that of regular SNeIa, a trend strengthened by the addition of similar events identified since then presented here.

Furthermore, our NCR analysis allows the locations of the transients \emph{within} their respective hosts to be investigated. More than simply saying the SN2002cx-like transients are found in hosts that display ongoing star formation, we quantitatively find a good match between SN2002cx-like events and SNeII-P with respect to association with recent star formation in their host galaxy. Such a match would indicate a similar progenitor age for SN2002cx-like transients and SNeII-P (i.e. a typical delay time of $30-50$ Myrs). From the NCR analysis we confirm that Ca-rich transients do not appear to follow recent star formation in their hosts and closely resemble the distribution of `normal' SNeIa, whose progenitors are expected to have significant life times ($\sim$~Gyrs).

The samples are, as mentioned previously, inherently eclectic and suffer many biases relative a volume-limited sample. Their fainter magnitudes compared to SNe in general would suggest that they will be difficult to detect on bright galaxy regions. We note, however, that SN2003dg (Ca-rich) and SN2005cc (SN2002cx-like), both typical of the mean brightness of their sample, were discovered on the brightest central regions of their respective hosts. The preference for discovery in fainter host locations would strengthen the arguement for SN2002cx-like events' association with star-formation, given it is plausible to miss some of these events if they are coincident with the brightest HII regions. The discovery magnitudes quoted in Tables \ref{tab:carich_props} and \ref{tab:2008ha_props} show there is no statistically significant difference between distribution of brightnesses in each sample, suggesting any bias from magnitude-limited searches will affect each sample similarly (although their faintness will possibly affect the comparison to `normal' SN types).

Our analysis provides new clues regarding the origin of these peculiar transient events, and can help constrain the suggested theoretical models. In the following we discuss these constraints in view of the suggested theoretical models for these transients.  

\emph{Ca-rich transients:} Several models were suggested for the origin of the Ca-rich events. The model of He-shell detonation on a CO WD, following He accretion from a He-WD, was first suggested by \citet{pere10} and gained additional support from the theoretical analysis by \cite{she09} and \citet{wald11}. Such a model points to a double degenerate origin for these types of transient. In particular \cite{wald11} suggested a low mass CO WD progenitor, which requires a long lived stellar origin, and possibly a low metallicity environment. An alternative model of a CC origin as suggested by \cite{kawa10} (see also discussion by \citeauthor{kasl12}) would require a young, star-forming environment. Our H$\alpha$-based analysis of the hosts of the Ca-rich transients makes clear that the majority of these are occurring a very long way from any detectable star formation.  This also strengthens the arguments of \citet{pere10} and \citet{kasl12} that even extremely high-velocity, high-mass runaway stars are implausible candidates as  progenitors of the Ca-rich transients. 
We therefore conclude that our analysis consistently points towards old progenitor systems, and a likely thermonuclear origin, for the Ca-rich transients (see additional support through the analysis of \citealt{yuan13}).
\setcitestyle{notesep={; }}

\emph{SN2002cx-like transients:}
Several models were also suggested for the origin of SN2002cx-like events. \citet{li03} and \citet{bra04} suggested they originate from the deflagration of a Chandrasekhar mass C/O WD. This model encounters difficulties explaining the diversity of such events and in particular the extremely low-mass and sub-luminous SN2008ha event. A more recent and detailed model by \citet[][see also \citealp{cal04,liv05,kro13}]{jor12} discusses a failed detonation model, in which a deflagration scenario fails to explode the WD, and only burns and ejects a fraction of the WD, leaving behind an intact (but now lower mass and polluted) WD remnant. This scenario can similarly explain the low velocities observed for SN2002cx-like events due to deflagration, but in addition provides a robust explanation for the diversity of the SN2002cx-like events and the possible production of extremely low-mass and low luminosity events. Both of these models begin with a Chandrasekhar mass WD, similar to the single-degenerate model suggested for type Ia SNe. WDs initially formed at high masses (which in turn form from  higher mass stellar progenitors with shorter lifetimes) and require less additional accretion in order to achieve the Chandrasekhar mass. This would generally point to their association with younger environments, where more massive stars and binaries evolve and transfer mass. However, the evolution towards the Chandrasekhar mass is still expected to be generally longer, and sometimes much longer, than the typical lifetimes of CCSN stellar progenitors ($>$8 M$_\odot$ stars). Although some SN2002cx-like transients have been found in old environments \citep{fole13}, our finding suggest a very young environment for the progenitors of these transients, comparable with that of type II-P SNe. The environmental constraints we find therefore do not exclude, but are less favorable for a Chandrasekhar mass C/O WD explosion.

\citet{fern13} suggest neutron star-WD mergers as a possible origin for SN2002cx-like events. Some of the properties of SN2002cx-like transients are qualitatively reproduced by the model, but more detailed studies are needed. This model would suggest a mixed distribution of old and young environments, due to the distribution of the gravitational wave inspiral time leading to the merger, in contrast with the strong bias to very young environment we find here. In addition, the total rate of neutron star-WD mergers is about 3 per cent of that of SNIa -- even if all such mergers resulted in an SN2002cx-like event, the expected rates would be an order of magnitude lower than those observed \citep{fole13}.

\citet{vale09} suggested SN2002cx-like transients arise from a variant of CCSNe with low ejecta velocity, although currently no detailed theoretical modeling of such events has been done and shown to produce such events. Our findings of similar environments for both these transients and those of CCSNe, are therefore consistent with the CC origin of SN2002cx-like transients. In particular, our detailed NCR statistics indicate that SN2002cx-like events share similar environments to those of SNeII-P, i.e. while they are evidently associated with star formation, a substantial fraction appear to outlive their natal HII regions, resulting in lower values of the NCR index than would be expected for the highest mass progenitors. 
In the context of this scenario, our analysis would therefore point to the lower-mass, 7-9~M$_{\odot}$ (with typical lifetimes of $30-50$ Myrs) progenitors discussed by \citet{vale09}, rather than the alternative high-mass Wolf-Rayet stars also discussed by them.  It is, however, still difficult to explain the complete lack of star-formation/young environment for one of the SN2002cx-like events, SN2008ge \citep{fole10a}. 


Two of the SN2002cx-like transients show spectral evidence for helium. Taken together with the young environment found for these events (beside SN2008ge), \citet{fole13} suggest this as possible evidence for their origin from a helium star accretion on to a WD. However, a helium layer may also form following hydrogen accretion and burning into helium on a WD \citep[][and references therein]{cas98}. We therefore conclude that although the existence of helium in even a small fraction of these events is a potentially important clue for their origin, its interpretation is still inconclusive. 

\section{Summary}

Our investigations of the environments and host types of Ca-rich transients show a lack of association with recent star formation (similar to that of SNeIa), and thus point to old progenitor systems, consistent with helium-shell detonation on low mass C/O WDs, and inconsistent with a CCSN origin. Conversely, we find the SN2002cx-like transients to be well matched by young progenitors (likely $<$50 Myrs lifetime) through an association to star-formation that is similar to that displayed by type II-P SNe. Such young progenitors are less favorable to failed detonations of Chandrasekhar mass C/O WDs, and more consistent with either the core-collapse of a 7-9 M$_\odot$ star, or a WD explosion following the accretion of helium star (note that one event, SN2008ge does not seem to fit with this conclusion). While the failed detonation model for these events appears to be consistent with the observable parameters of SN2002cx-like events themselves, the latter two models currently lack an actual detailed study. Therefore, they can not yet be adequately compared with observations, beyond the generally consistent aspects of their expected environments as studied here.

\section*{Acknowledgements}

This research has made use of the NASA/IPAC Extragalactic Database (NED) which is operated by the Jet Propulsion Laboratory, California Institute of Technology, under contract with the National Aeronautics and Space Administration. We acknowledge the usage of the HyperLeda database. The Liverpool Telescope is operated on the island of La Palma by Liverpool John Moores University in the Spanish Observatorio del Roque de los Muchachos of the Instituto de Astrof\'{i}sica de Canarias with financial support from the UK Science and Technology Facilities Council. The Isaac Newton Telescope is operated on the island of La Palma by the Isaac Newton Group in the Spanish Observatorio del Roque de los Muchachos of the Instituto de Astrof\'{i}sica de Canarias. This research also uses observations made with the ESO 2.2-m telescope at the La Silla Observatory (proposal ID 084.D-0195). JDL acknowledges the UK Science and Technology Facilities Council for research studentship support. HBP is supported by the I-CORE Program of the Planning and Budgeting Committee and The Israel Science Foundation (grant No 1829/12). J.~A. acknowledges support from CONICYT through FONDECYT grant 3110142, and by the Millenium Center for Supernova Science (P10-064-F), with input from `Fondo de Innovacin para La Competitividad, del Ministerio de Economa, Fomento y Turismo de Chile'. A.~G. was supported by grants from the ISF, the EU/FP7/ERC, the Minerva and ARCHES programs, and the Helen and Martin Kimmel Award for Innovative Investigation.


\label{lastpage}

\end{document}